\documentclass{llncs}\usepackage[]{graphicx}\usepackage[]{color}
\makeatletter
\def\maxwidth{ %
  \ifdim\Gin@nat@width>\linewidth
    \linewidth
  \else
    \Gin@nat@width
  \fi
}
\makeatother

\definecolor{fgcolor}{rgb}{0.345, 0.345, 0.345}

\usepackage{framed}
\makeatletter
 {\par\unskip\endMakeFramed%
 \at@end@of@kframe}
\makeatother

\definecolor{shadecolor}{rgb}{.97, .97, .97}
\definecolor{messagecolor}{rgb}{0, 0, 0}
\definecolor{warningcolor}{rgb}{1, 0, 1}
\definecolor{errorcolor}{rgb}{1, 0, 0}

\usepackage{alltt} 

\usepackage{epsfig,endnotes,url}

\usepackage{xspace}

\usepackage[english]{babel}
\usepackage[latin1]{inputenc}
\usepackage{amsmath, amssymb, latexsym} 

\usepackage{wasysym}

\usepackage[ShowComments]{mycomment}

\usepackage{graphicx}
\usepackage[dvipsnames,table]{xcolor}
\usepackage{rotating}

\usepackage[disable,colorinlistoftodos]{todonotes}   

\newcommand{\todoB}[1]{\todo[color=Yellow, inline]{\textbf{B}---#1}}

\newcommand{\pretty}[1]{} 

\newcommand{\extend}[1]{\todo[color=GreenYellow, inline]{\textbf{EXT}---#1}}

\definecolor{viriviolet}{HTML}{351042}
\definecolor{virigreen}{HTML}{317F79}
\definecolor{viriyellow}{HTML}{FCE528}

\usepackage{multirow}
\usepackage{booktabs}
\usepackage{rotating}

\usepackage{paralist}

\usepackage{balance}

\usepackage{versions}
\includeversion{DocumentVersionConference}
\excludeversion{DocumentVersionTR}
\includeversion{DocumentVersionLNCS}
\excludeversion{RedundantContent}

\usepackage[caption=false]{subfig}

\newtheorem{researchquestion}{RQ}

\newcommand{\vari}[1]{\ensuremath{\mathit{#1}\xspace}}
\newcommand{\const}[1]{\ensuremath{\mathsf{#1}\xspace}}

\newcommand{\CASCAde}{ERC Starting Grant CASCAde (GA n\textsuperscript{o}716980)}
\IfFileExists{upquote.sty}{\usepackage{upquote}}{}
\begin{document}

















\newcommand{\sampleBoxplot}{
\begin{figure}[tb]

\includegraphics[width=\maxwidth]{figure/data_sample_sizes-1} 
\caption{Boxplot of the sample sizes of the SLR sample}
\label{fig:sampleBoxplot}
\end{figure}
}

\newcommand{\sampleTests}{
\begin{table}[ht]
\centering
\caption{Sample Refinement on Extracted Effect Sizes (Adapted from Gro{\ss}~\cite{Gross2020})} 
\label{tab:sampleTests}
\begingroup\footnotesize
\begin{tabular}{lrr}
  \toprule
\textbf{Phase} & Excluded & Retained \\ 
  \midrule
Total effects extracted & 0 & 650 \\ 
   \midrule
 \quad \textsf{statcheck} extraction &  &  252 \\
 \quad Test statistic manual coding &   &  89 \\
 \quad Means \& SD manual coding    &   &  309 \\
 \midrule
 \textit{Refinement in this study}\\
Independent-samples test on dependent sample & 46 & 604 \\ 
  Treated proportion as $t$-distribution & 8 & 596 \\ 
  Reported dependent-samples test w/o correlation & 62 & 534 \\ 
  Reported $\chi^2$ without $\vari{df}$ & 5 & 529 \\ 
  $\chi^2$ with $\vari{df} > 1$ without contingency & 72 & 457 \\ 
  Multi-way $F$-test & 22 & 435 \\ 
  Yielded infinite ES or variance & 3 & 432 \\ 
  Duplicate of other coded test & 1 & 431 \\ 
   \bottomrule
\end{tabular}
\endgroup
\end{table}
}




\newcommand{\ppvHeatmap}{
\begin{figure*}[tbp]
\centering

\includegraphics[width=\maxwidth]{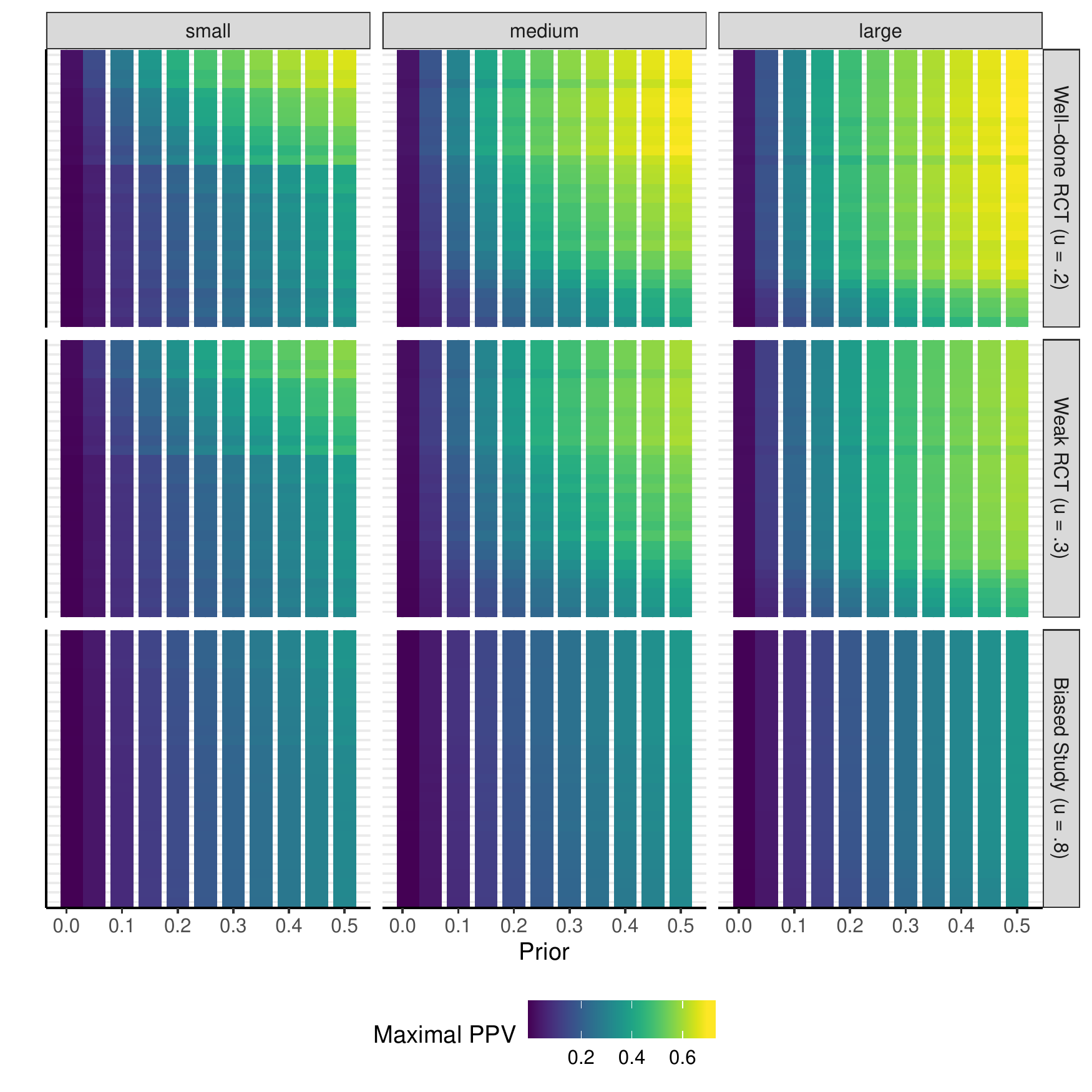} 
\caption{Heatmap of Positive Predictive Value (PPV) by effect size threshold and bias.}
\label{fig:ppvHeatmap}
\end{figure*}
}

\newcommand{\ppvDensityMCCsubfig}{
\begin{figure*}[tbp]
\centering\captionsetup{position=bottom}
\begin{minipage}{0.30\textwidth}%
\subfloat[SRCT/Small]{
\label{fig:ppvDensityMCCsmall}
\centering\includegraphics[keepaspectratio,width=\columnwidth]{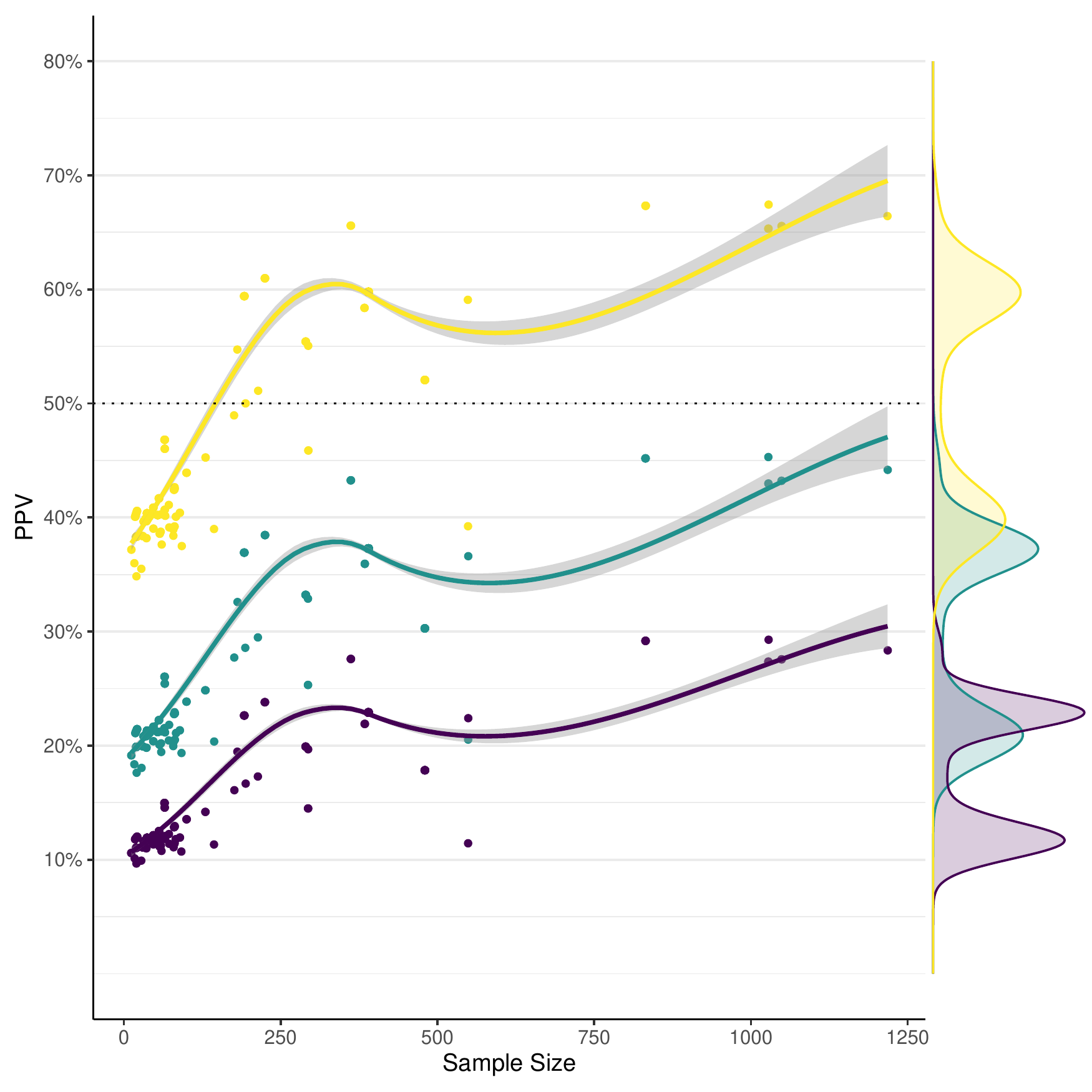}
}
\end{minipage}~
\begin{minipage}{0.30\textwidth}%
\subfloat[SRCT/Medium]{%
\label{fig:ppvDensityMCCmedium}
\centering\includegraphics[keepaspectratio,width=\columnwidth]{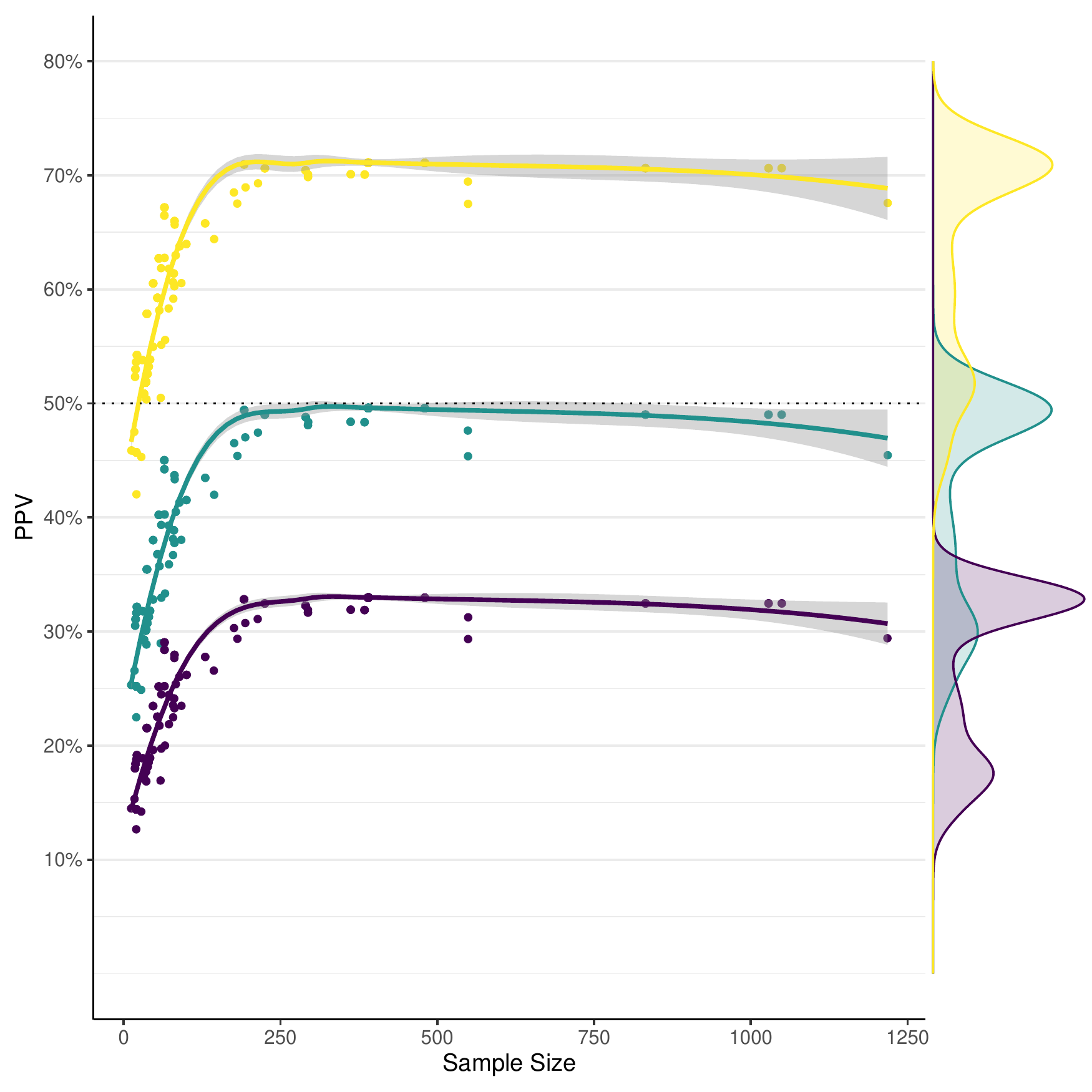}
}
\end{minipage}~
\begin{minipage}{0.30\textwidth}%
\subfloat[SRCT/Large]{%
\label{fig:ppvDensityMCClarge_SRCT}
\centering\includegraphics[keepaspectratio,width=\columnwidth]{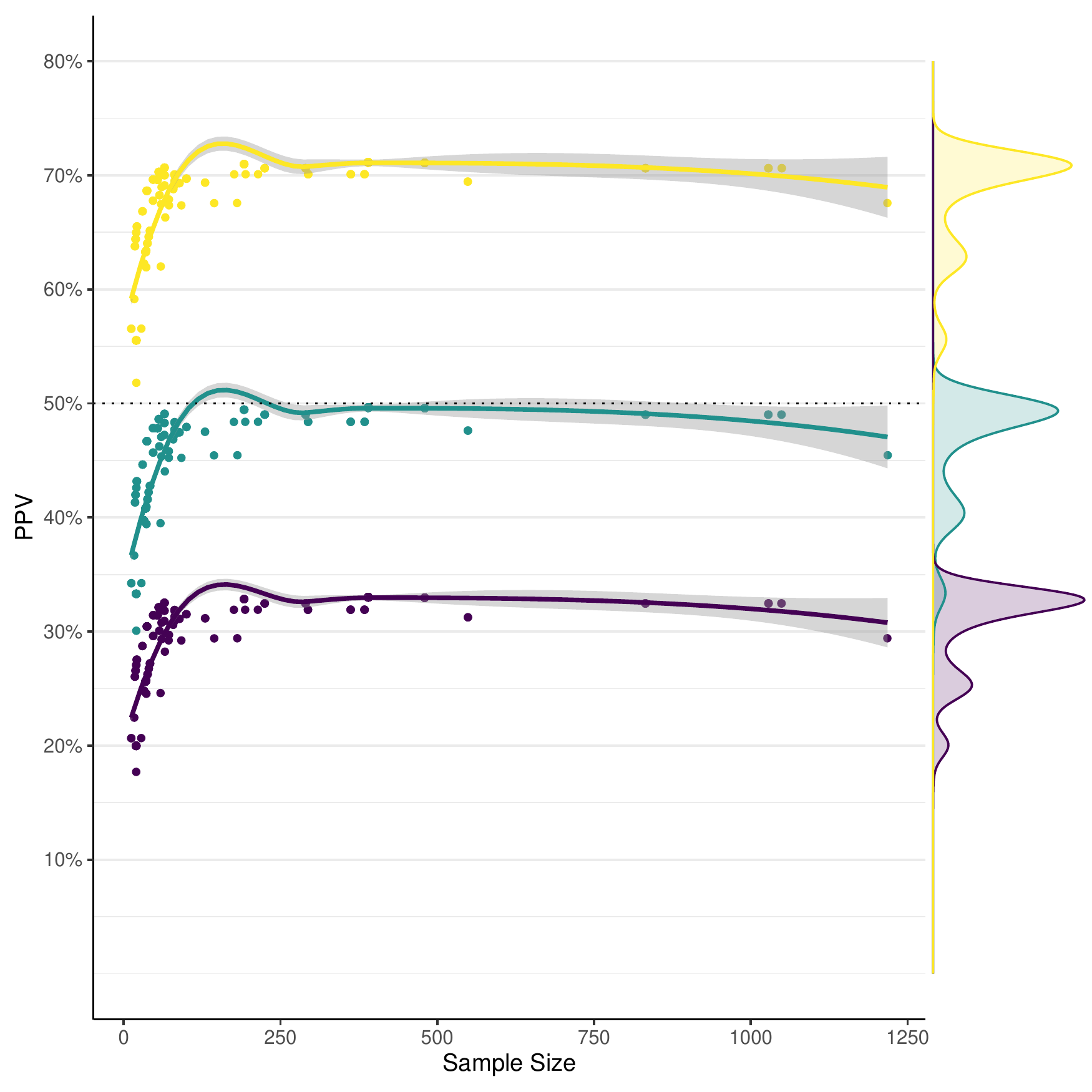}
}
\end{minipage}

\begin{minipage}{0.30\textwidth}%
\subfloat[WRCT/Small]{
\label{fig:ppvDensityMCCsmall}
\centering\includegraphics[keepaspectratio,width=\columnwidth]{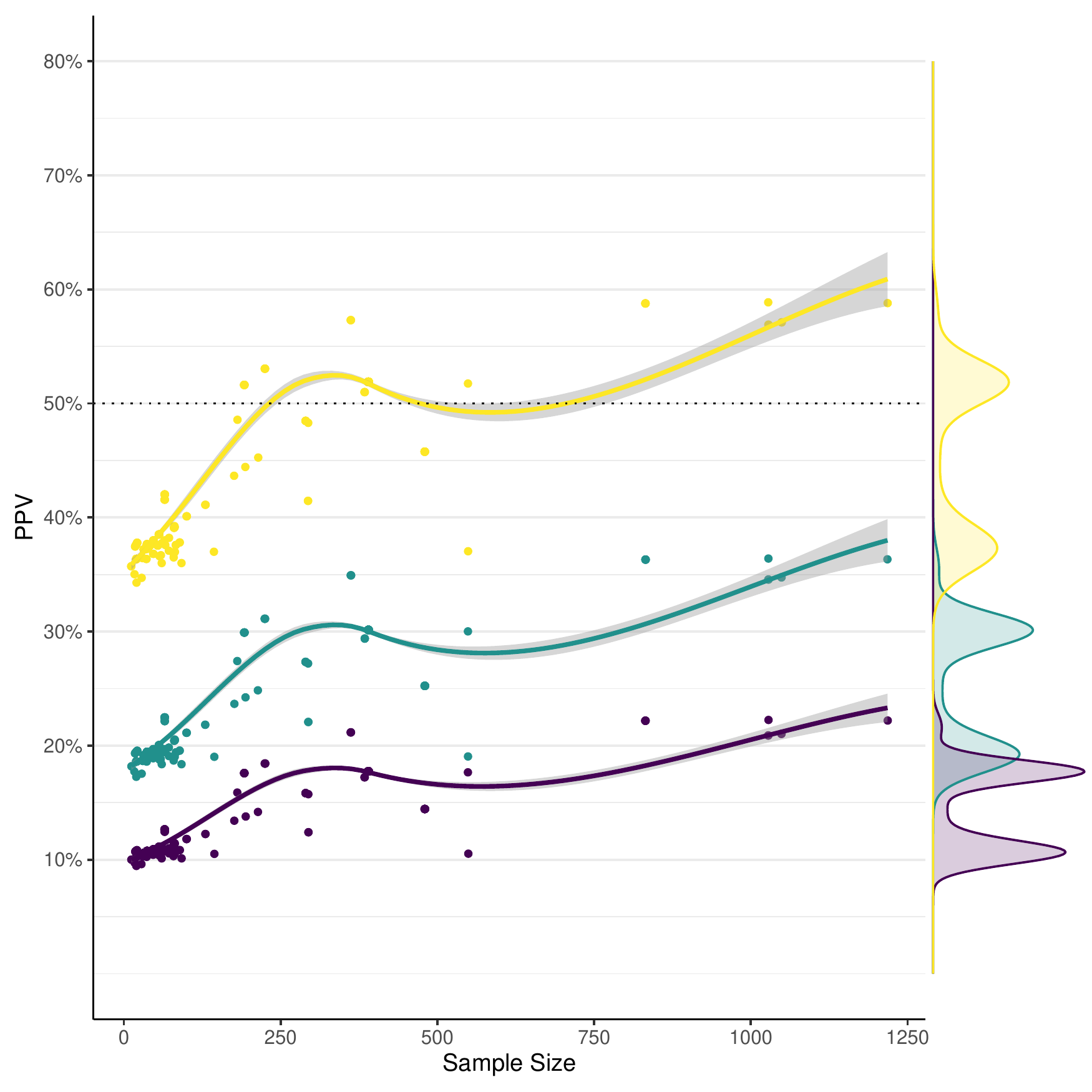}
}
\end{minipage}~
\begin{minipage}{0.30\textwidth}%
\subfloat[WRCT/Medium]{%
\label{fig:ppvDensityMCCmedium}
\centering\includegraphics[keepaspectratio,width=\columnwidth]{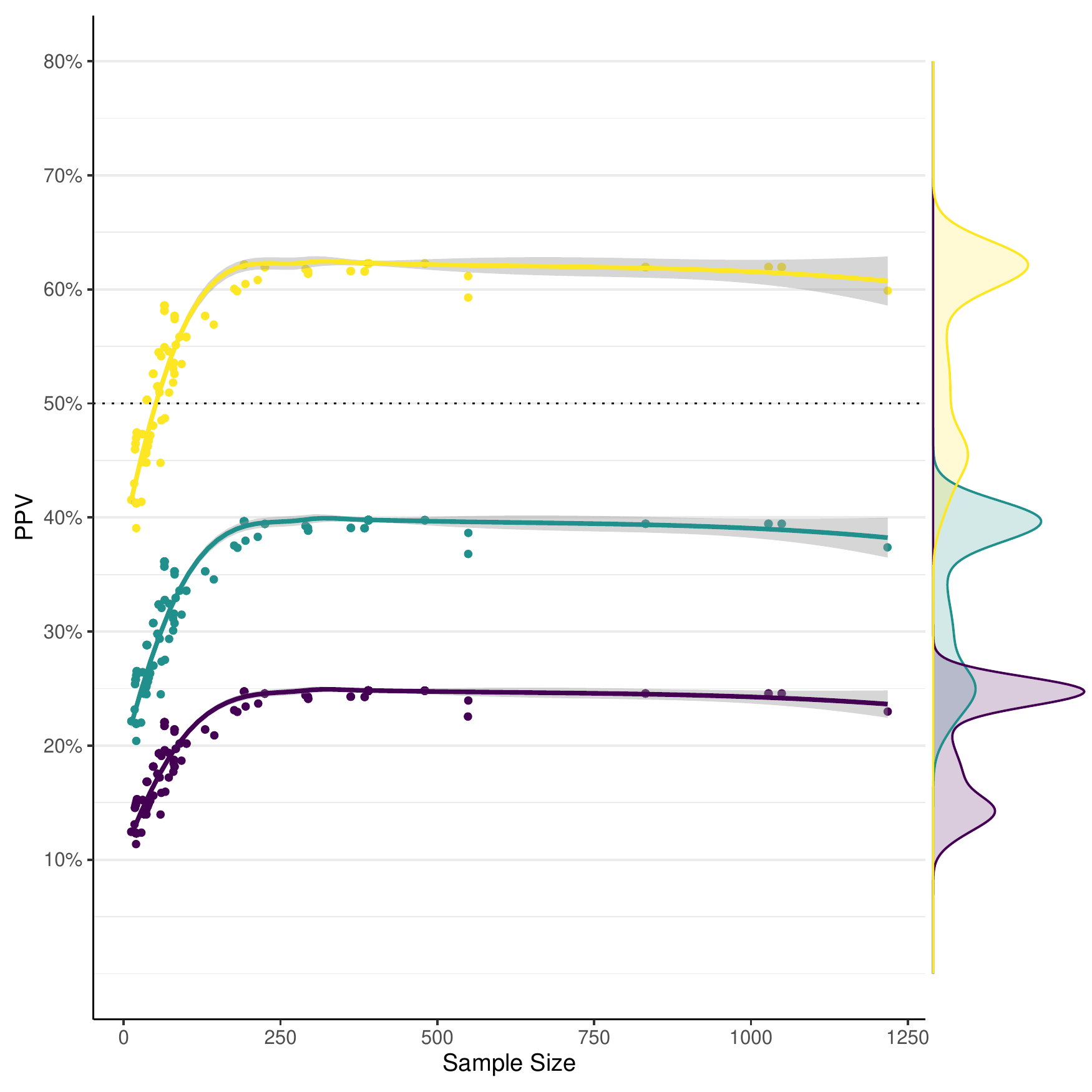}
}
\end{minipage}~
\begin{minipage}{0.30\textwidth}%
\subfloat[WRCT/Large]{%
\label{fig:ppvDensityMCClarge_SRCT}
\centering\includegraphics[keepaspectratio,width=\columnwidth]{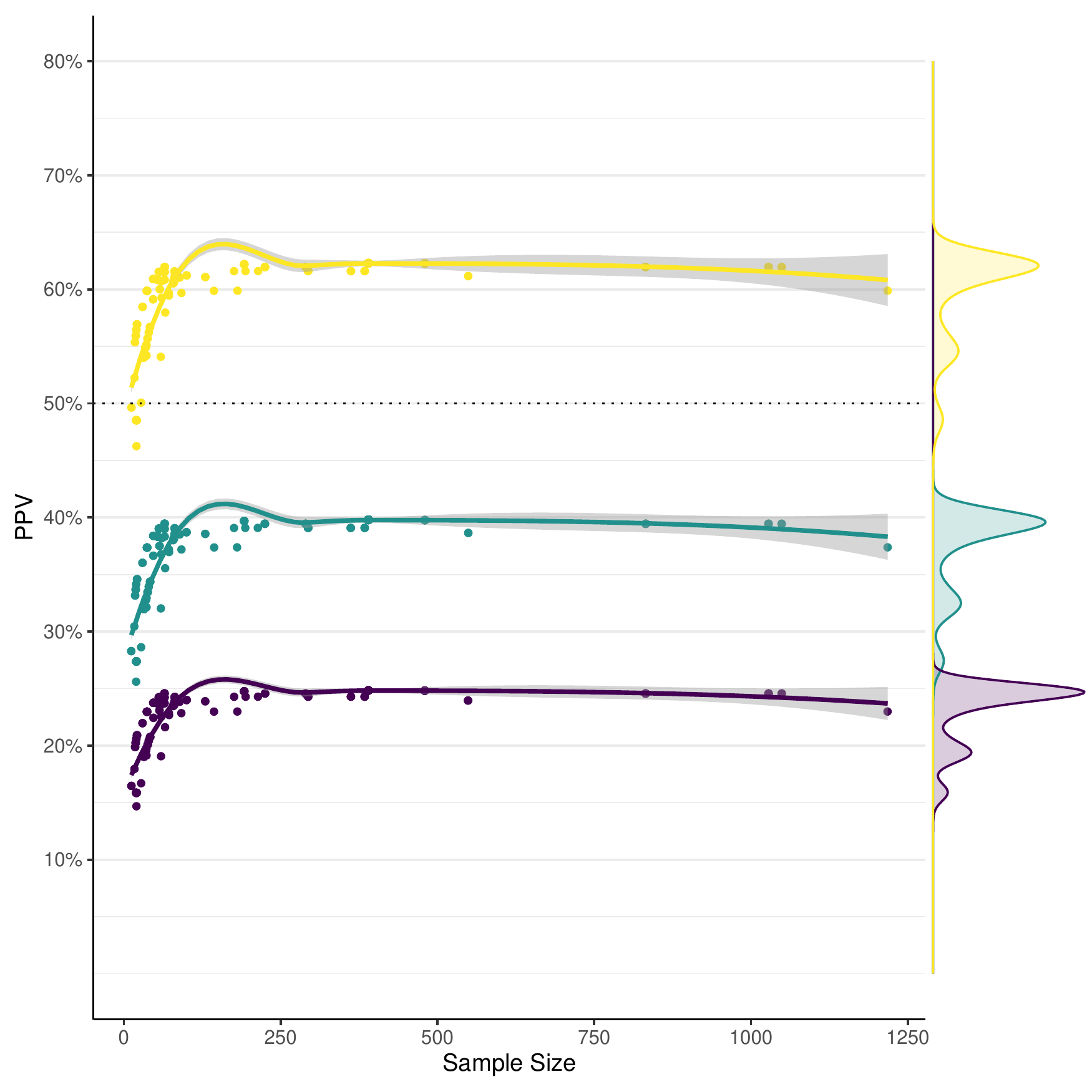}
}
\end{minipage}

\begin{minipage}{0.30\textwidth}%
\subfloat[WS/Small]{
\label{fig:ppvDensityMCCsmall}
\centering\includegraphics[keepaspectratio,width=\columnwidth]{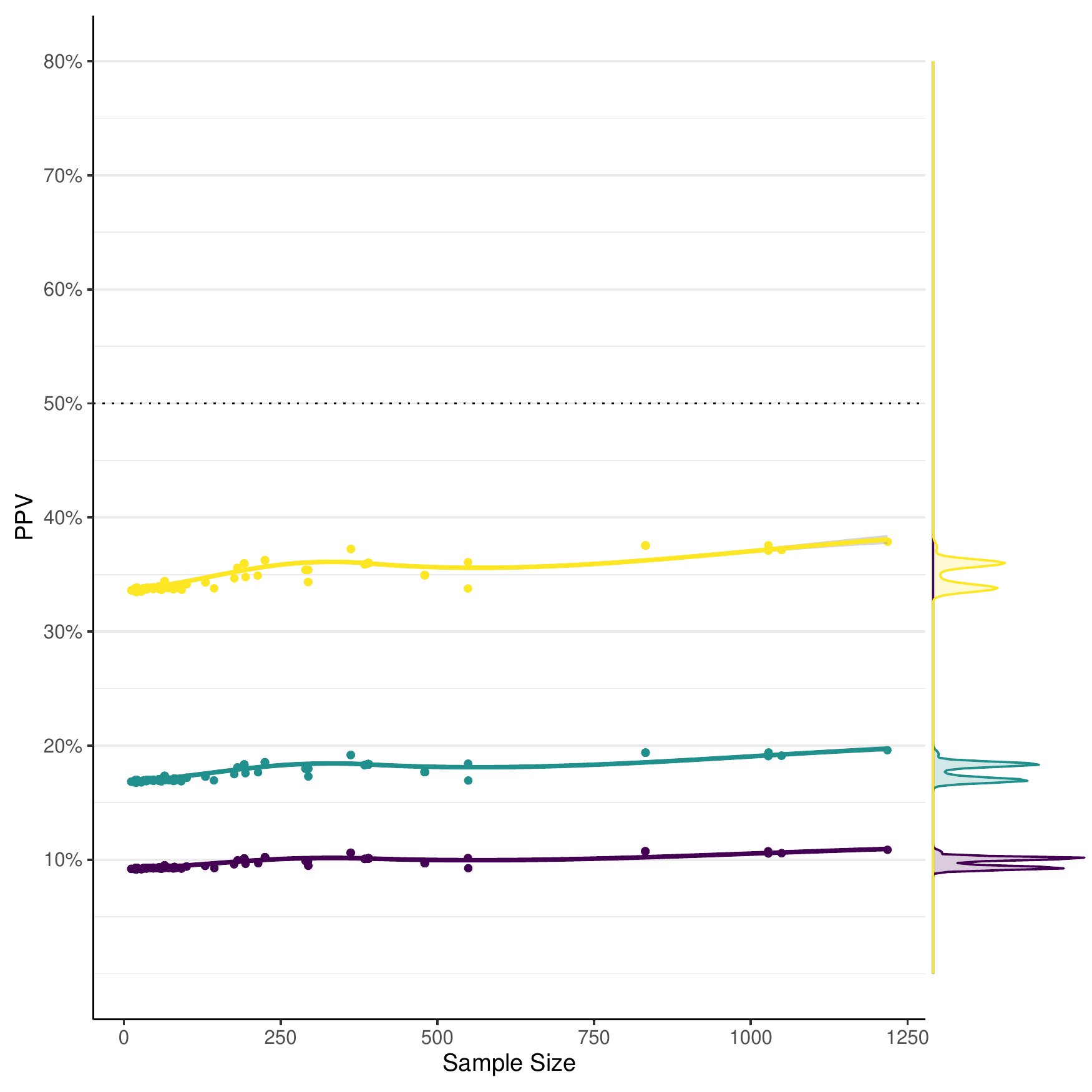}
}
\end{minipage}~
\begin{minipage}{0.30\textwidth}%
\subfloat[WS/Medium]{%
\label{fig:ppvDensityMCCmedium}
\centering\includegraphics[keepaspectratio,width=\columnwidth]{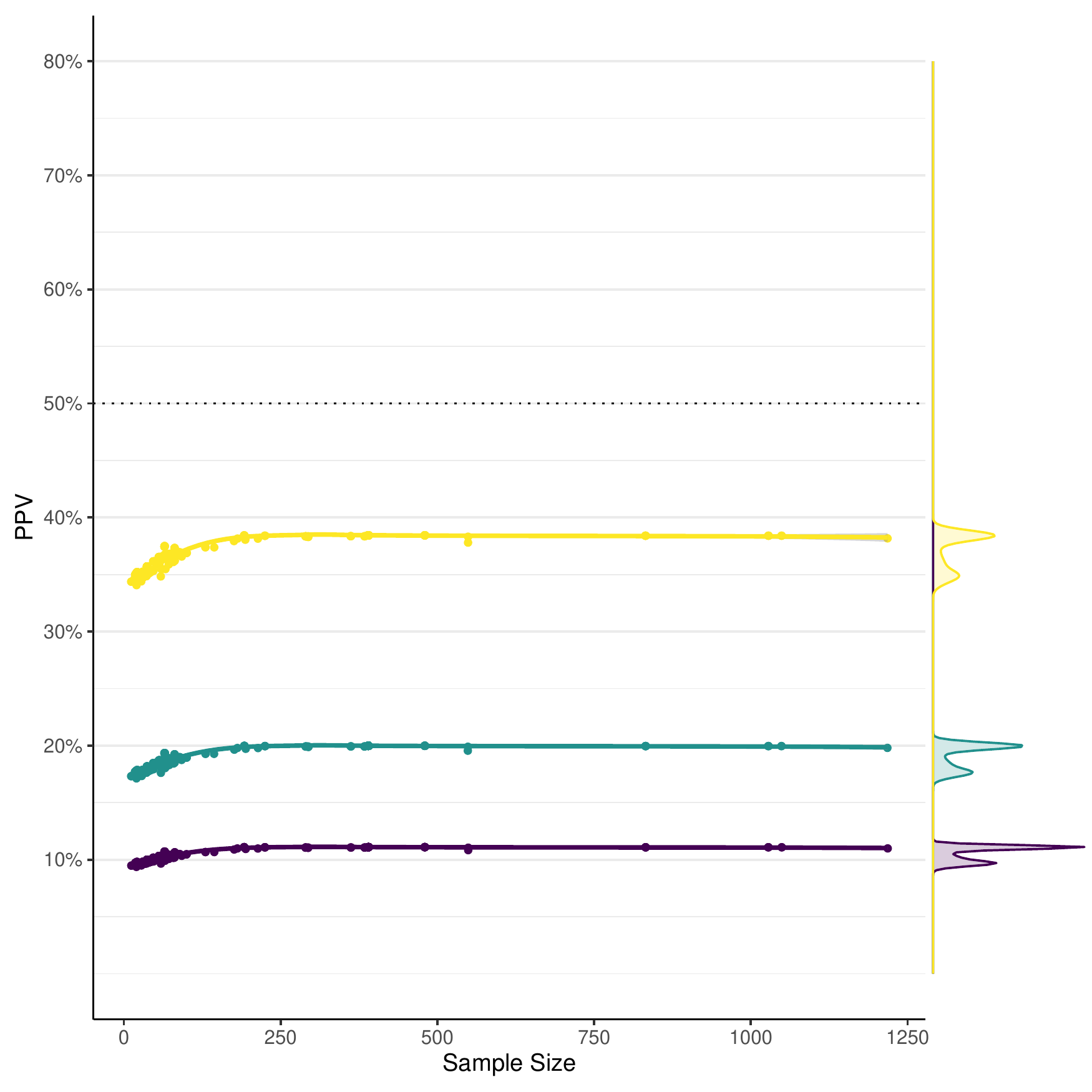}
}
\end{minipage}~
\begin{minipage}{0.30\textwidth}%
\subfloat[WS/Large]{%
\label{fig:ppvDensityMCClarge_SRCT}
\centering\includegraphics[keepaspectratio,width=\columnwidth]{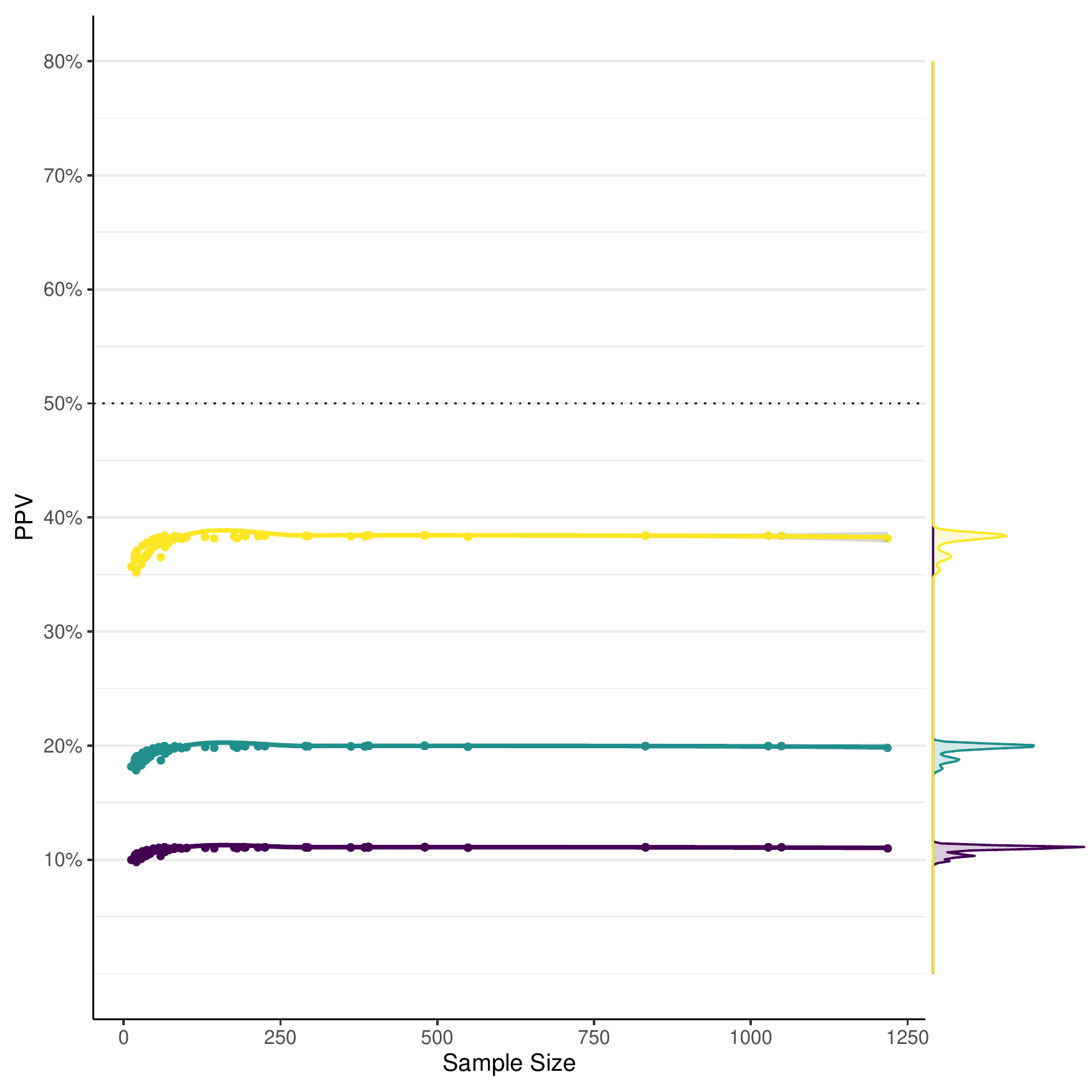}
}
\end{minipage}
\caption{Positive Predictive Value (PPV) versus effect size classes and biases}
\label{fig:ppvDensityMCC}
\end{figure*}
}
\newcommand*\rot{\rotatebox{90}}\newcommand{\ppvDensityMCCBiasByThreshold}{
\begin{figure*}[tbp]
\centering\captionsetup{position=bottom}
\begin{tabular}{llccc}
\toprule
&& \multicolumn{3}{c}{Effect Size Threshold}\\
\midrule
&& Small & Medium & Large\\
\cmidrule(lr){3-3}\cmidrule(lr){4-4}\cmidrule(lr){5-5}
\multirow{21}{*}{
\rot{Study Bias}} &
\multirow{7}{*}{
\rot{\rlap{Well-run RCT ($u = .2$)}}} &
\begin{minipage}{0.30\textwidth}%
\includegraphics[keepaspectratio,width=\columnwidth]{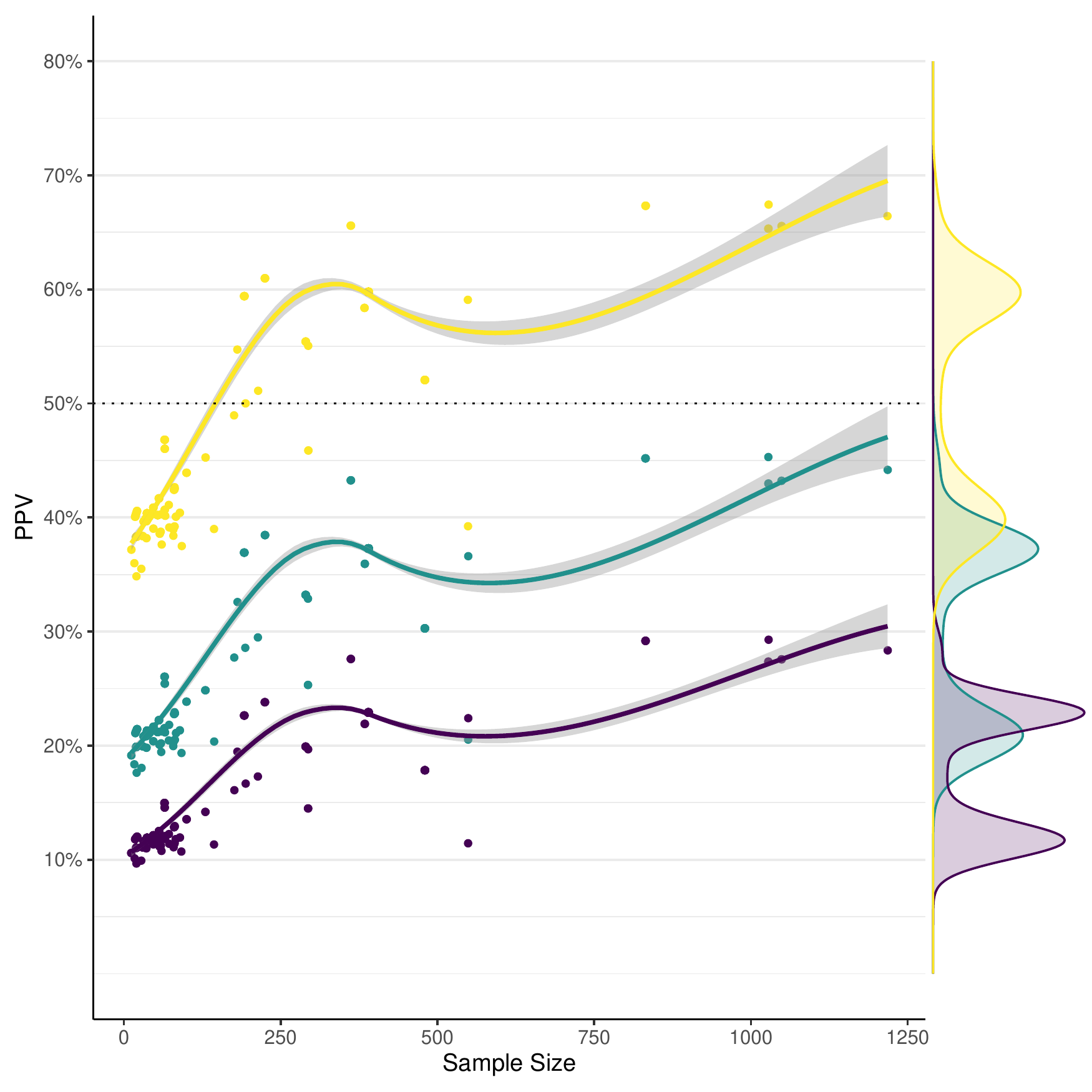}
\end{minipage} &
\begin{minipage}{0.30\textwidth}%
\includegraphics[keepaspectratio,width=\columnwidth]{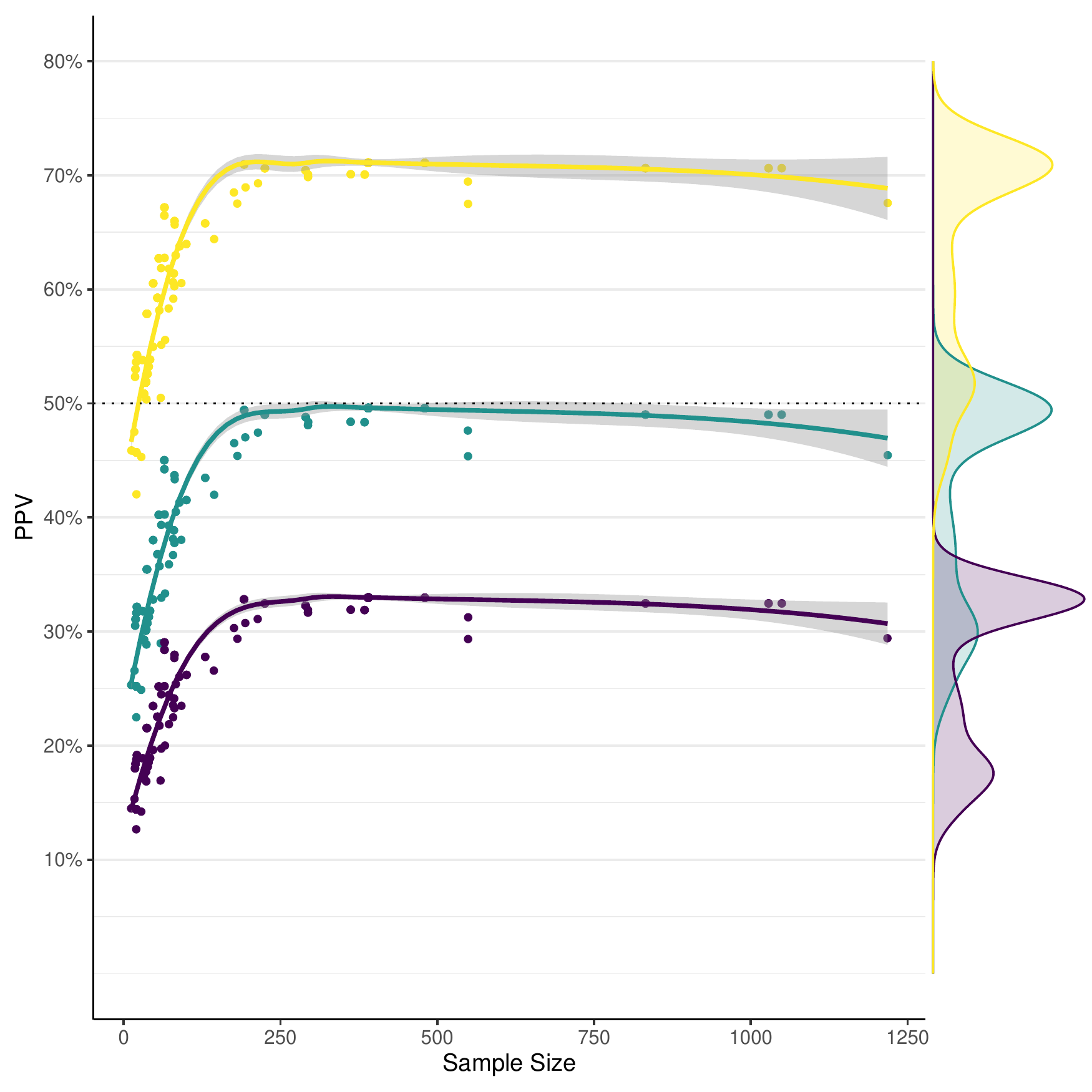}
\end{minipage} & 
\begin{minipage}{0.30\textwidth}%
\includegraphics[keepaspectratio,width=\columnwidth]{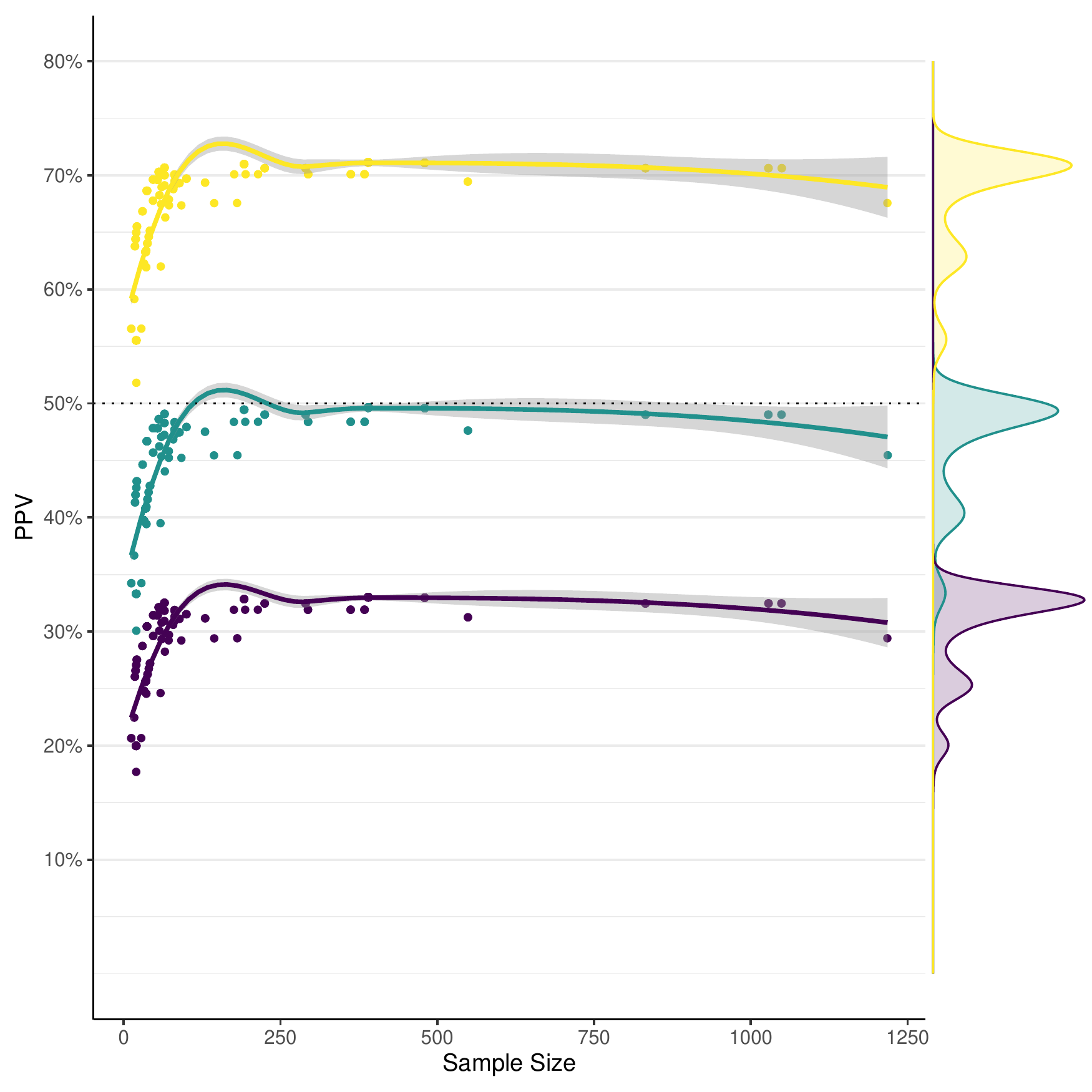}
\end{minipage}\\
&
\multirow{7}{*}{
\rot{\rlap{Weak RCT ($u = .3$)}}} &
\begin{minipage}{0.30\textwidth}%
\includegraphics[keepaspectratio,width=\columnwidth]{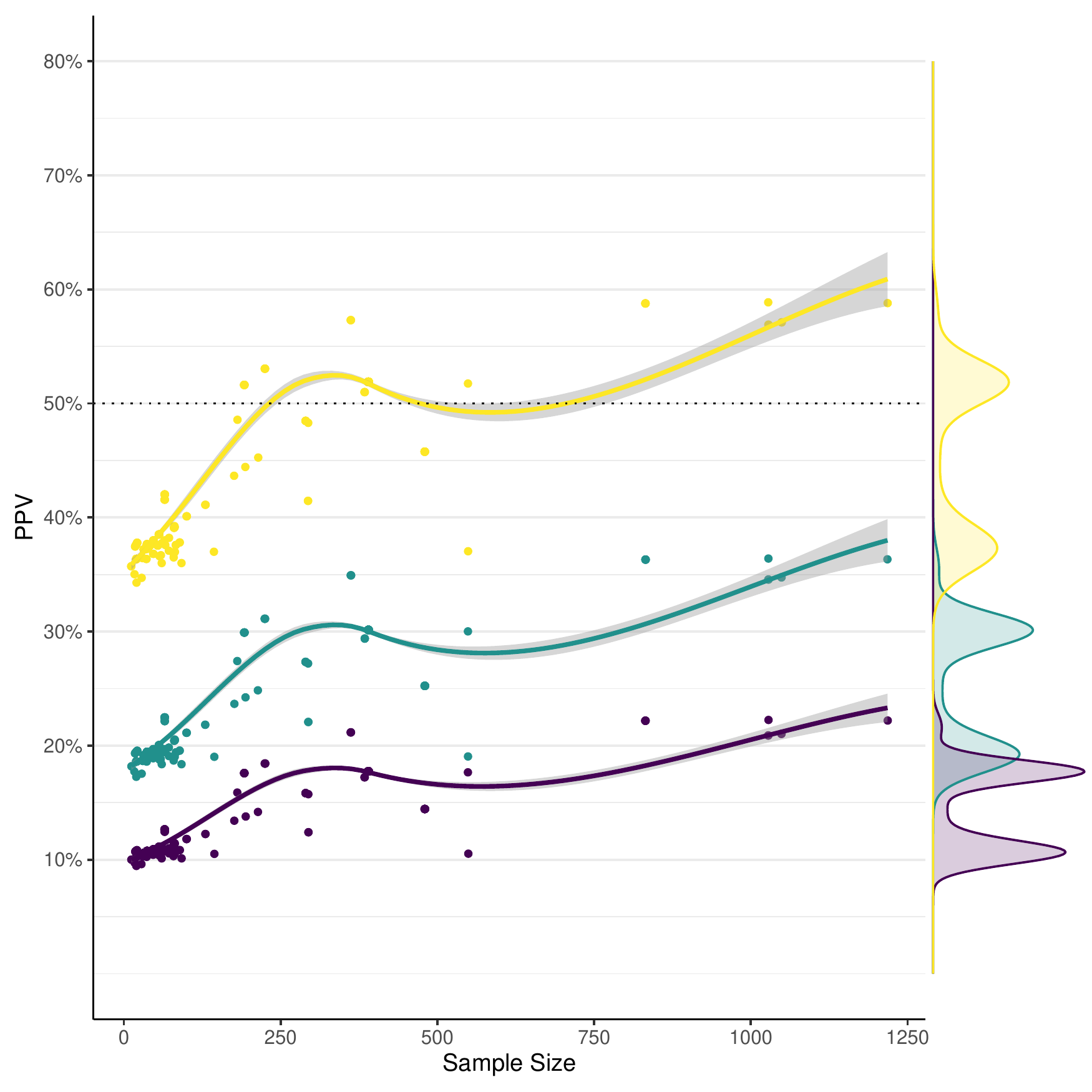}
\end{minipage} &
\begin{minipage}{0.30\textwidth}%
\includegraphics[keepaspectratio,width=\columnwidth]{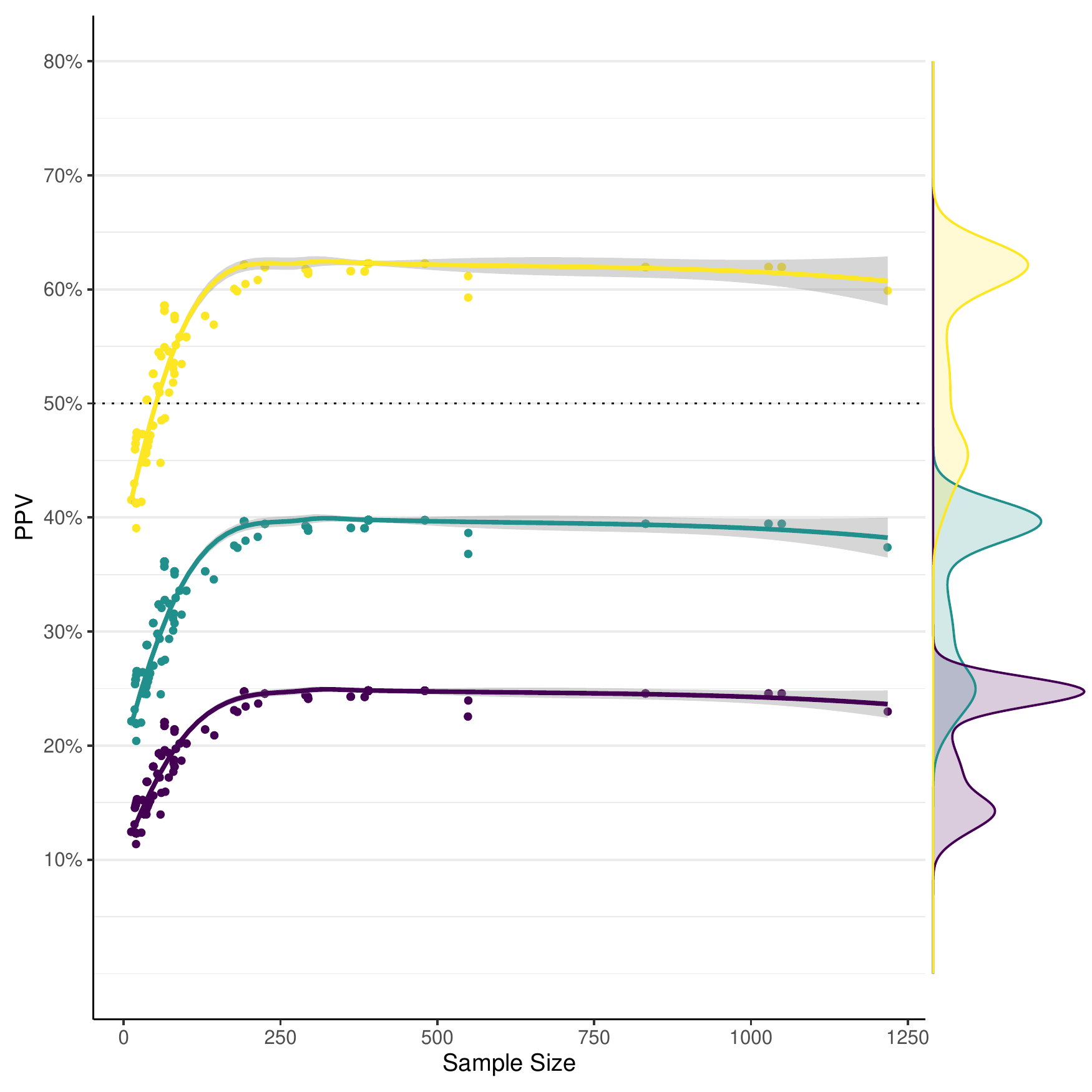}
\end{minipage} & 
\begin{minipage}{0.30\textwidth}%
\includegraphics[keepaspectratio,width=\columnwidth]{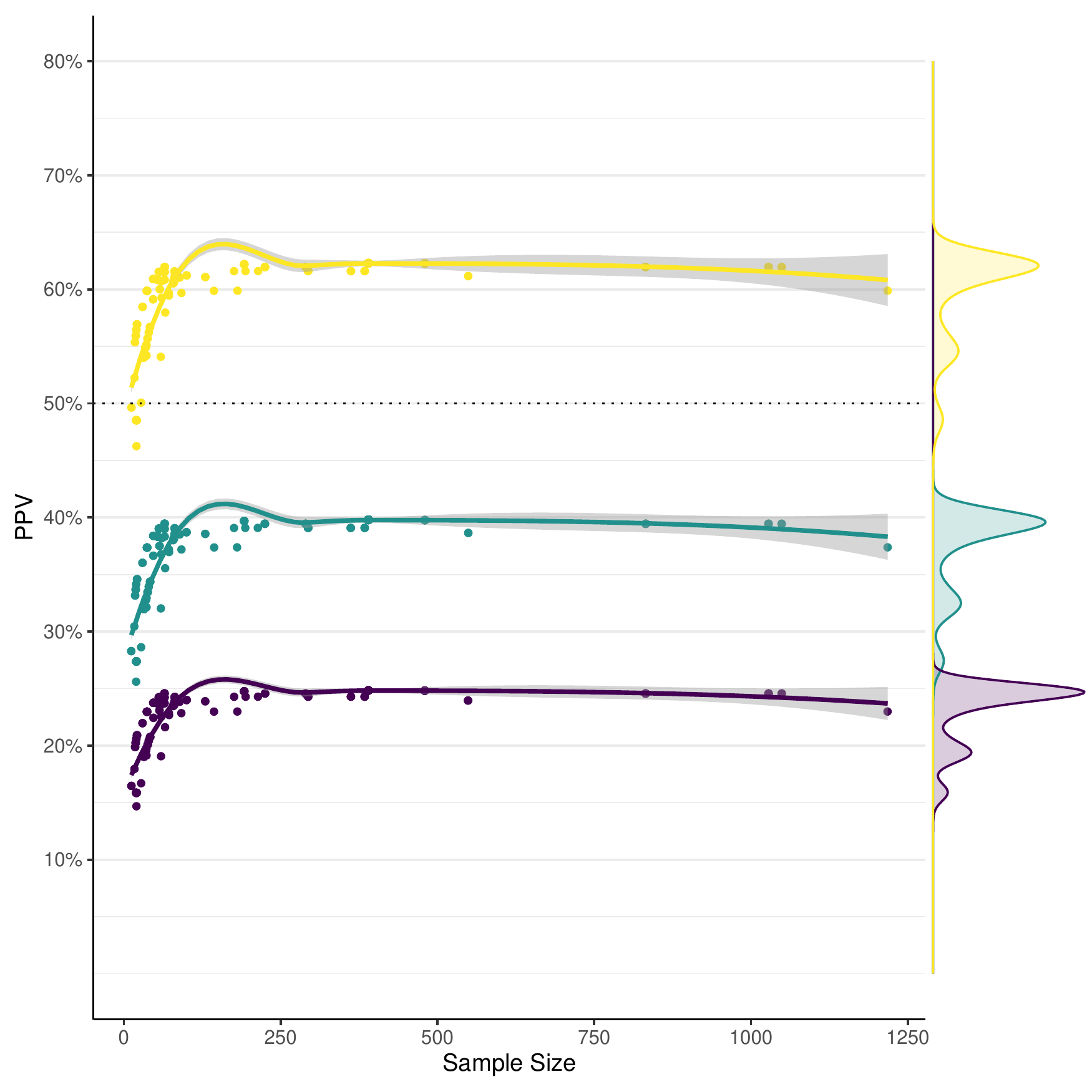}
\end{minipage}\\
&
\multirow{7}{*}{
\rot{\rlap{Biased Study ($u = .8$)}}} &
\begin{minipage}{0.30\textwidth}%
\includegraphics[keepaspectratio,width=\columnwidth]{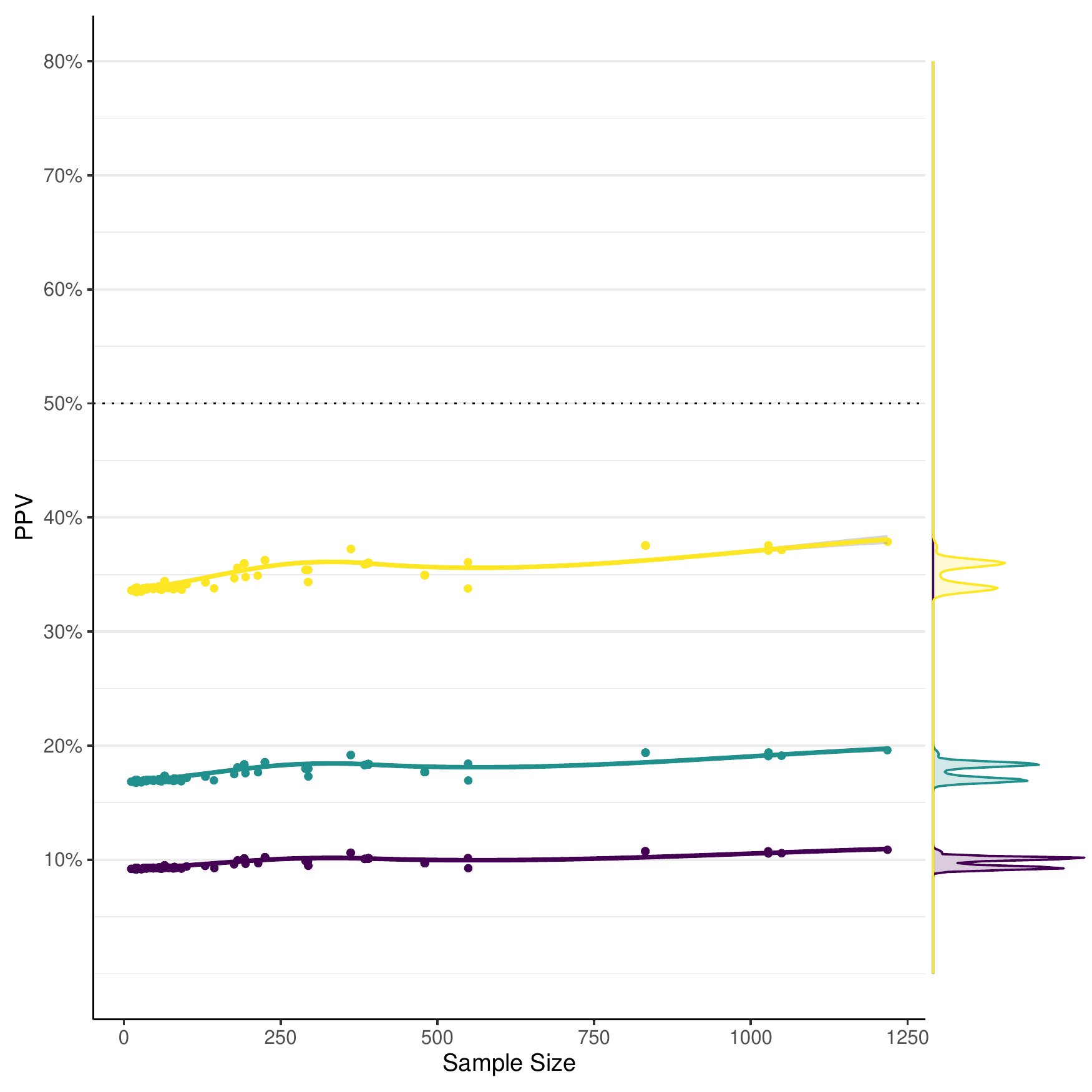}
\end{minipage} &
\begin{minipage}{0.30\textwidth}%
\includegraphics[keepaspectratio,width=\columnwidth]{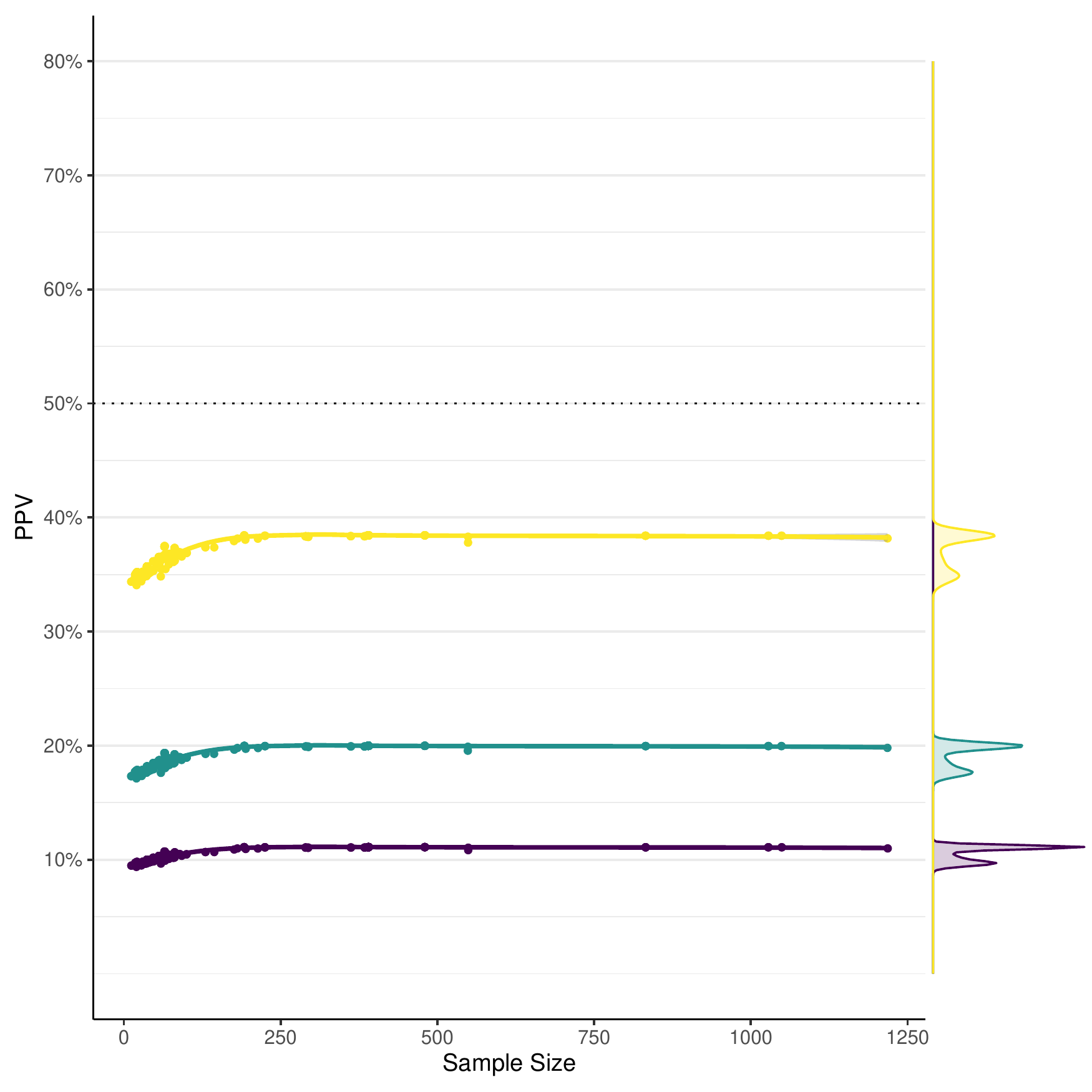}
\end{minipage} &
\begin{minipage}{0.30\textwidth}%
\includegraphics[keepaspectratio,width=\columnwidth]{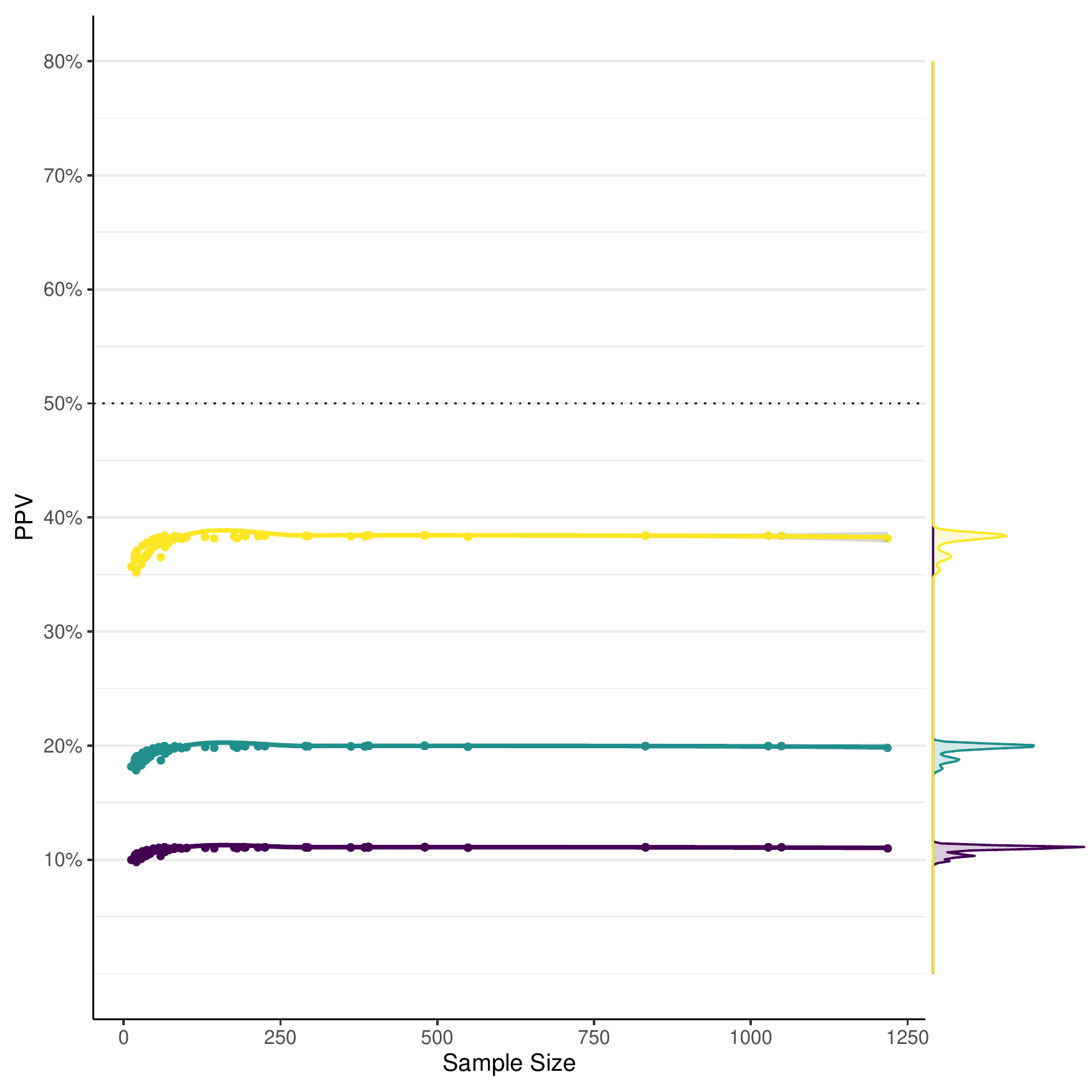}
\end{minipage}\\
\bottomrule
\end{tabular}
\caption{Positive Predictive Value (PPV) versus effect size classes and biases}
\label{fig:ppvDensityMCC}
\end{figure*}
}

\newcommand{\ppvDensityMCCBiasByPrior}{
\begin{figure*}[tbp]
\centering\captionsetup{position=bottom}
\begin{tabular}{llccc}
\toprule
&& \multicolumn{3}{c}{Prior Probability (Confirmatoriness)}\\
\midrule
&& Exploratory (prior = 0.1) & Intermediate (prior = 0.2) & Confirmatory (prior = 0.5)\\
\cmidrule(lr){3-3}\cmidrule(lr){4-4}\cmidrule(lr){5-5}
\multirow{21}{*}{
\rot{Study Bias}} &
\multirow{7}{*}{
\rot{\rlap{Well-run RCT ($u = .2$)}}} &
\begin{minipage}{0.30\textwidth}%
\includegraphics[keepaspectratio,width=\columnwidth]{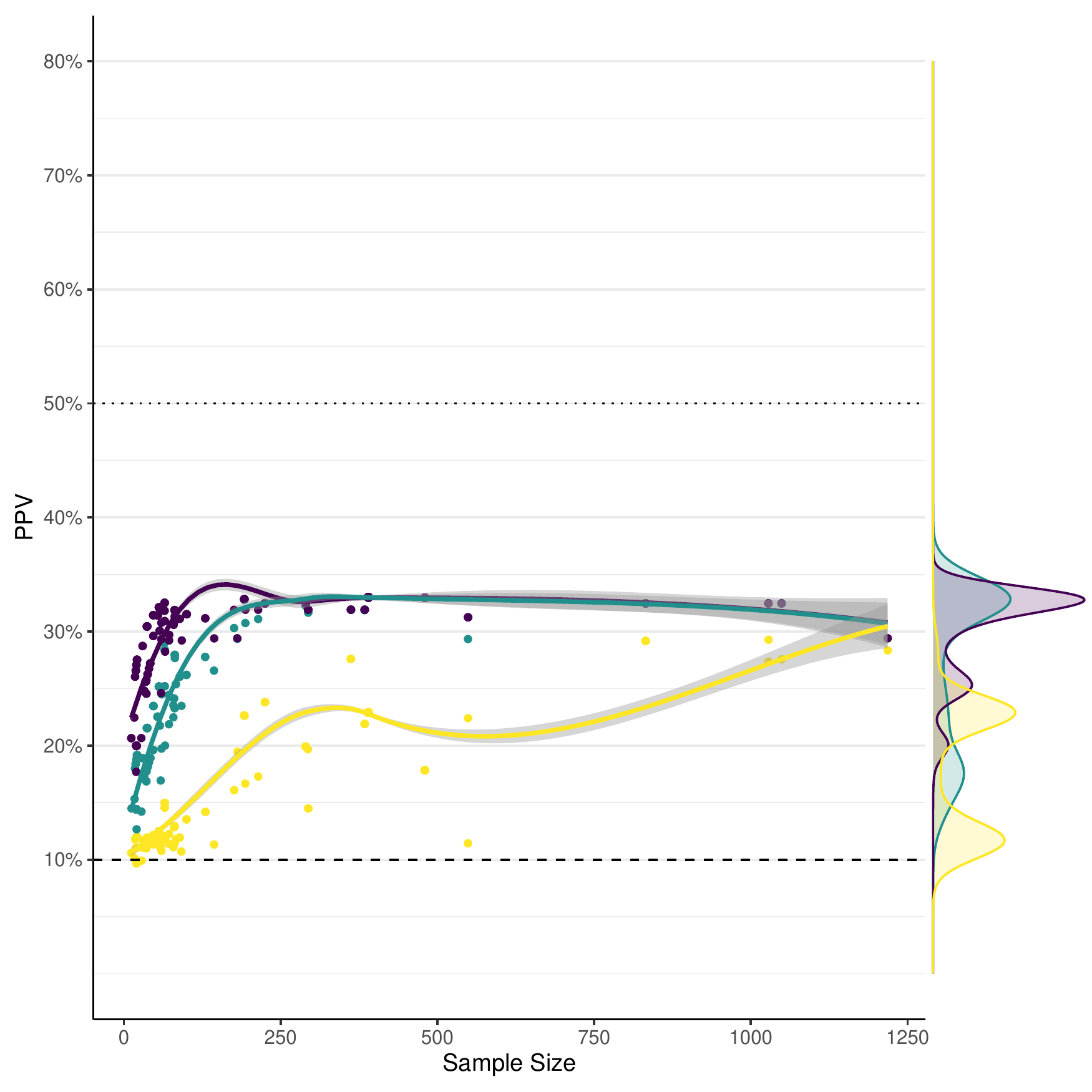}
\end{minipage} &
\begin{minipage}{0.30\textwidth}%
\includegraphics[keepaspectratio,width=\columnwidth]{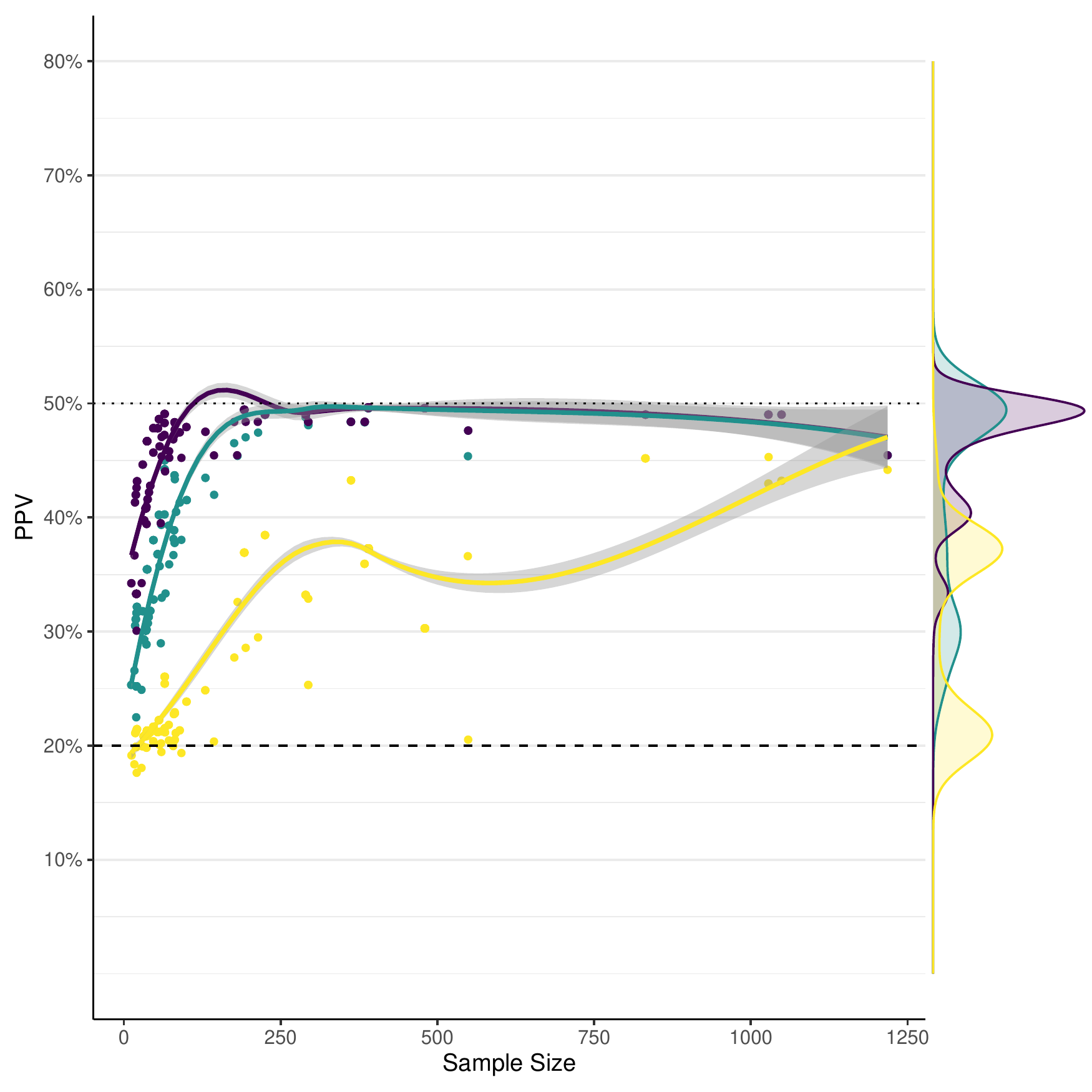}
\end{minipage} & 
\begin{minipage}{0.30\textwidth}%
\includegraphics[keepaspectratio,width=\columnwidth]{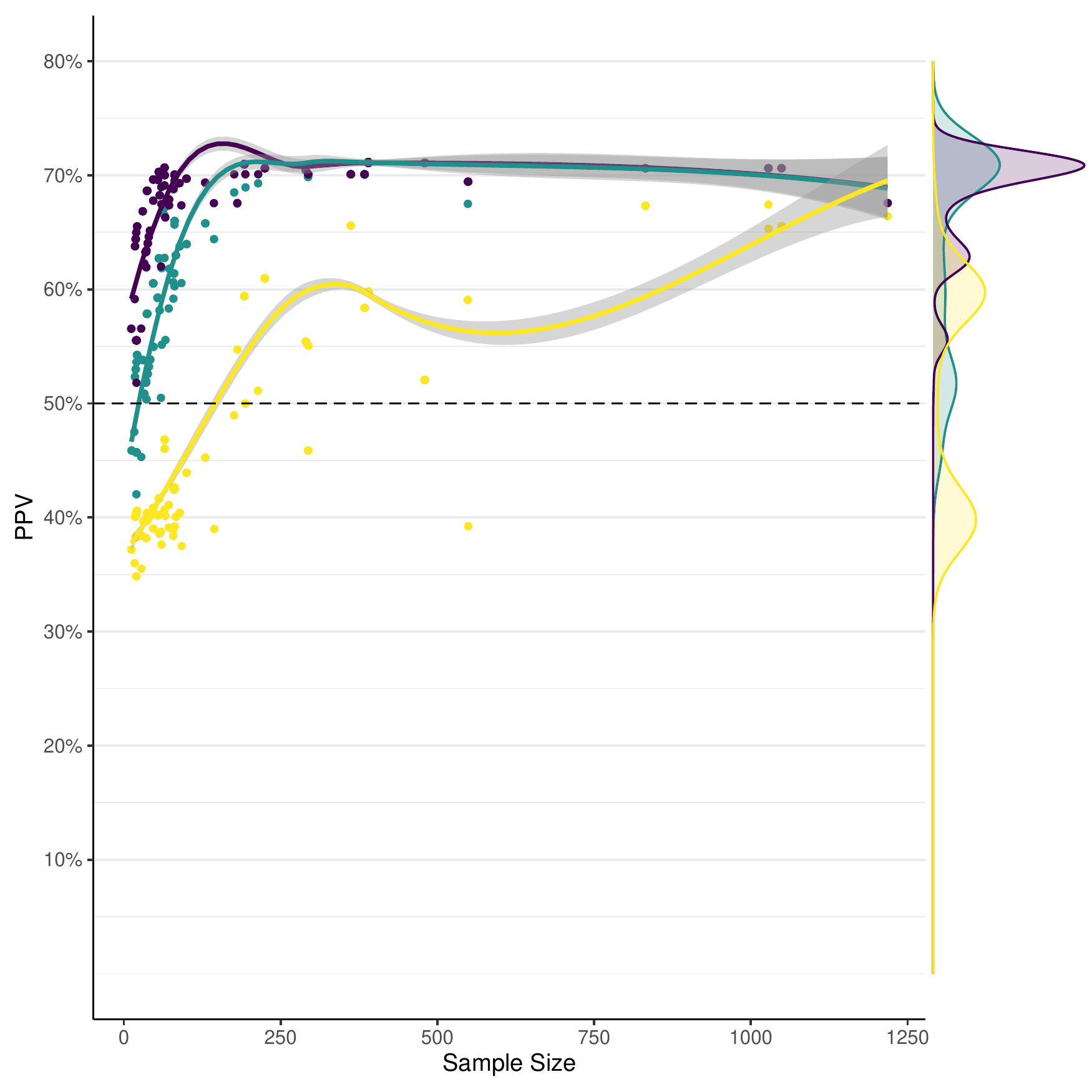}
\end{minipage}\\
&
\multirow{7}{*}{
\rot{\rlap{Weak RCT ($u = .3$)}}} &
\begin{minipage}{0.30\textwidth}%
\includegraphics[keepaspectratio,width=\columnwidth]{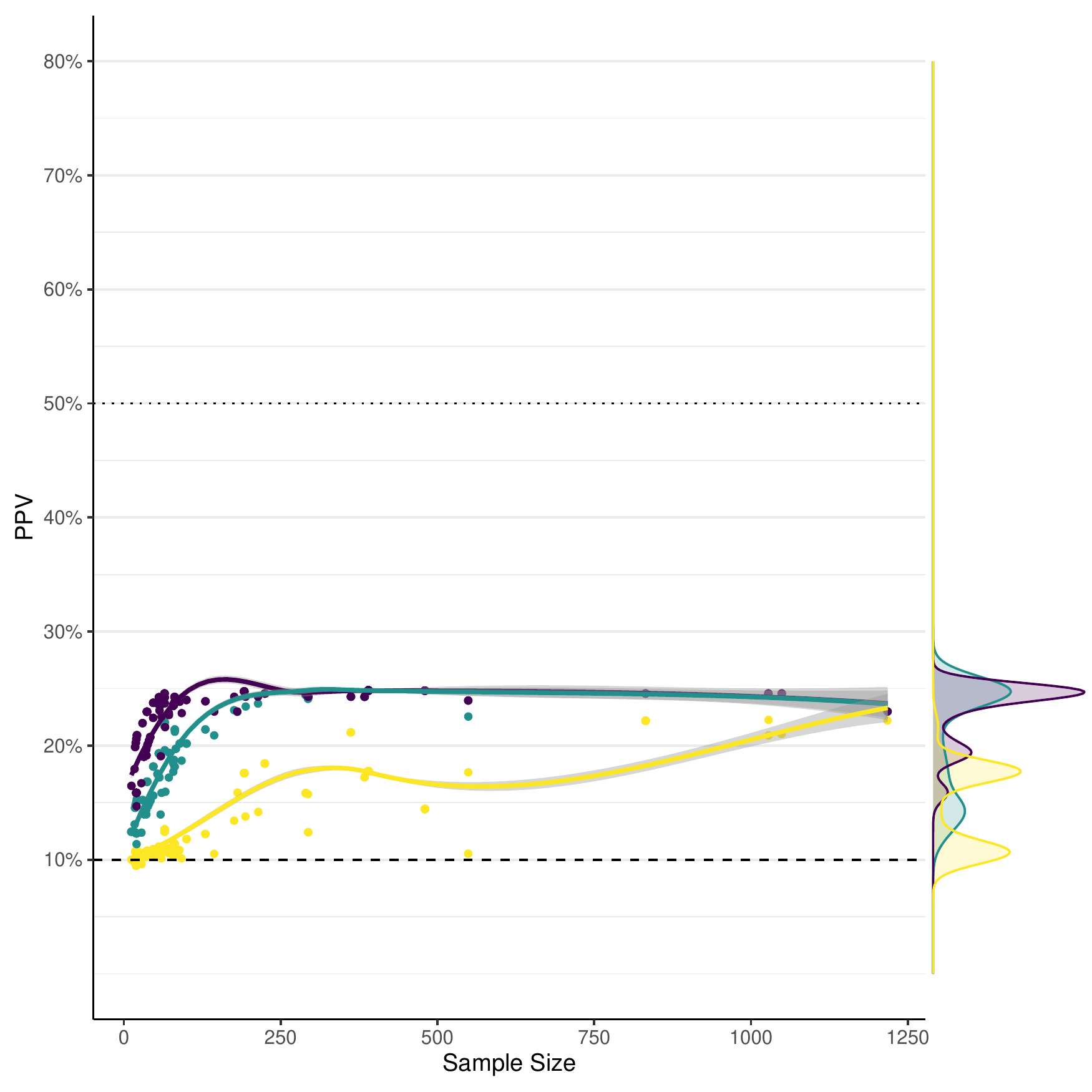}
\end{minipage} &
\begin{minipage}{0.30\textwidth}%
\includegraphics[keepaspectratio,width=\columnwidth]{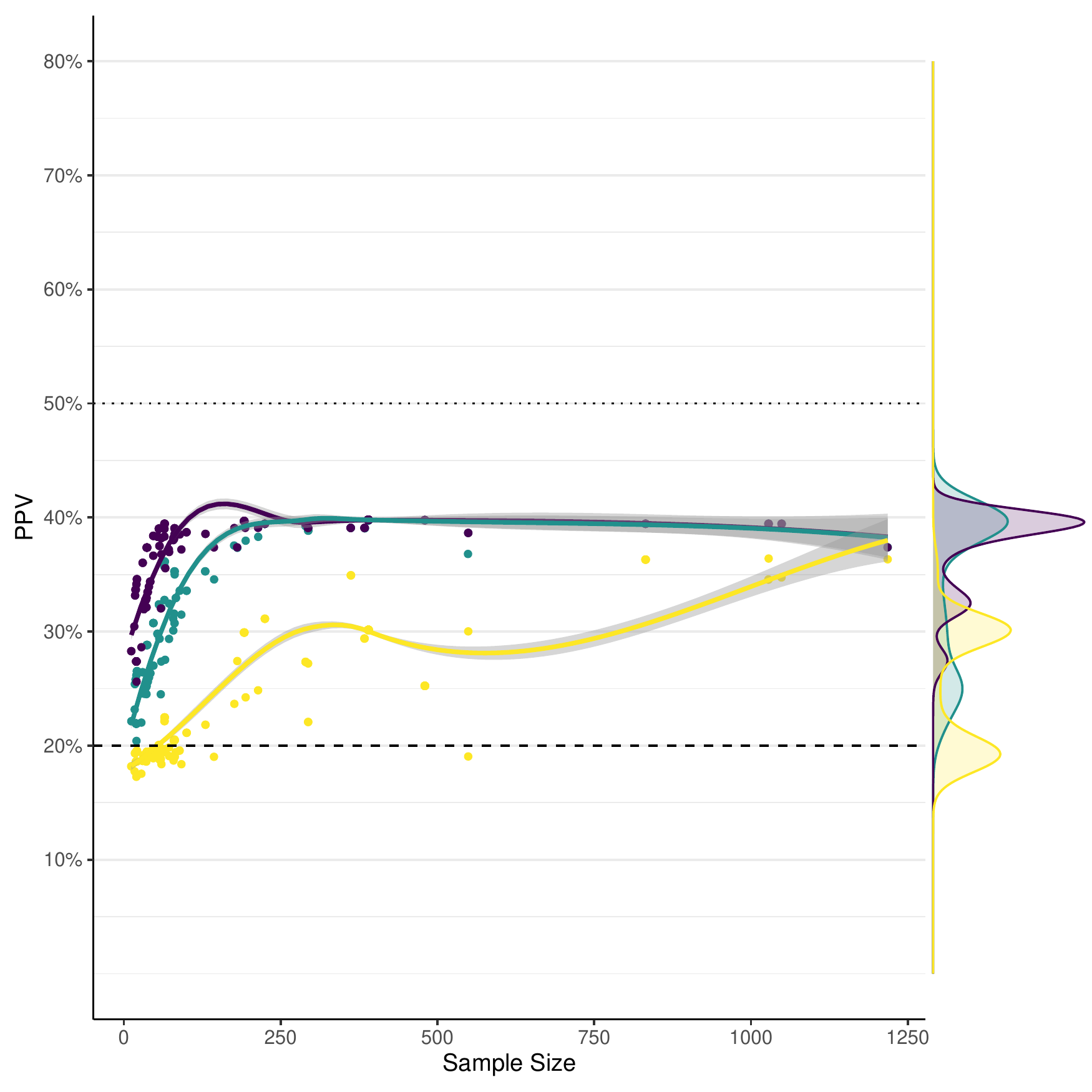}
\end{minipage} & 
\begin{minipage}{0.30\textwidth}%
\includegraphics[keepaspectratio,width=\columnwidth]{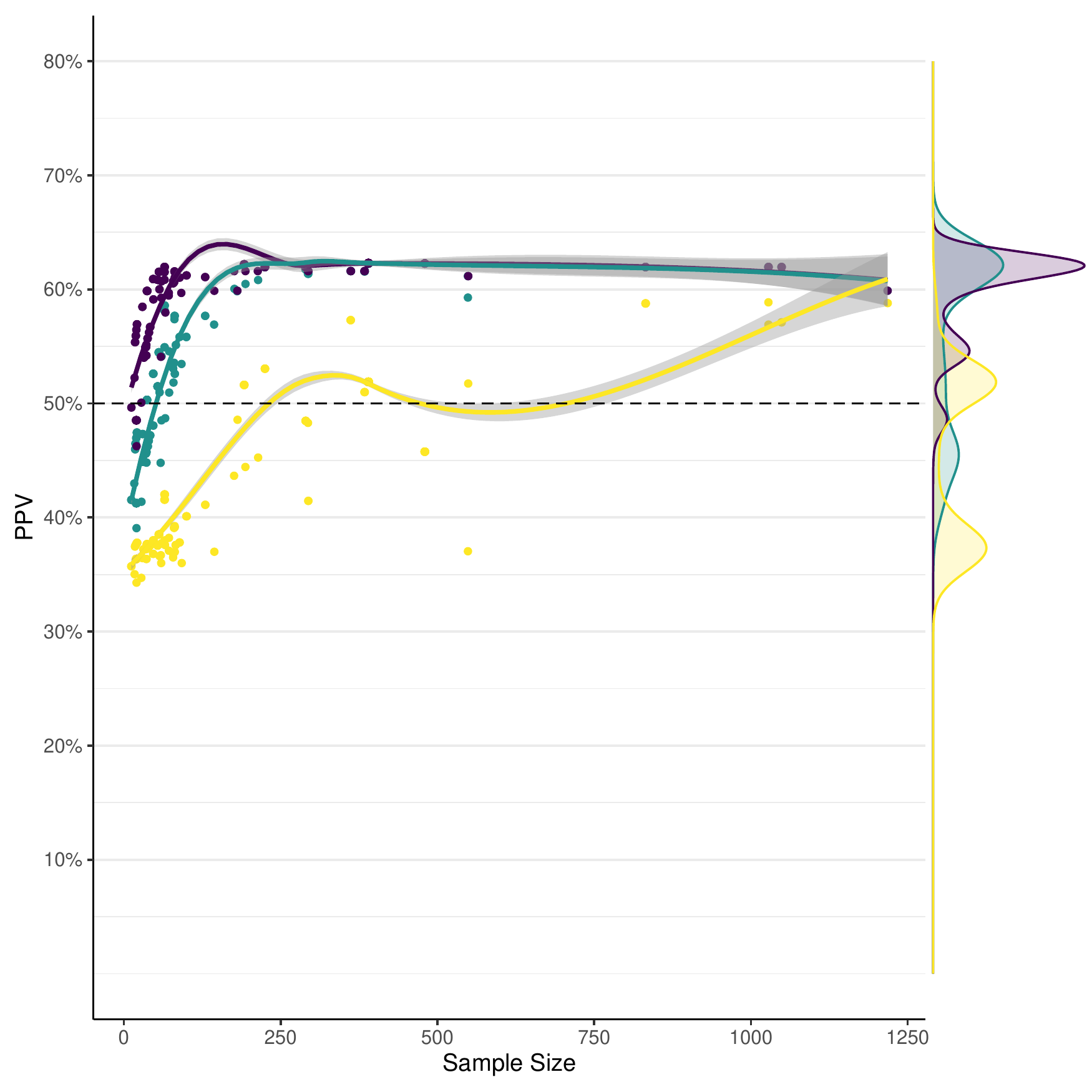}
\end{minipage}\\
&
\multirow{7}{*}{
\rot{\rlap{Biased Study ($u = .8$)}}} &
\begin{minipage}{0.30\textwidth}%
\includegraphics[keepaspectratio,width=\columnwidth]{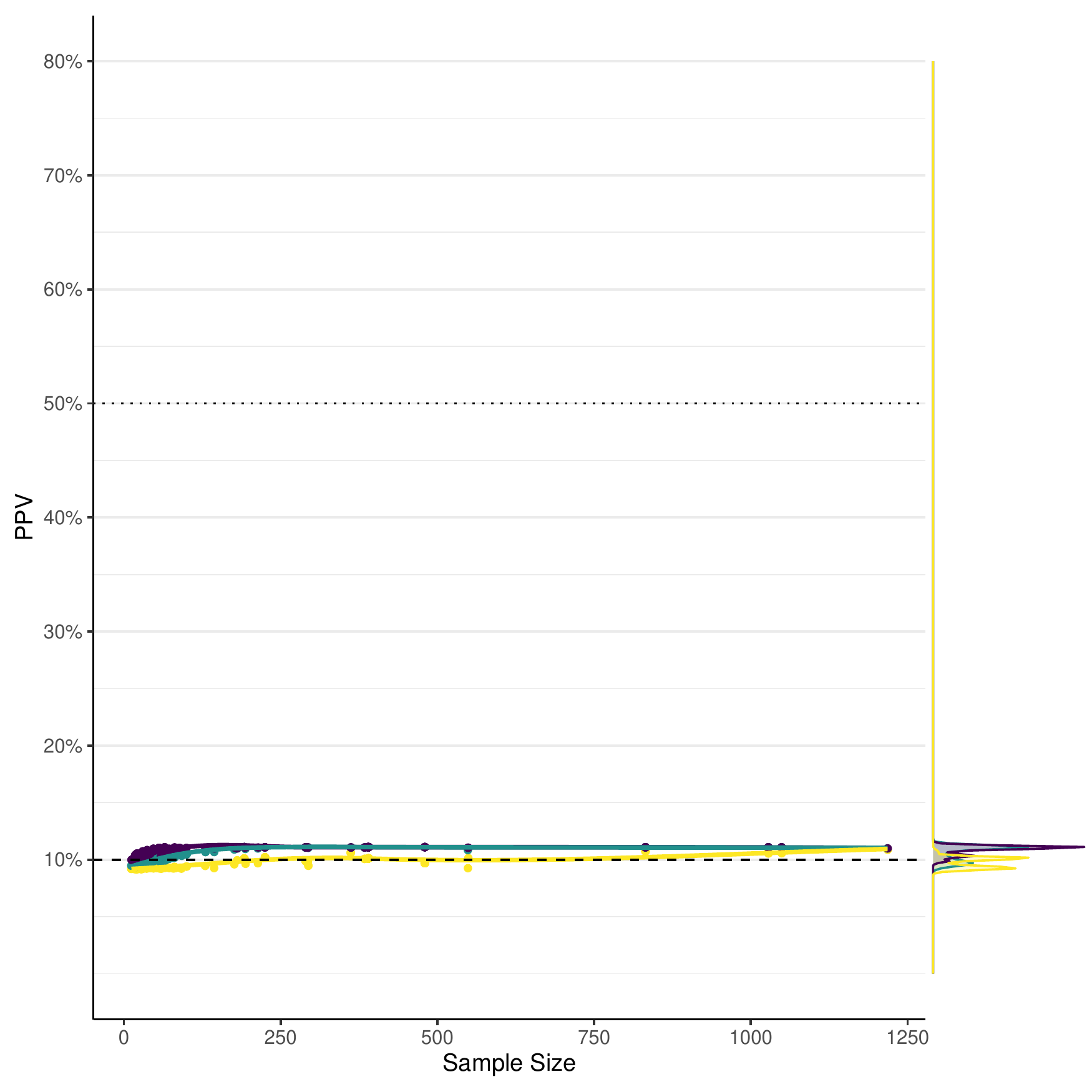}
\end{minipage} &
\begin{minipage}{0.30\textwidth}%
\includegraphics[keepaspectratio,width=\columnwidth]{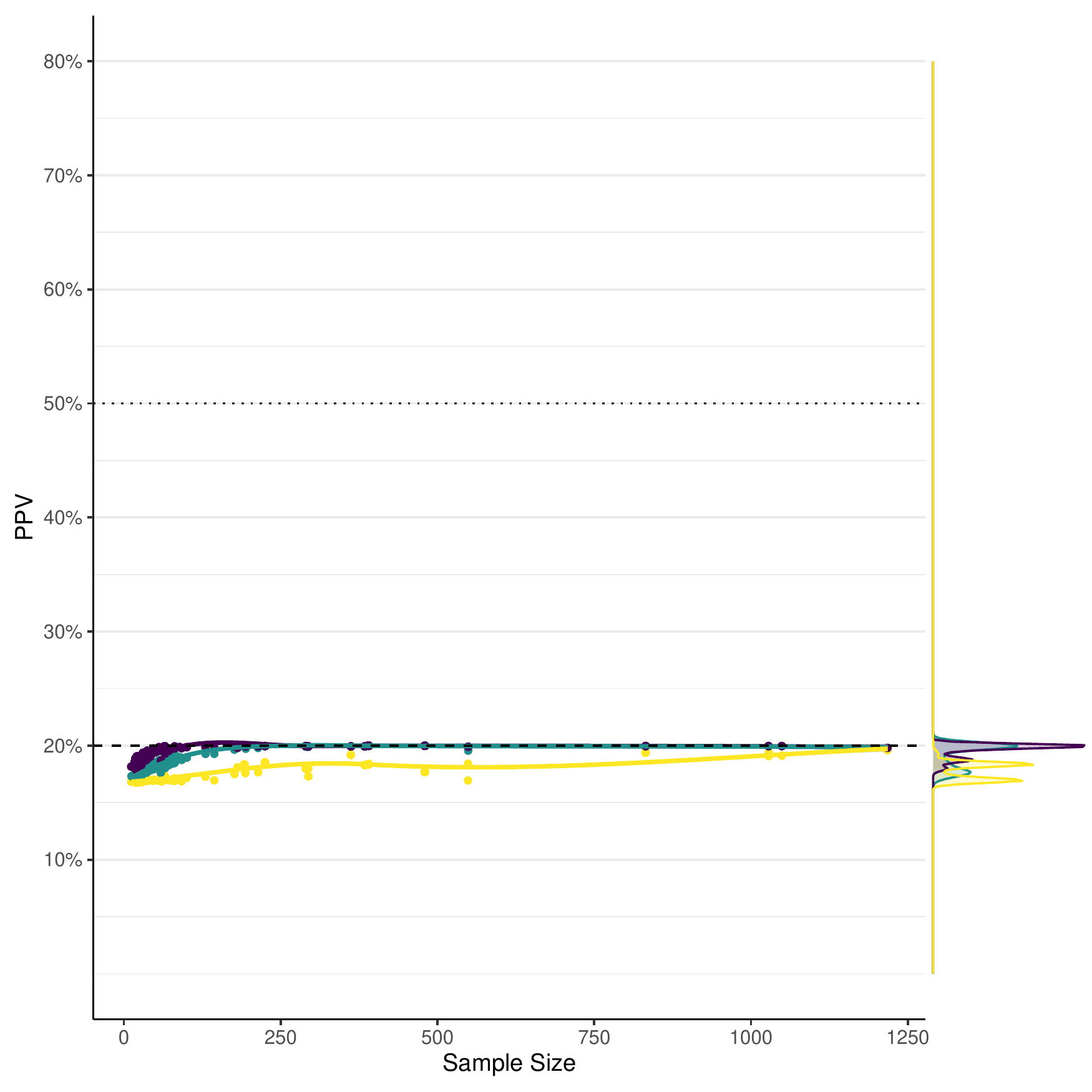}
\end{minipage} &
\begin{minipage}{0.30\textwidth}%
\includegraphics[keepaspectratio,width=\columnwidth]{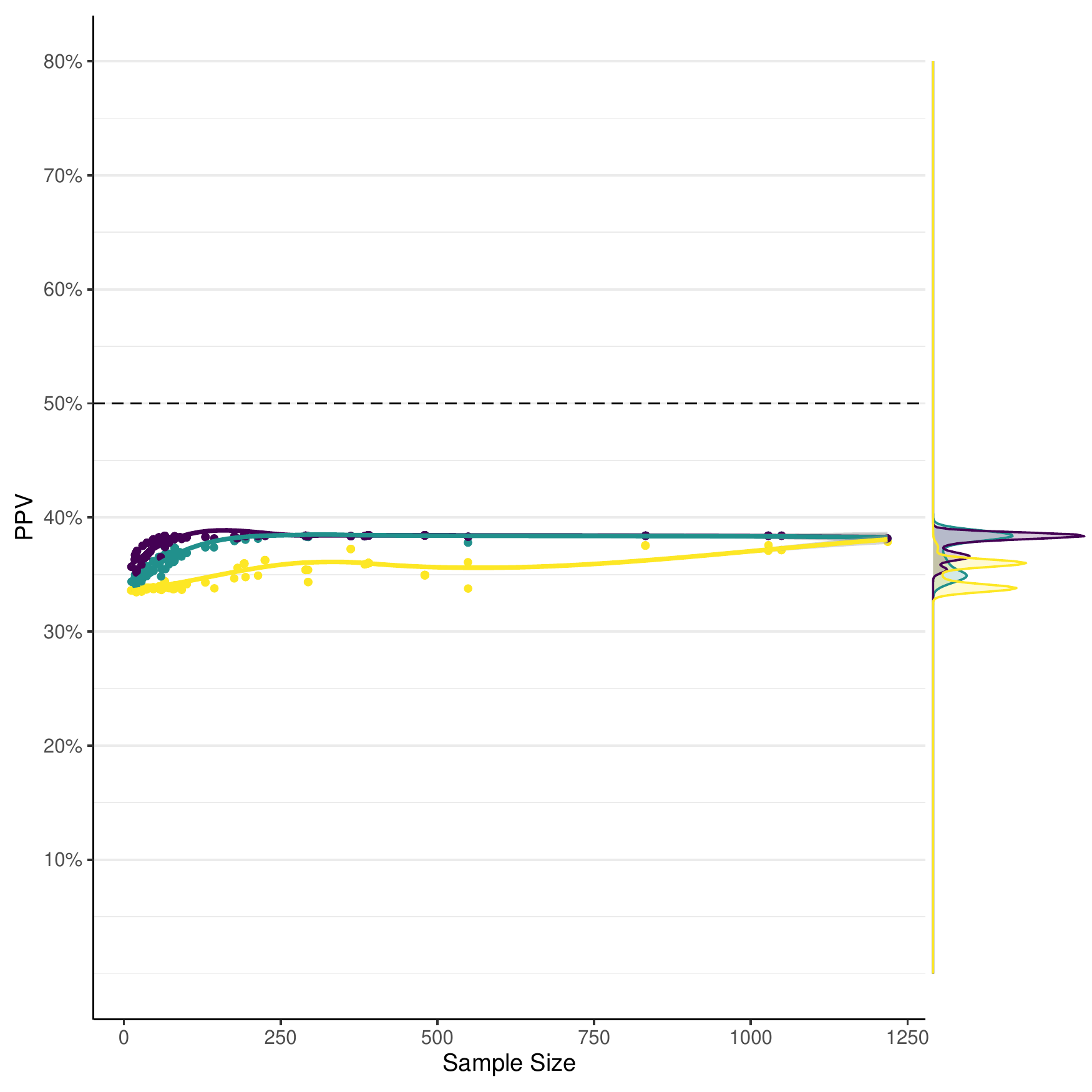}
\end{minipage}\\
\bottomrule
\end{tabular}
\caption{Positive Predictive Value (PPV) versus bias and prior}
\label{fig:ppvDensityMCC}
\end{figure*}
}

\newcommand{\fprpDensityMCCBiasByPrior}{
\begin{figure*}[tbp]
\centering\captionsetup{position=bottom}
\begin{tabular}{llccc}
\toprule
&& \multicolumn{3}{c}{Prior Probability (Confirmatoriness)}\\
\midrule
&& Exploratory (prior = 0.1) & Intermediate (prior = 0.2) & Confirmatory (prior = 0.5)\\
\cmidrule(lr){3-3}\cmidrule(lr){4-4}\cmidrule(lr){5-5}
\multirow{21}{*}{
\rot{Study Bias}} &
\multirow{7}{*}{
\rot{\rlap{Well-run RCT ($u = .2$)}}} &
\begin{minipage}{0.30\textwidth}%
\includegraphics[keepaspectratio,width=\columnwidth]{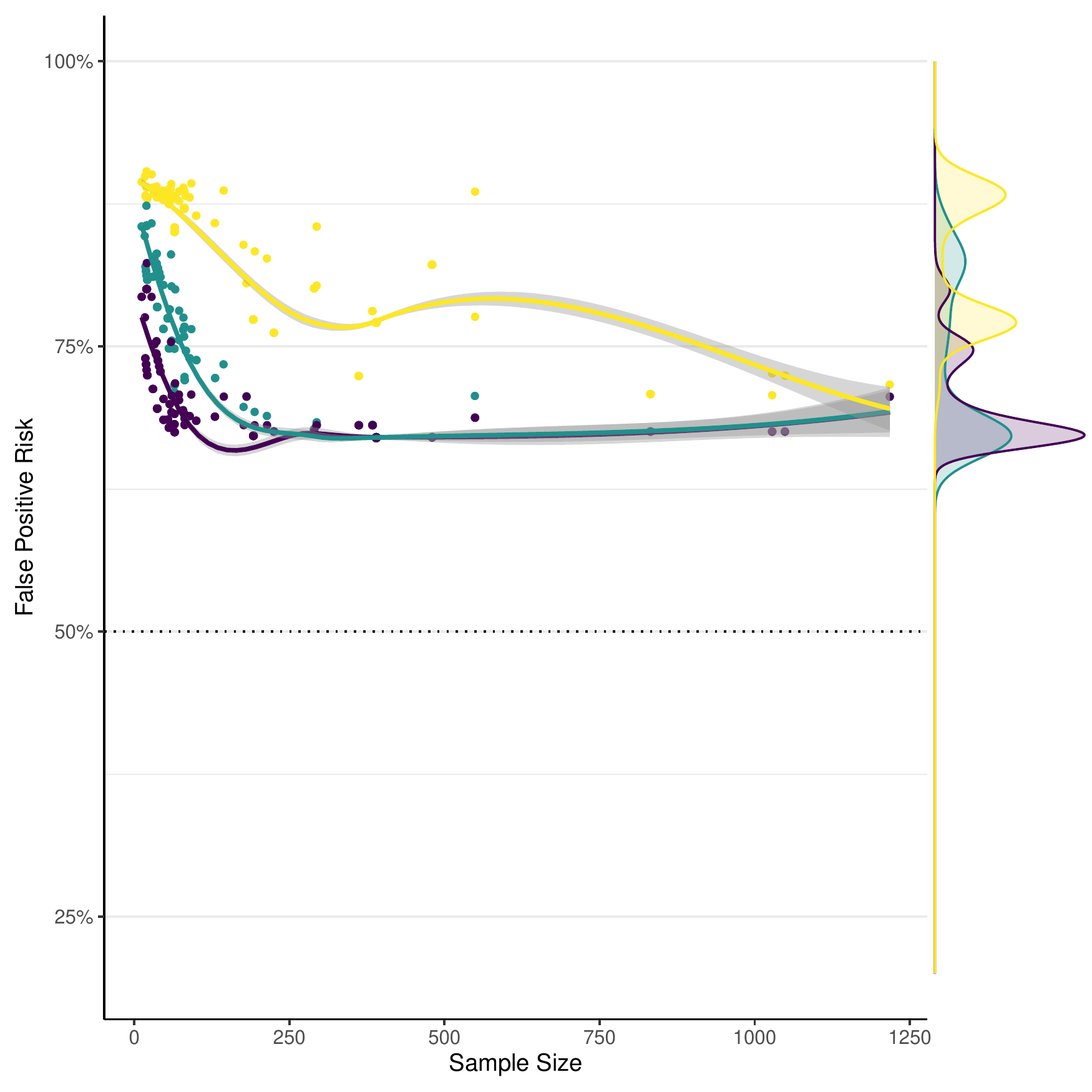}
\end{minipage} &
\begin{minipage}{0.30\textwidth}%
\includegraphics[keepaspectratio,width=\columnwidth]{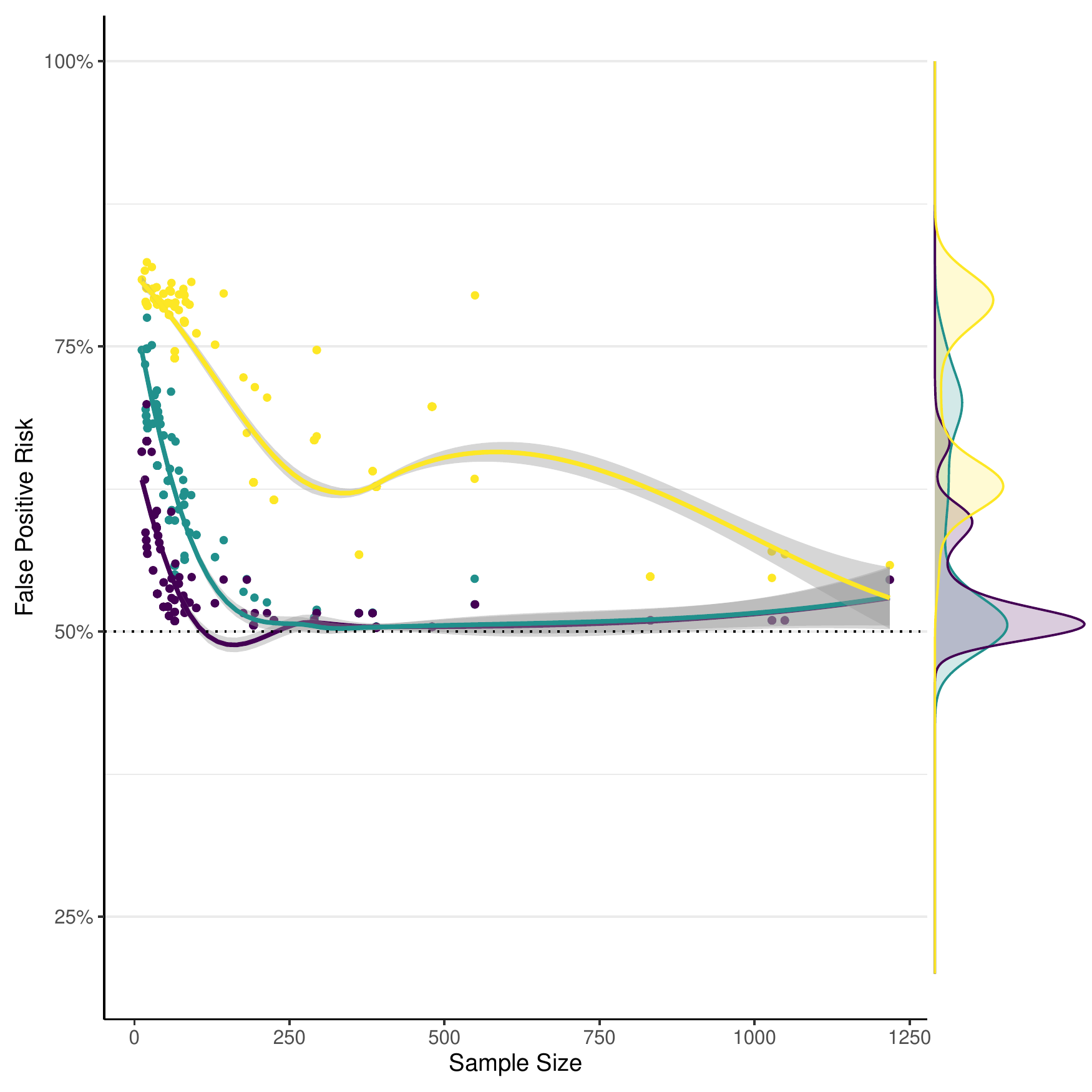}
\end{minipage} & 
\begin{minipage}{0.30\textwidth}%
\includegraphics[keepaspectratio,width=\columnwidth]{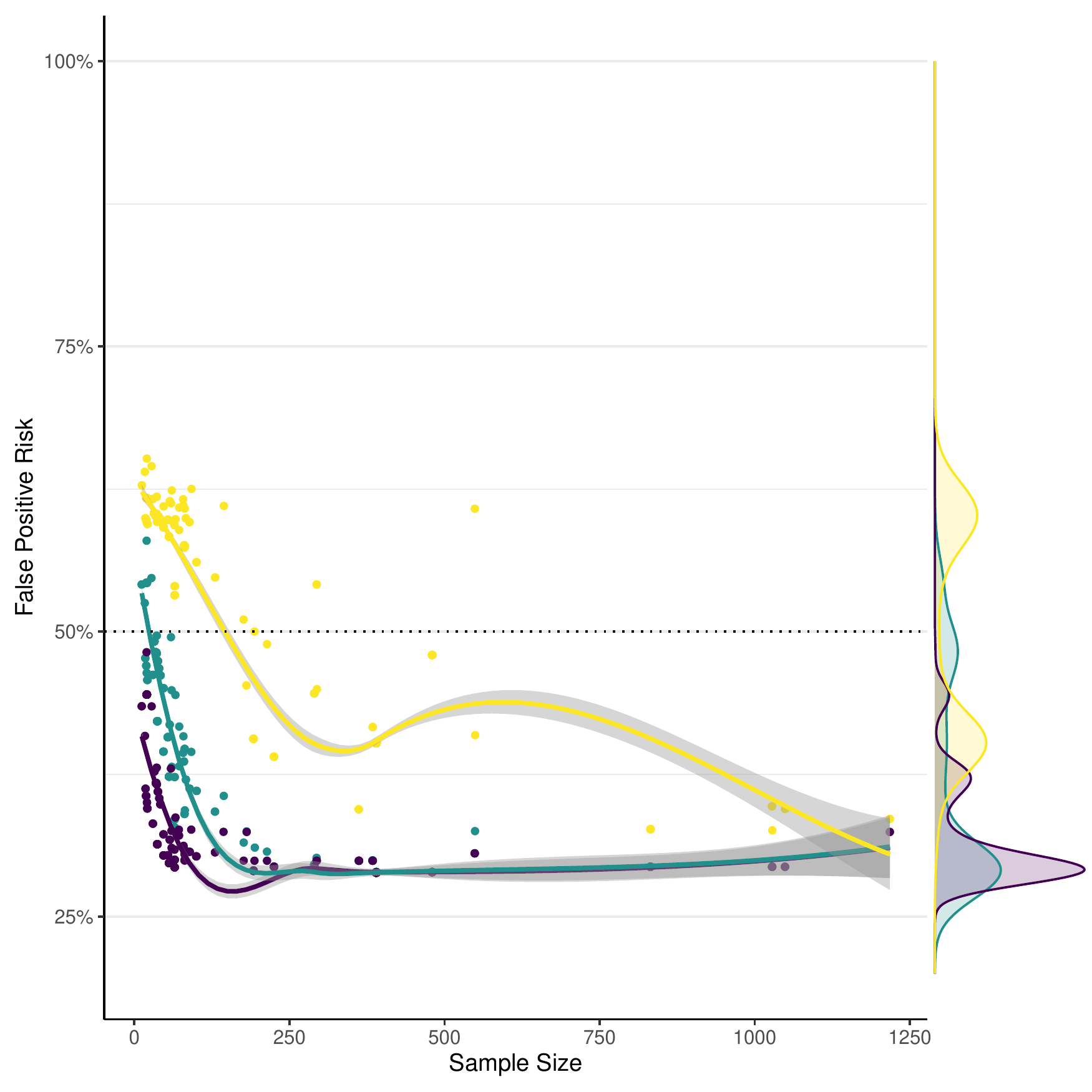}
\end{minipage}\\
&
\multirow{7}{*}{
\rot{\rlap{Weak RCT ($u = .3$)}}} &
\begin{minipage}{0.30\textwidth}%
\includegraphics[keepaspectratio,width=\columnwidth]{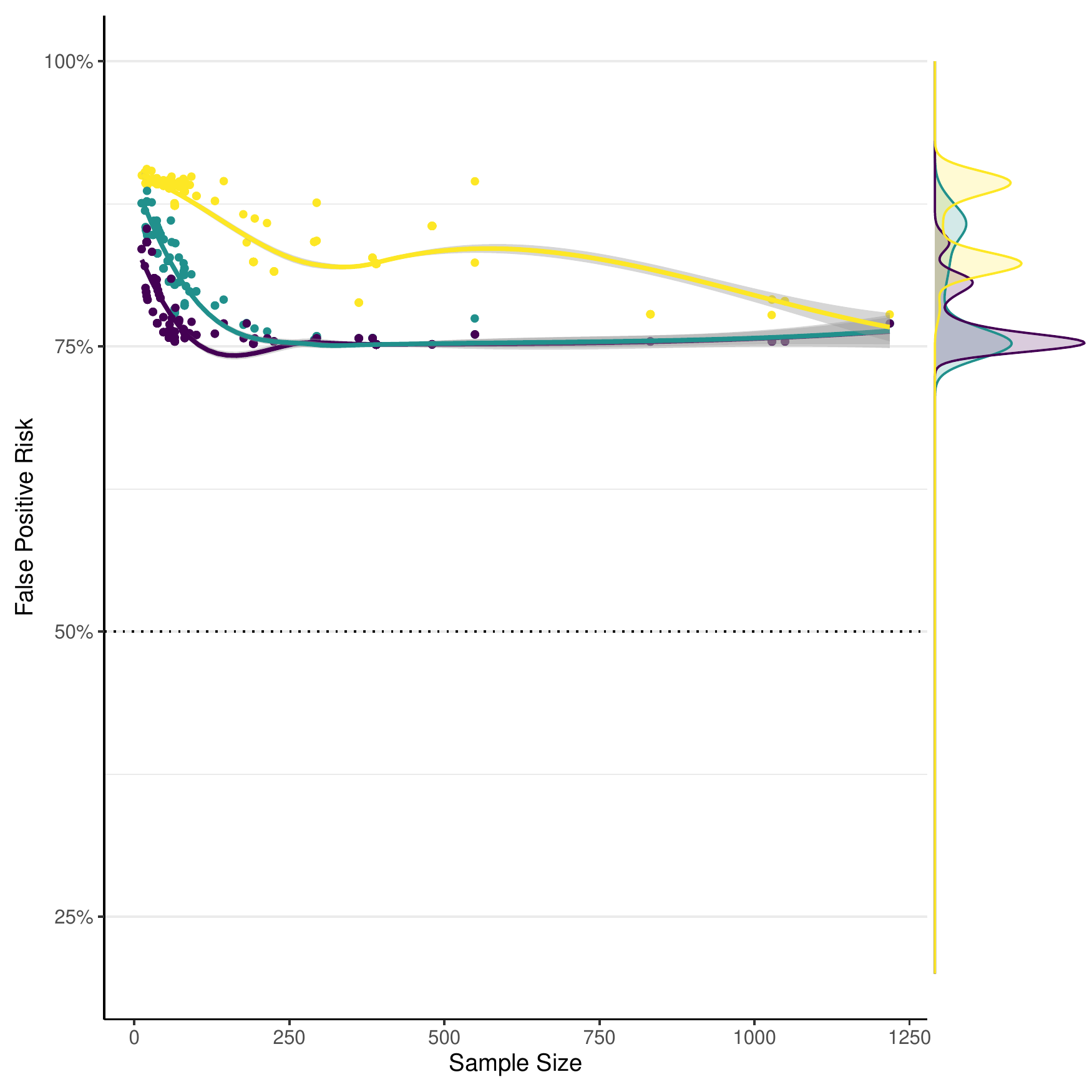}
\end{minipage} &
\begin{minipage}{0.30\textwidth}%
\includegraphics[keepaspectratio,width=\columnwidth]{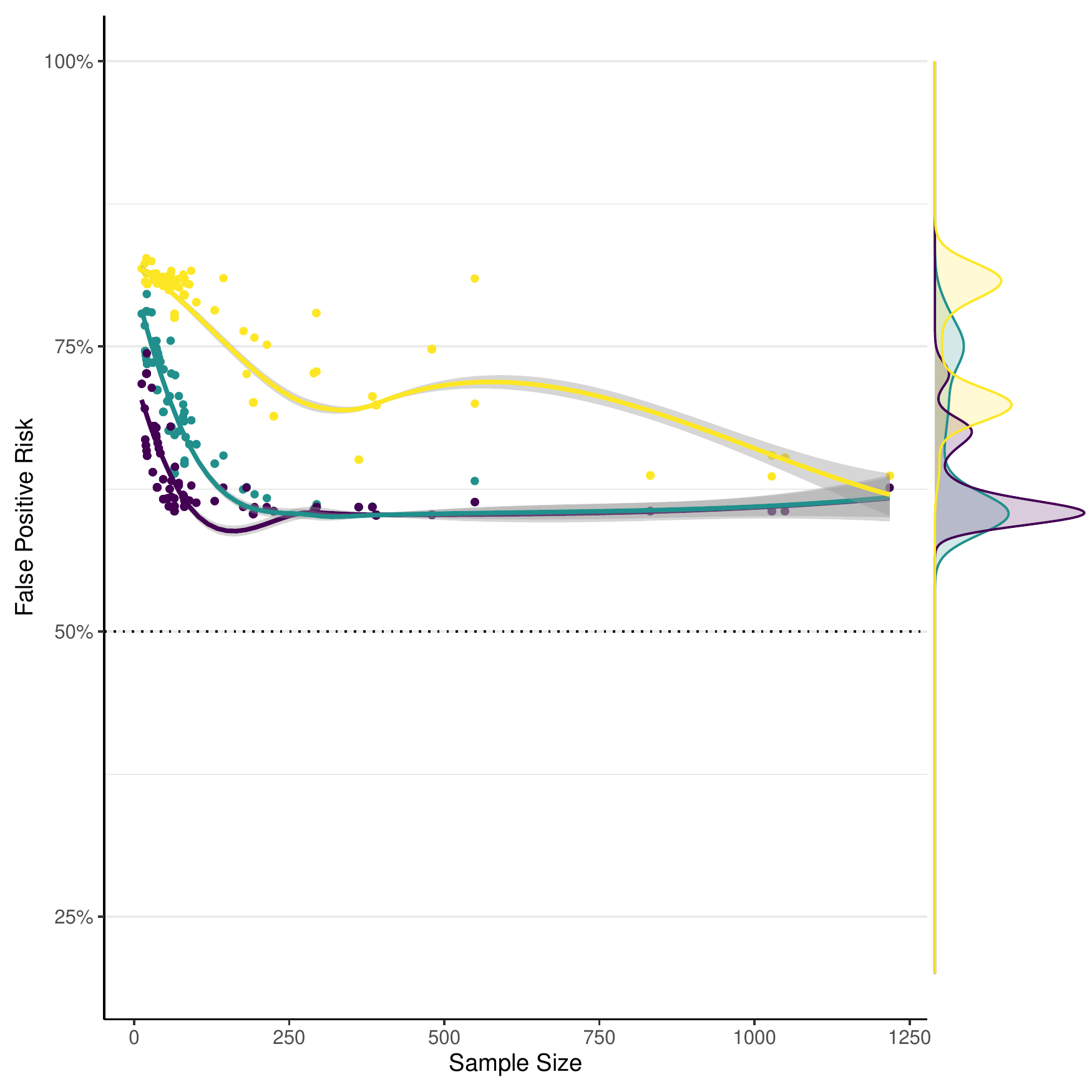}
\end{minipage} & 
\begin{minipage}{0.30\textwidth}%
\includegraphics[keepaspectratio,width=\columnwidth]{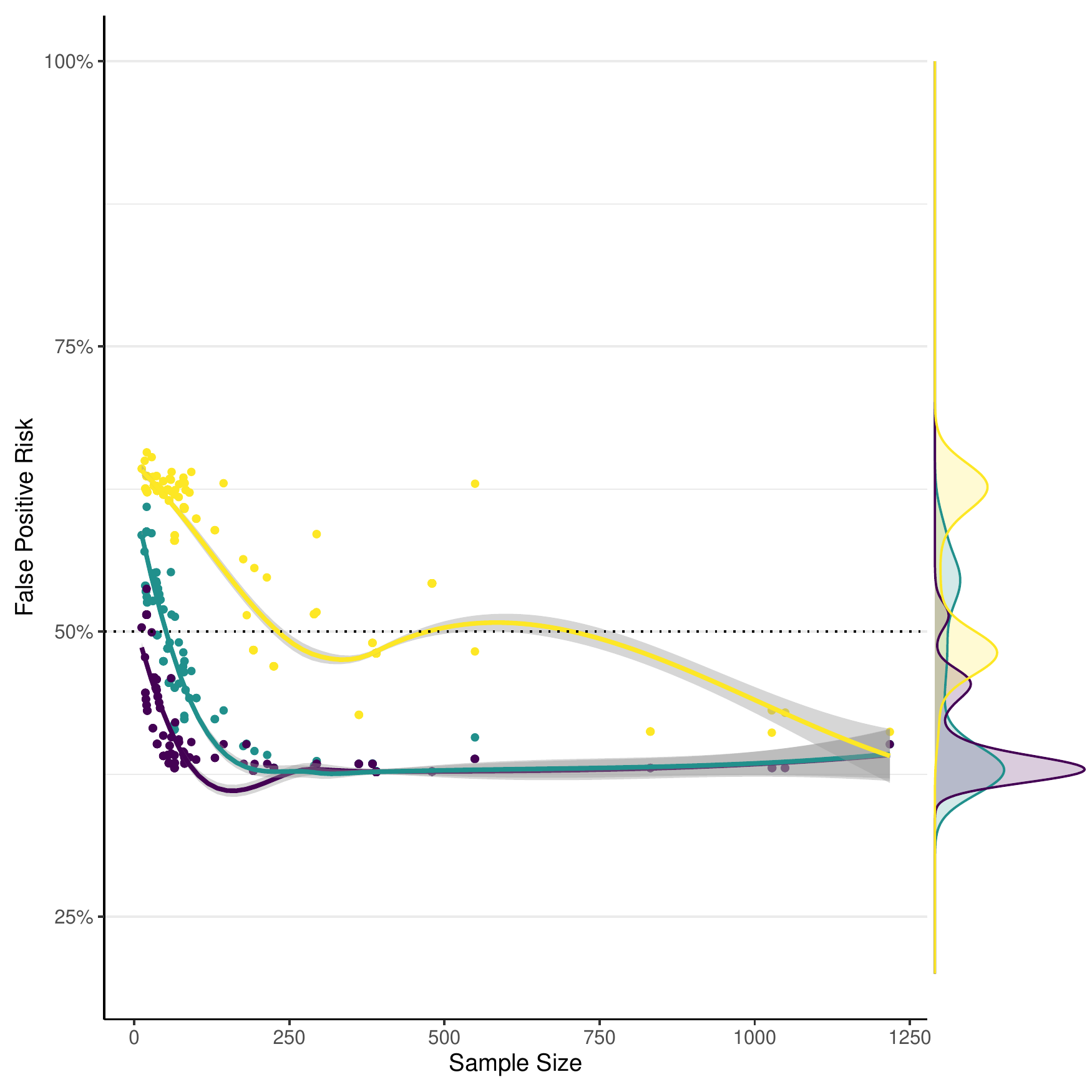}
\end{minipage}\\
&
\multirow{7}{*}{
\rot{\rlap{Biased Study ($u = .8$)}}} &
\begin{minipage}{0.30\textwidth}%
\includegraphics[keepaspectratio,width=\columnwidth]{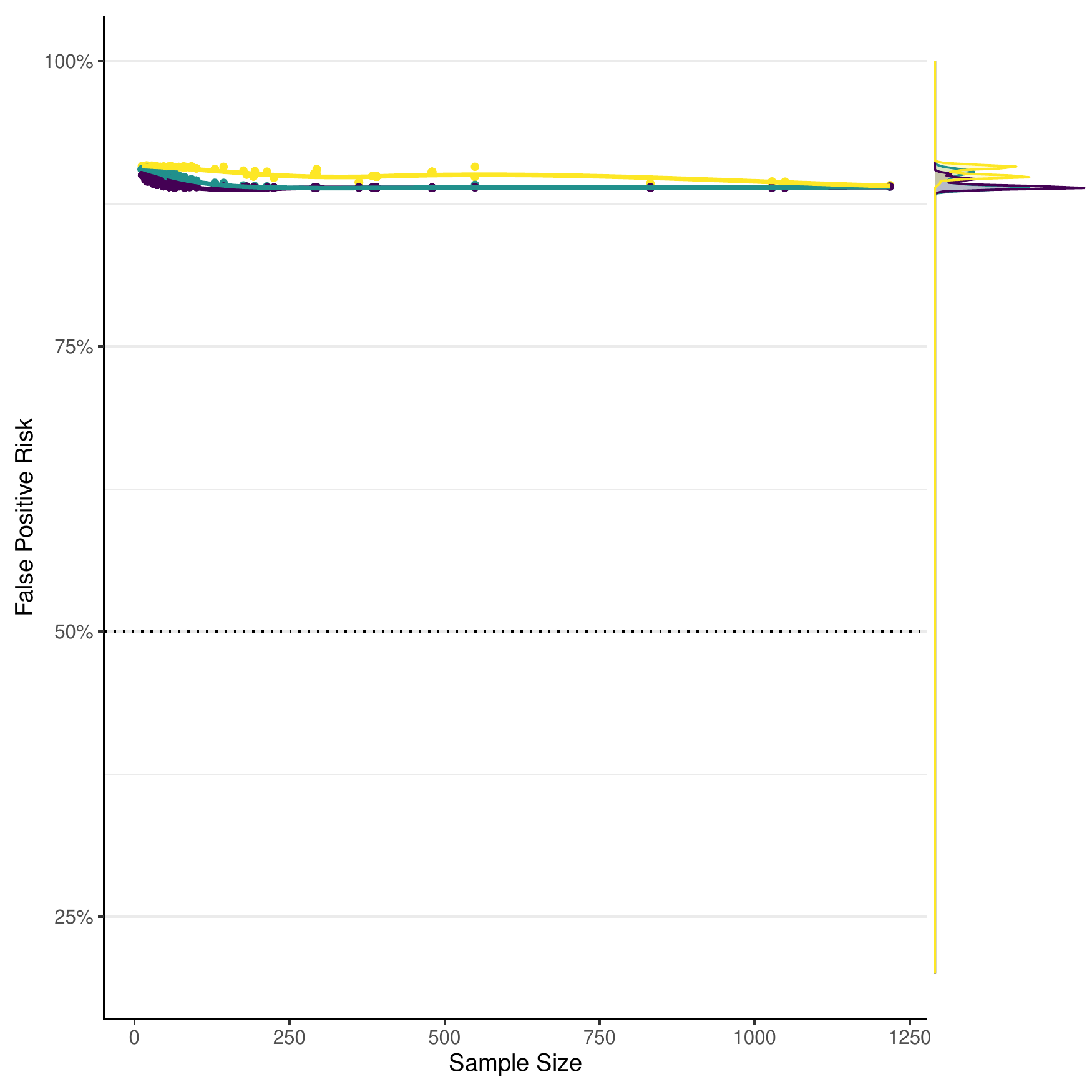}
\end{minipage} &
\begin{minipage}{0.30\textwidth}%
\includegraphics[keepaspectratio,width=\columnwidth]{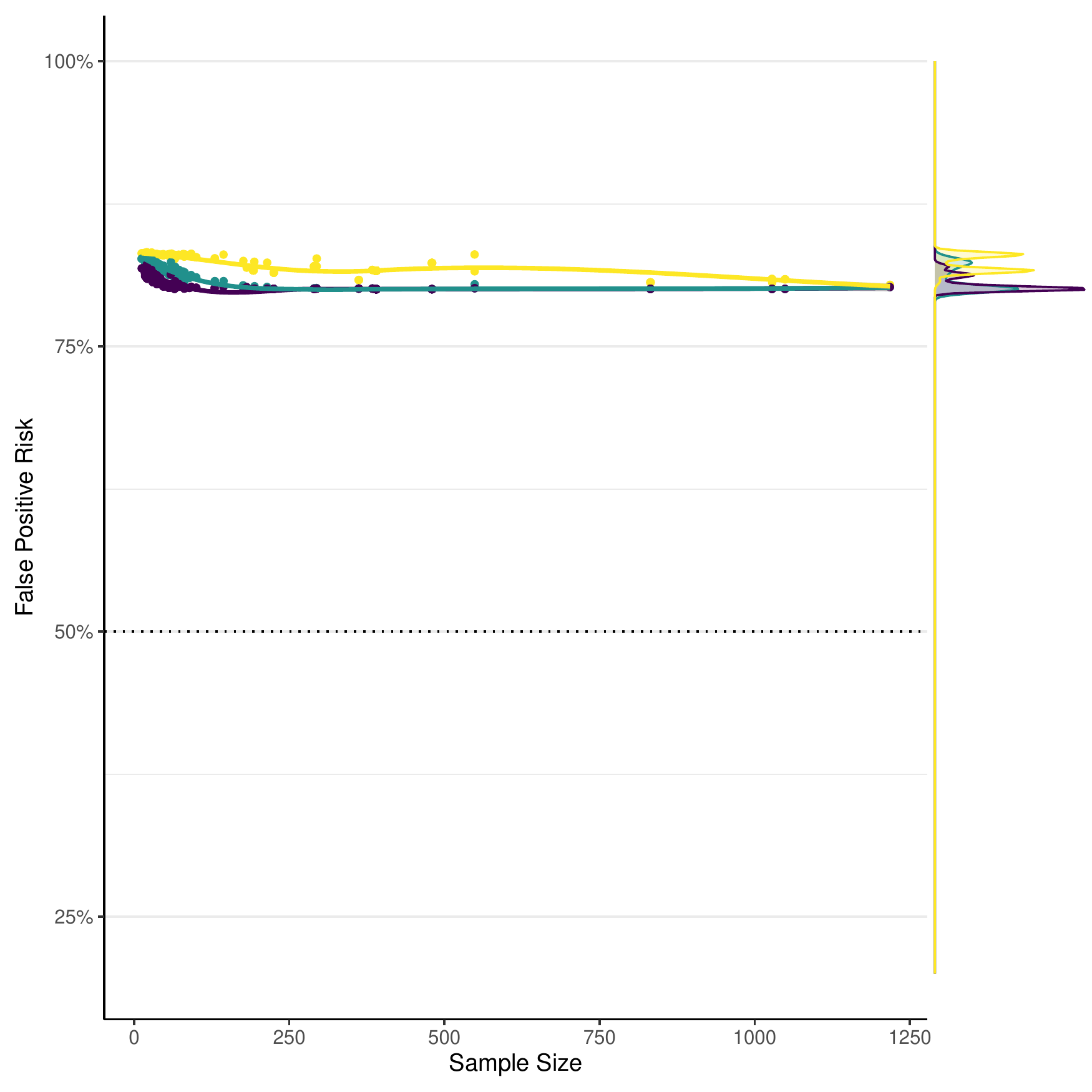}
\end{minipage} &
\begin{minipage}{0.30\textwidth}%
\includegraphics[keepaspectratio,width=\columnwidth]{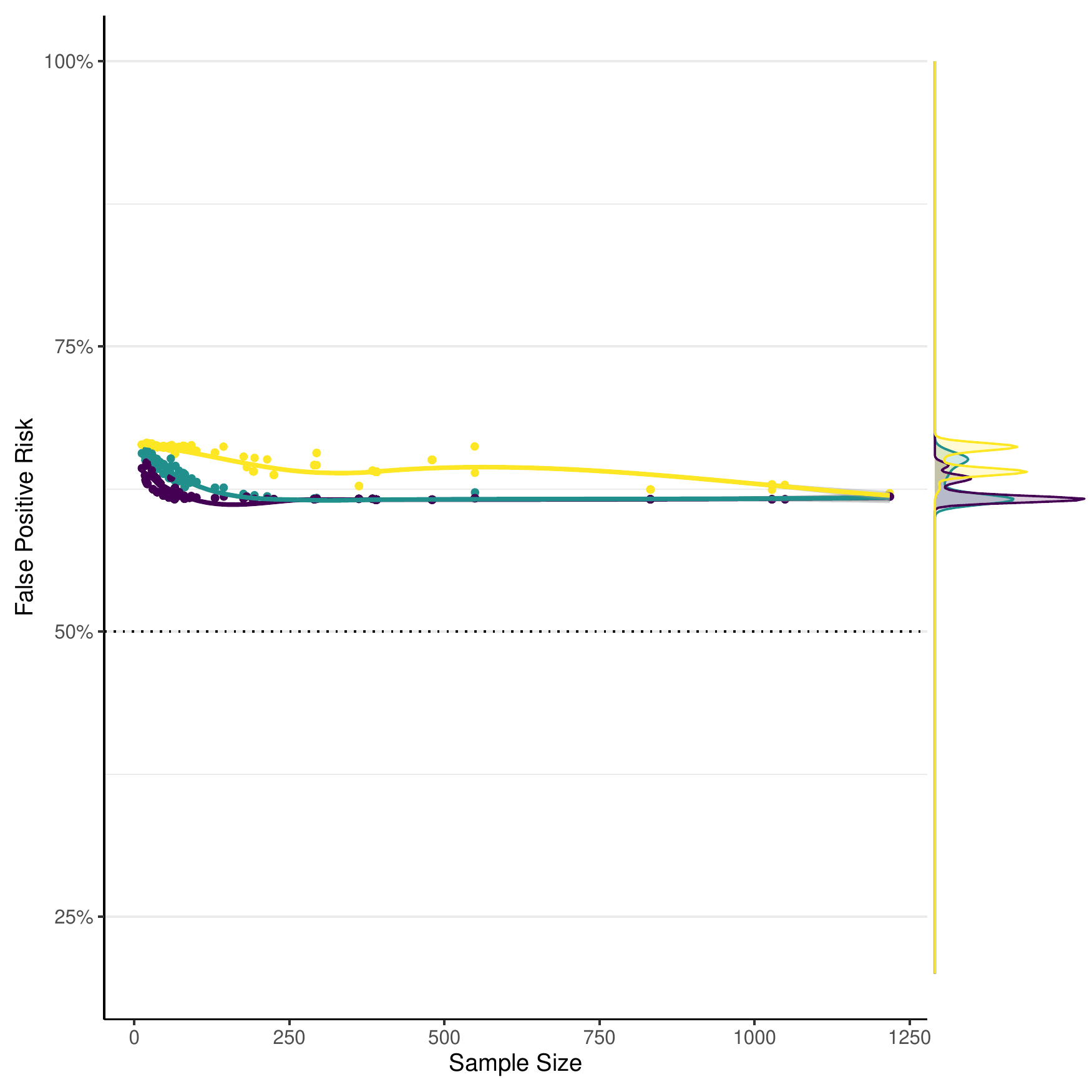}
\end{minipage}\\
\bottomrule
\end{tabular}
\caption{False Positive Risk (FPR) versus bias and prior. \emph{Note:} Statistics are based on family-wise multiple comparison corrections. Effect size thresholds are
    \textcolor{viriyellow}{small} ($d = .2$),
    \textcolor{virigreen}{medium} ($d = .5$), and 
    \textcolor{viriviolet}{large} ($d = .8$).}
\label{fig:fprpDensityMCCBiasByPrior}
\end{figure*}
}

\newcommand{\scatterFPRPintermediateWRCT}{
\begin{figure*}[tbp]
\centering

\includegraphics[width=\maxwidth]{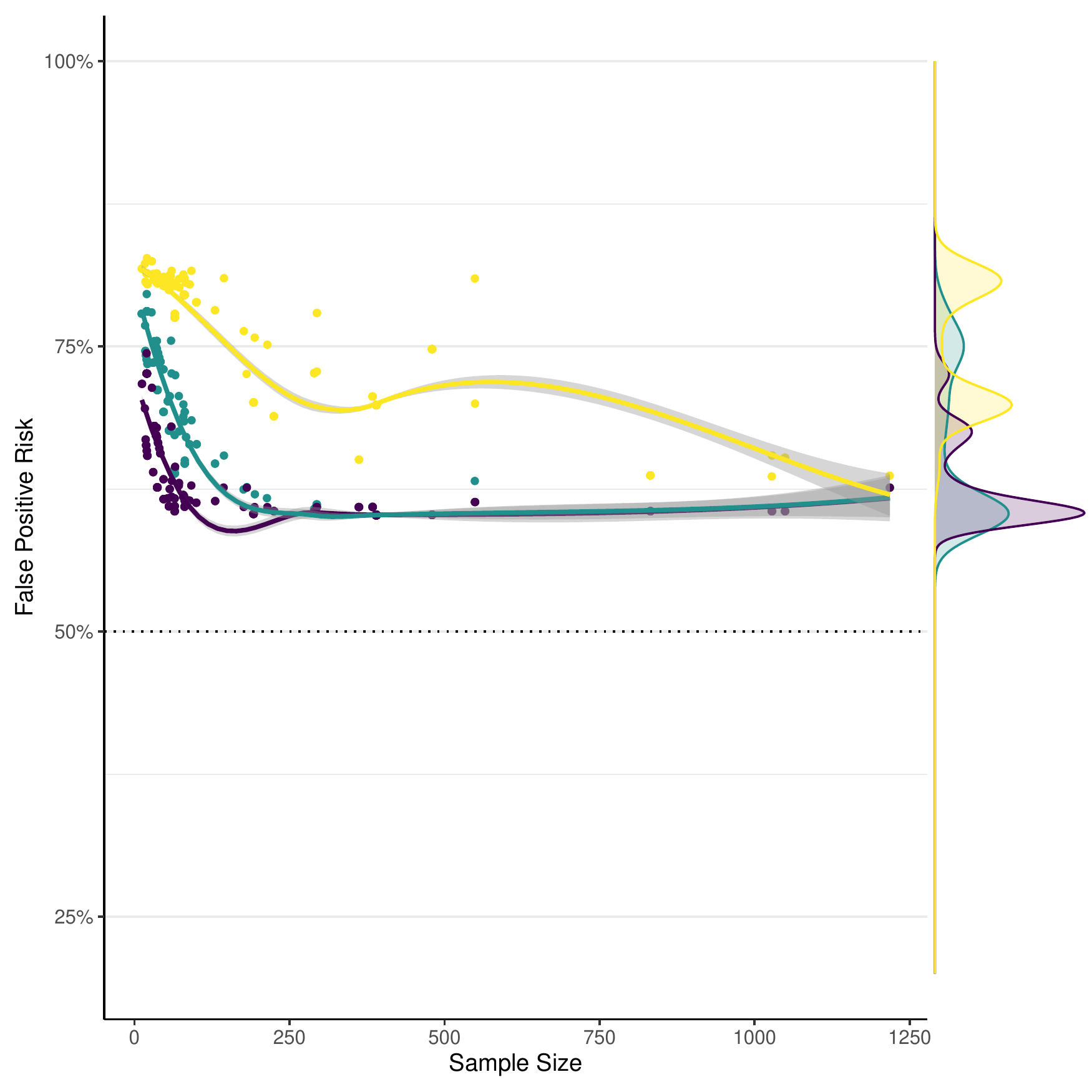} 
\caption{False Positive Risk for a weak Random Controlled Trial (RCT). \emph{Note:} Parameters are fixed to bias $u = 0.3$, $\vari{prior} = 0.2$. The statistics are multiple-comparison corrected. Effect size thresholds are
    \textcolor{viriyellow}{small} ($d = .2$),
    \textcolor{virigreen}{medium} ($d = .5$), and 
    \textcolor{viriviolet}{large} ($d = .8$).}
\label{fig:fprp_WRCT}
\end{figure*}
}

\newcommand{\scatterLRMCC}{
\begin{figure*}[tbp]
\centering

\includegraphics[width=\maxwidth]{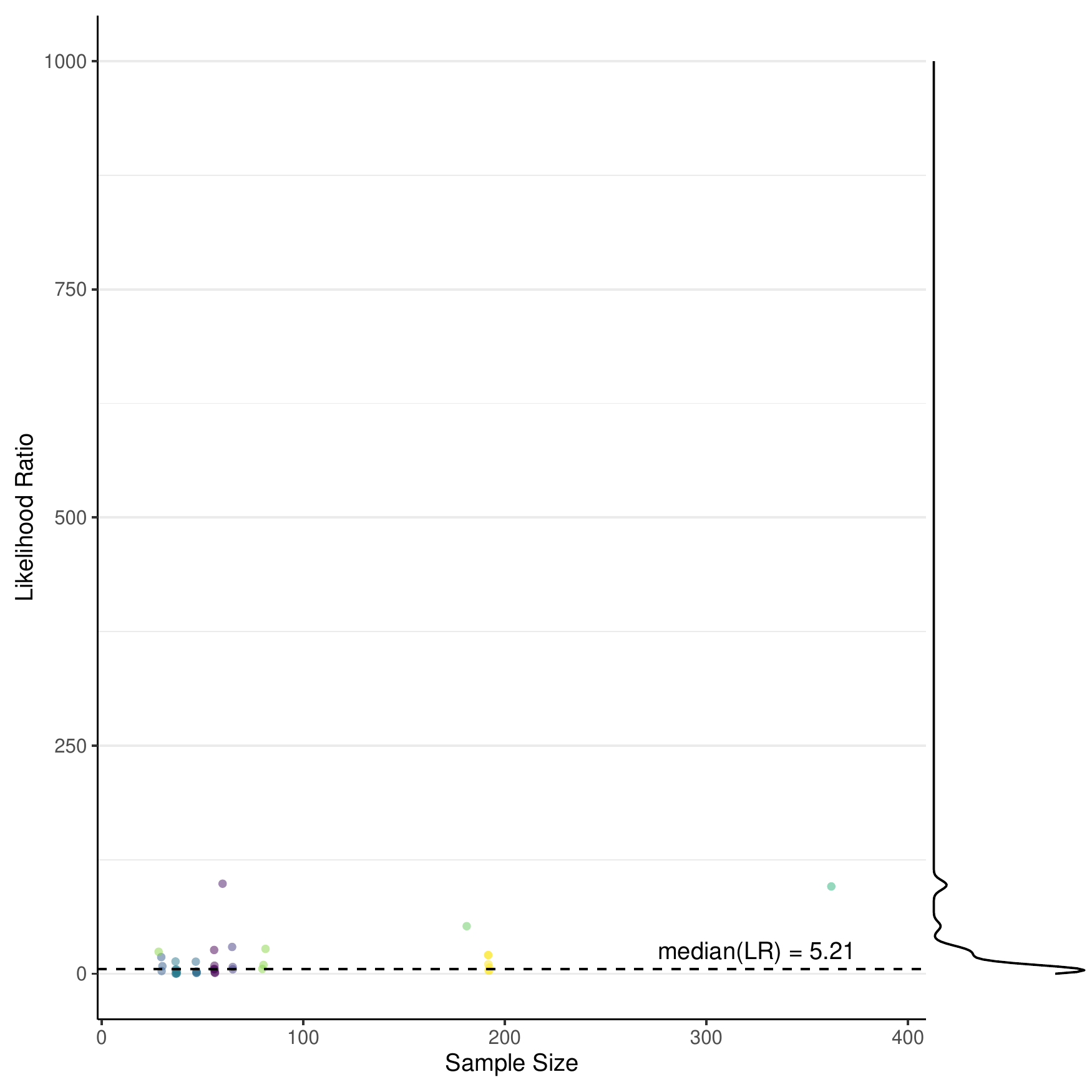} 
\caption{Distribution of the likelihood ratio by sample size. \emph{Note:} The effect size threshold is fixed at medium ($d = 0.5$). We limited the displayed Likelihood Ratio to $\vari{LR} < 1000$ for visual clarity.}
\label{fig:scatterLR}
\end{figure*}
}

\newcommand{\scatterLR}{
\begin{figure*}[tbp]
\centering

\includegraphics[width=\maxwidth]{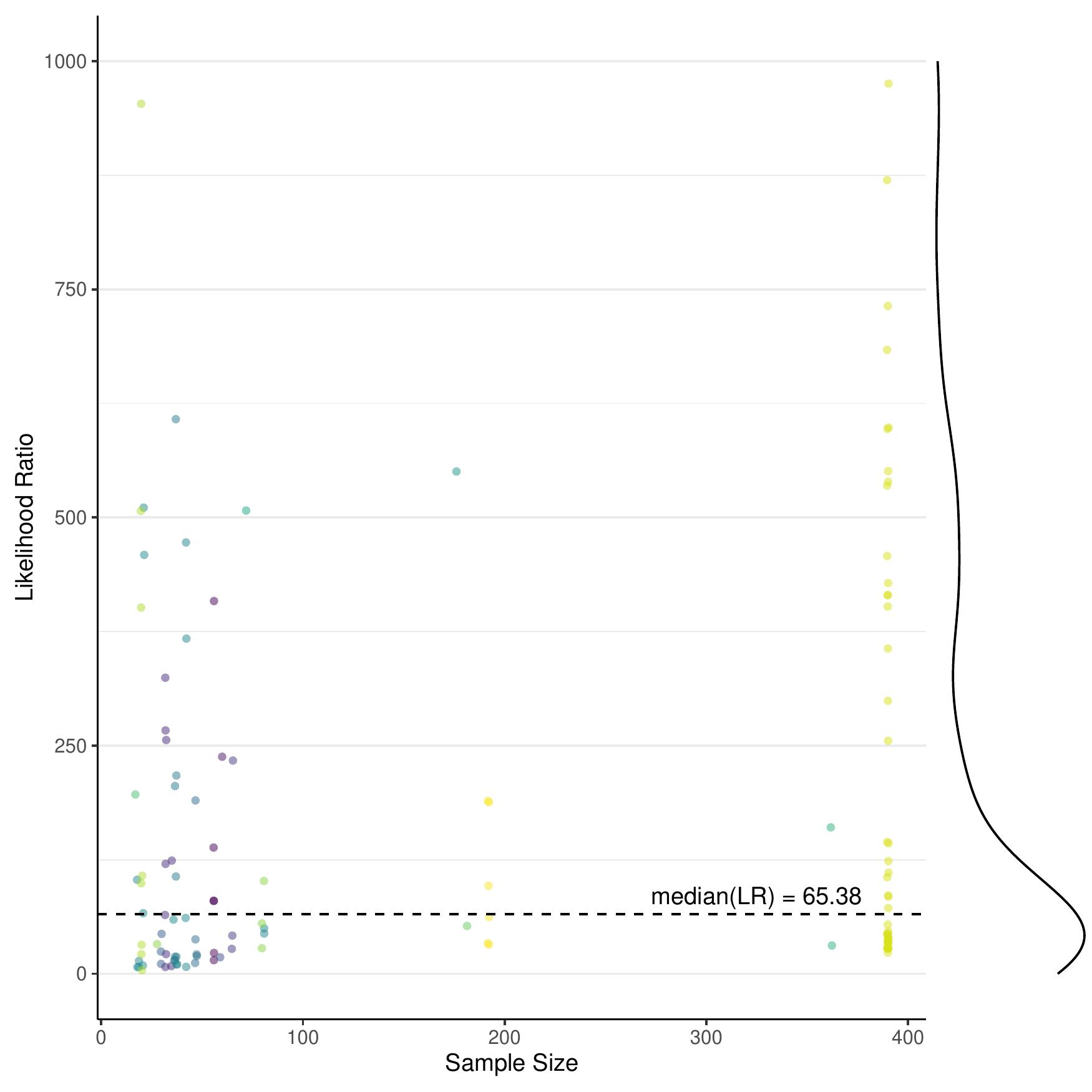} 
\caption{Distribution of the likelihood ratio by sample size. \emph{Note:} The effect size threshold is fixed at medium ($d = 0.5$). We limited the displayed Likelihood Ratio to $\vari{LR} < 1000$ for visual clarity.}
\label{fig:scatterLR}
\end{figure*}
}

\newcommand{\lrScatterCombined}{
\begin{figure*}[tbp]
\centering\captionsetup{position=bottom}
\begin{minipage}{0.49\textwidth}%
\subfloat[Without MCC]{
\label{fig:scatterLR}
\centering\includegraphics[keepaspectratio,width=\columnwidth]{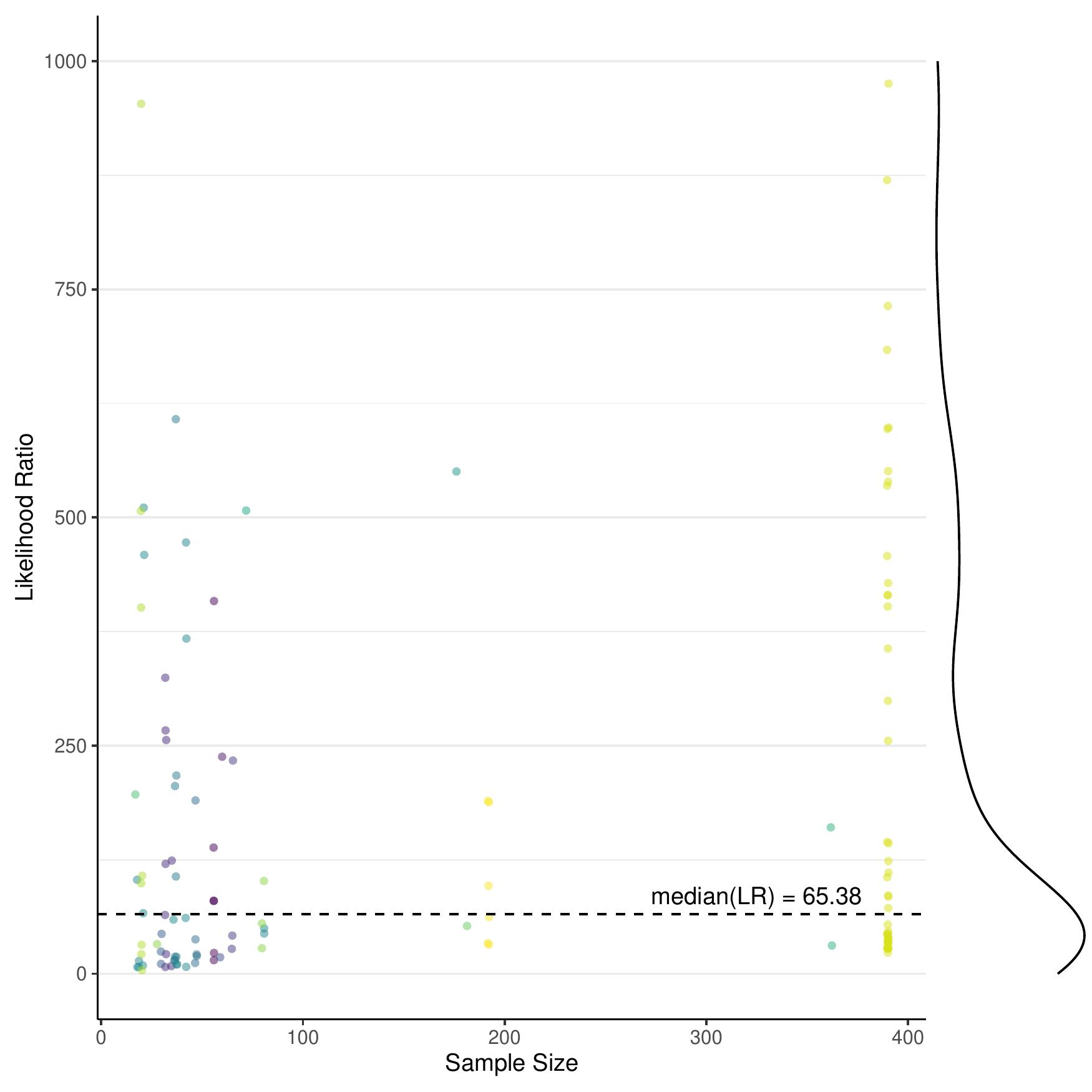}
}
\end{minipage}~
\begin{minipage}{0.49\textwidth}%
\subfloat[With MCC]{%
\label{fig:scatterLRMCC}
\centering\includegraphics[keepaspectratio,width=\columnwidth]{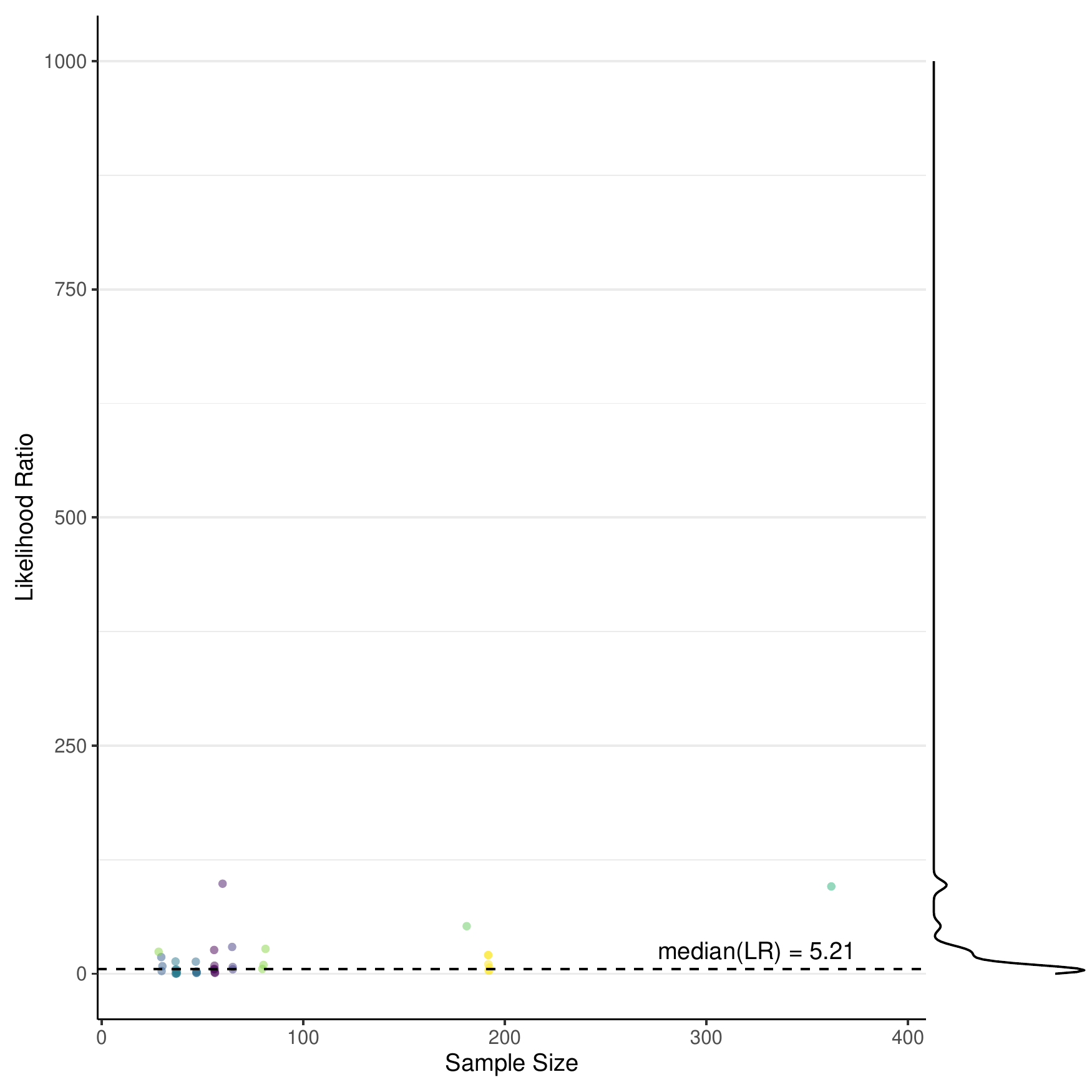}
}
\end{minipage}
\caption{Distribution of the likelihood ratio by sample size. \emph{Note:} The effect size threshold is fixed at medium ($d = 0.5$). We limited the displayed Likelihood Ratio to $\vari{LR} < 1000$ for visual clarity.}
\label{fig:lrScatterCombined}
\end{figure*}
}

\newcommand{\scatterRBPMCC}{
\begin{figure*}[tbp]
\centering

\includegraphics[width=\maxwidth]{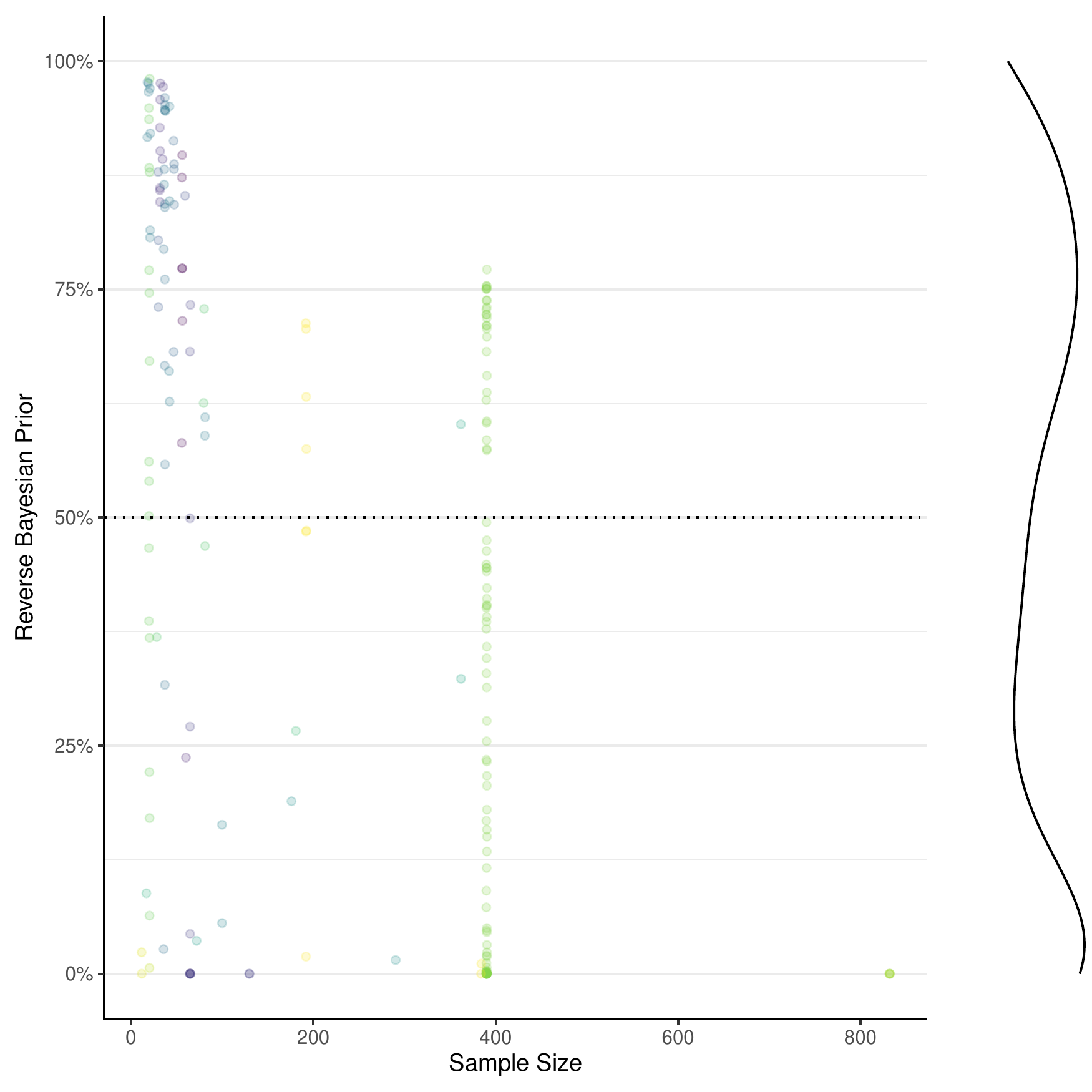} 
\caption{Reverse Bayesian Prior. \emph{Note:} The dot colors denote different studies.}
\label{fig:scatterRBPMCC}
\end{figure*}
}

\newcommand{\scatterRBP}{
\begin{figure*}[tbp]
\centering

\includegraphics[width=\maxwidth]{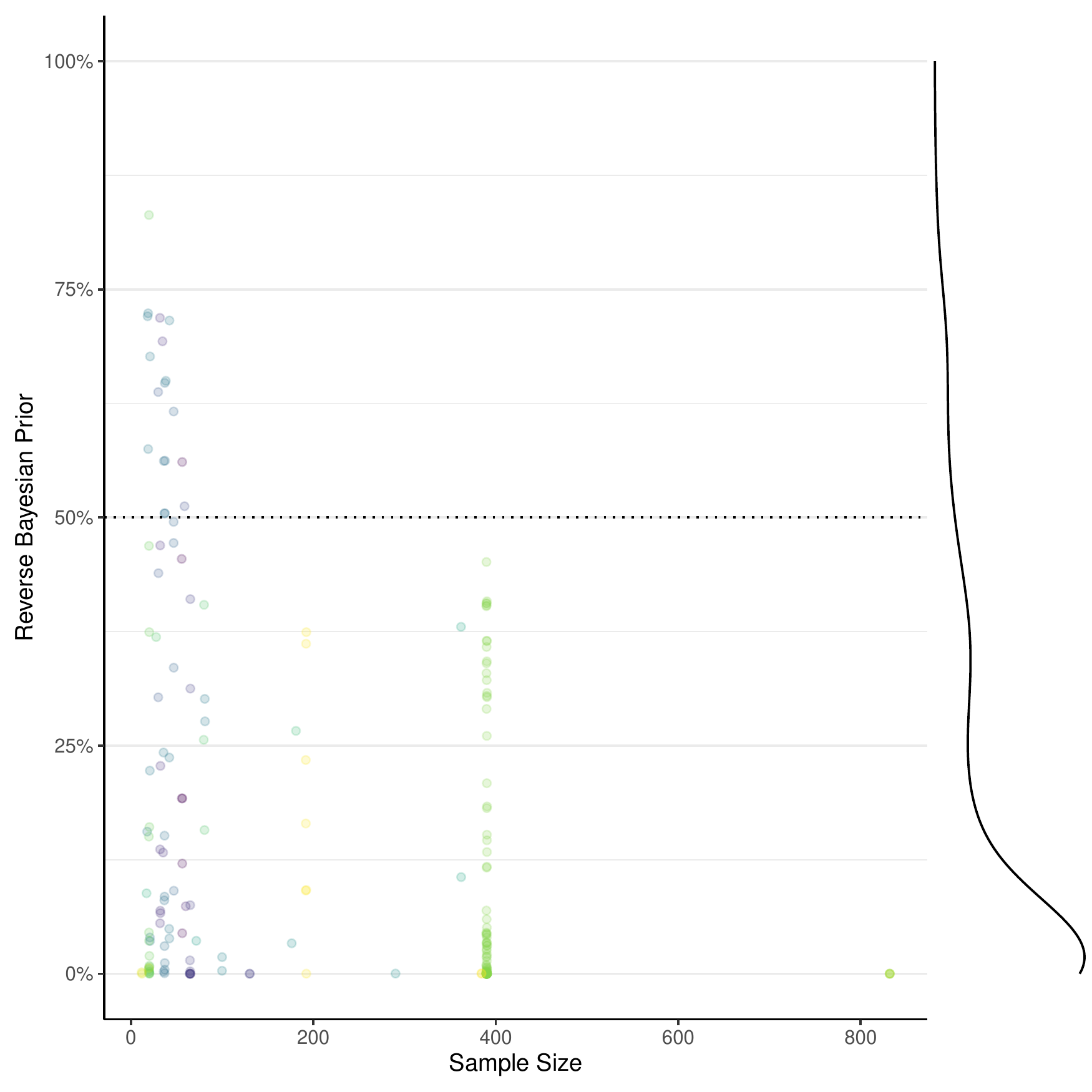} 
\caption{Reverse Bayesian Prior. \emph{Note:} The dot colors denote different studies.}
\label{fig:scatterRBP}
\end{figure*}
}

\newcommand{\rbpScatterCombined}{
\begin{figure*}[tbp]
\centering\captionsetup{position=bottom}
\begin{minipage}{0.49\textwidth}%
\subfloat[Without MCC]{
\label{fig:scatterRPB}
\centering\includegraphics[keepaspectratio,width=\columnwidth]{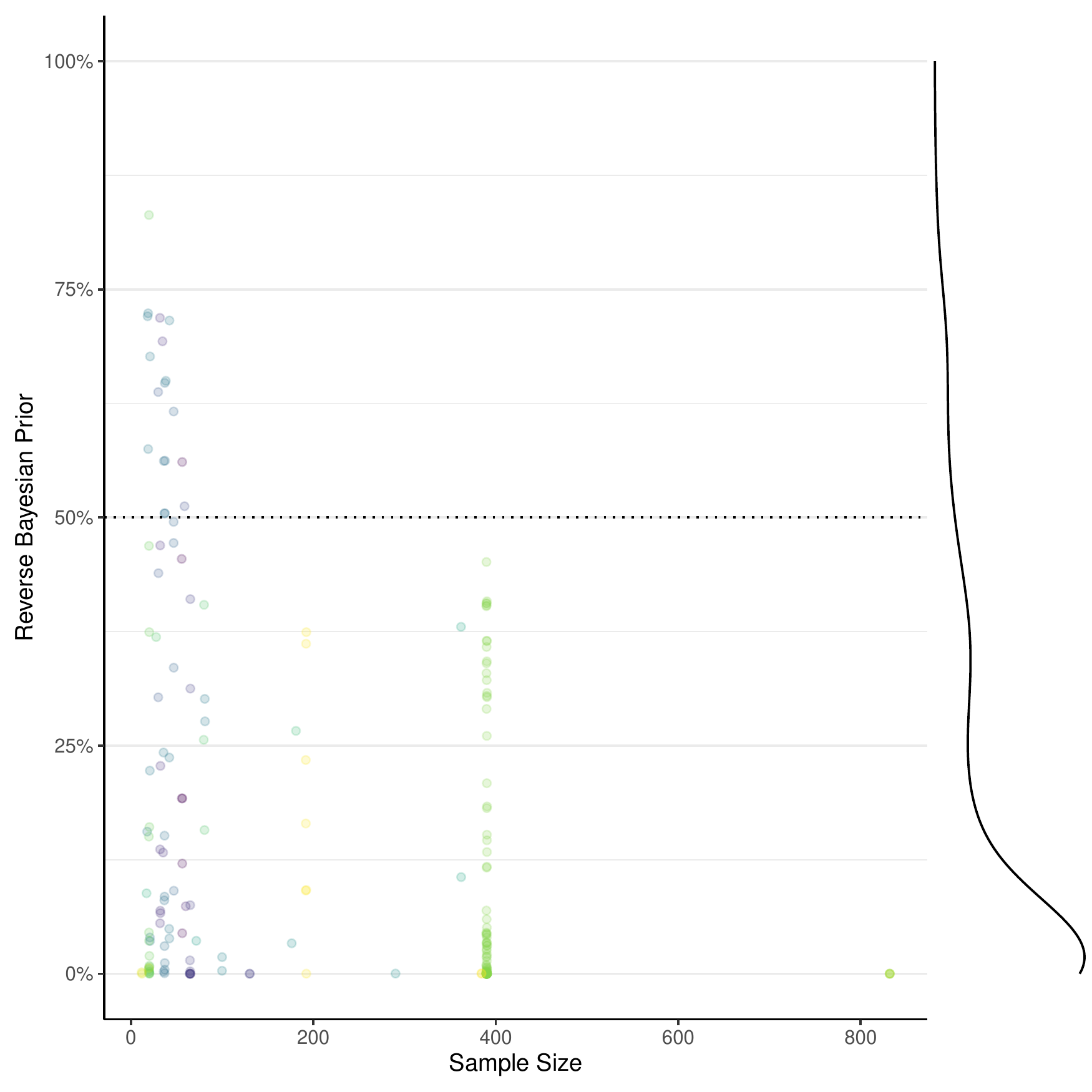}
}
\end{minipage}~
\begin{minipage}{0.49\textwidth}%
\subfloat[With MCC]{%
\label{fig:scatterRBPMCC}
\centering\includegraphics[keepaspectratio,width=\columnwidth]{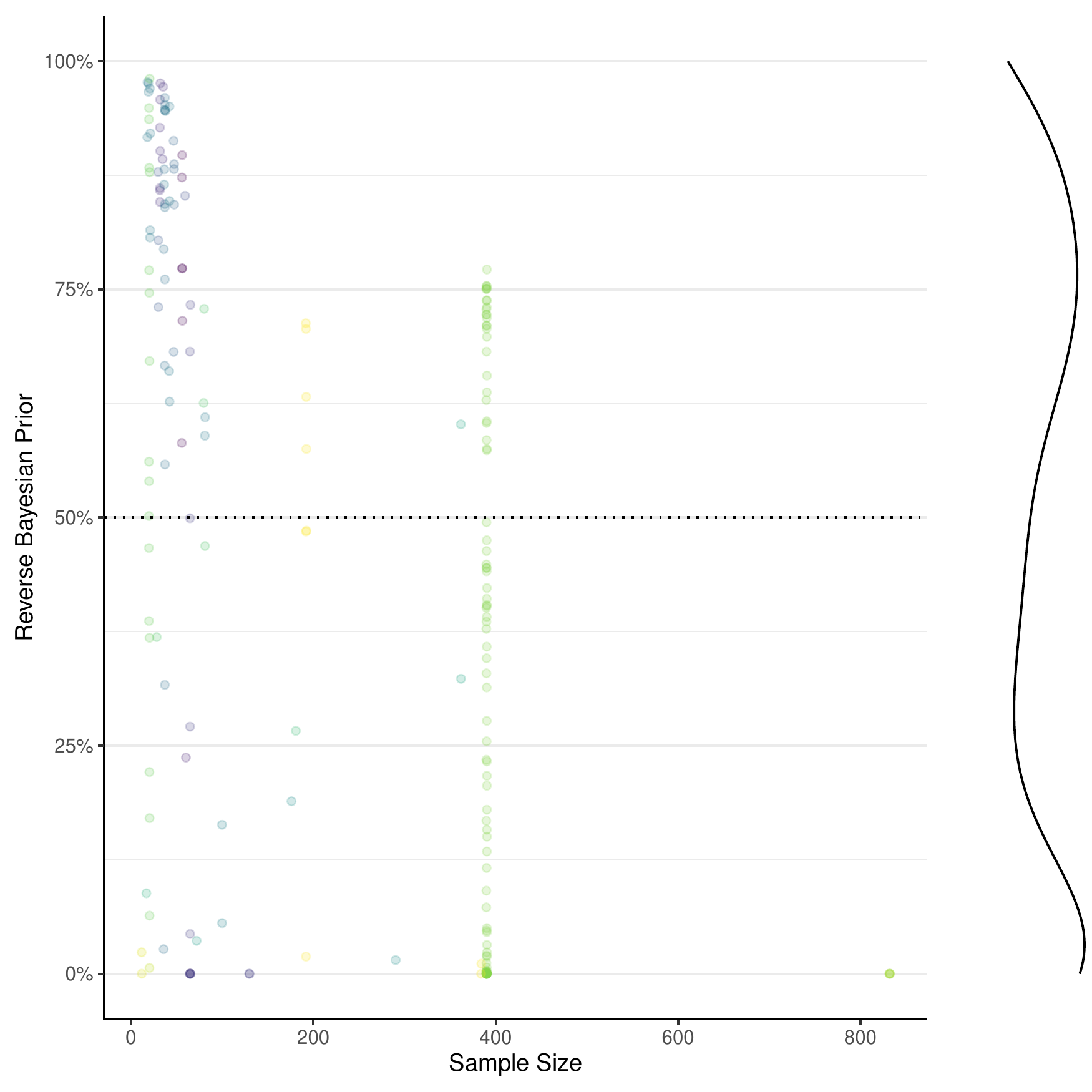}
}
\end{minipage}
\caption{Reverse Bayesian prior without and with multiple-comparison corrections.}
\label{fig:rbpScatterCombined}
\end{figure*}
}

\newcommand{\scatterACPARBP}{
\begin{figure*}[tbp]
\centering

\includegraphics[width=\maxwidth]{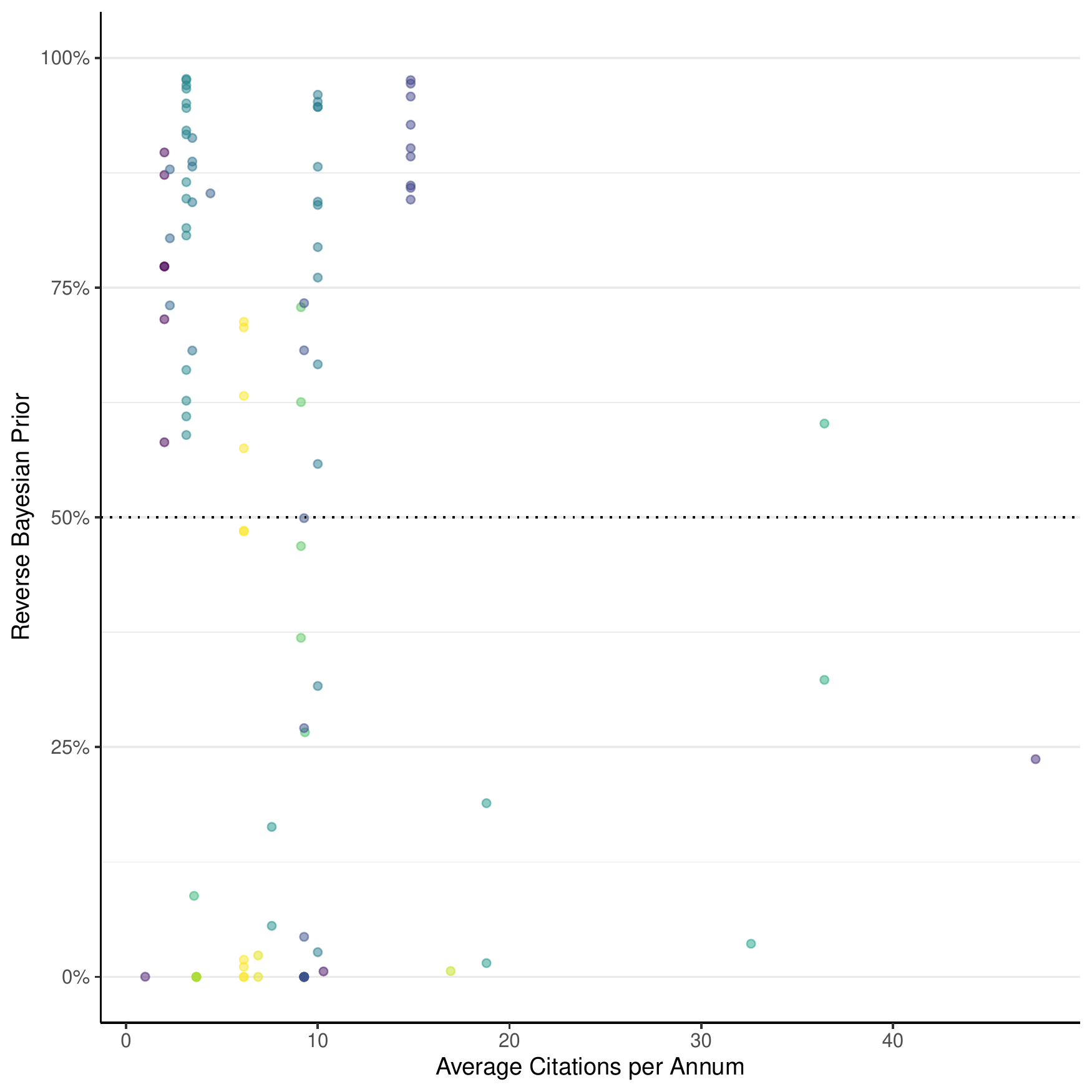} 
\caption{Reverse Bayesian Prior by Average Citations per Annum. \emph{Note:} The dot colors denote different studies.}
\label{fig:scatterACPARBP}
\end{figure*}
}


\date{}

\begin{DocumentVersionConference}
\title{Why Most Results of Socio-Technical Security User Studies Are False\thanks{Open Science Framework: \protect\url{https://osf.io/7gv6h/}.}}
\end{DocumentVersionConference}

\author{Thomas Gro{\ss}}
\institute{School of Computing\\Newcastle University, UK\\
\email{thomas.gross@newcastle.ac.uk}}

\maketitle

\begin{abstract}
\noindent\textbf{Background.} 
 In recent years, cyber security user studies have been scrutinized for their reporting completeness, statistical reporting fidelity, statistical reliability and biases. It remains an open question what strength of evidence positive reports of such studies actually yield. We focus on the extent to which positive reports indicate relation true in reality, that is, a probabilistic assessment.

\noindent\textbf{Aim.} 
 This study aims at establishing the overall strength of evidence in cyber security user studies, with the dimensions 
 \begin{inparaenum}[(a)] 
    \item Positive Predictive Value (PPV) and its complement False Positive Risk (FPR), 
    \item Likelihood Ratio (LR), and
    \item Reverse-Bayesian Prior (RBP) for a fixed tolerated False Positive Risk.
 \end{inparaenum}

\noindent\textbf{Method.}
Based on $431$ coded statistical inferences in $146$ cyber security user studies from a published SLR covering the years 2006--2016, we first compute a simulation of the \emph{a posteriori} false positive risk based on assumed prior and bias thresholds. 
Second, we establish the observed likelihood ratios for positive reports. Third, we compute the reverse Bayesian argument on the observed positive reports by computing the prior required for a fixed \emph{a posteriori} false positive rate. 

\noindent\textbf{Results.}
We obtain a comprehensive analysis of the strength of evidence including an account of appropriate multiple comparison corrections. The simulations show that even in face of well-controlled conditions and high prior likelihoods, only few studies achieve good \emph{a posteriori} probabilities. 

\noindent\textbf{Conclusions.}
Our work shows that the strength of evidence of the field is weak and that most positive reports are likely false. From this, we learn what to watch out for in studies to advance the knowledge of the field.

\end{abstract}


\section{Introduction}

Empirical user studies heave an important place in studying socio-technical security. They investigate user attitude and behaviors in face of security technologies as well as the impact of different interventions. They affect a wide range of topics in the field.

Aiming at advancing the quality of evidence in the field, cyber security user studies have been appraised for a number of factors in recent years:
\begin{inparaenum}[(i)]
  \item their reporting completeness~\cite{SLR2017},
  \item their statistical reporting fidelity~\cite{Gross2019,Gross2019a}, and
  \item their statistical reliability~\cite{Gross2020,Gross2020a}.
\end{inparaenum}

Most recently, Gro{\ss}~\cite{Gross2020} estimated effect sizes from reported statistical tests, simulated statistical power estimates from effect size thresholds, and showed a range of overall biases of the field. Specifically, that prior study investigated the estimated statistical power of studies in question and observed that few studies achieved recommended power thresholds. Thereby, the study observed a power failure in the field. These observations, however,  do not quantify the strength of evidence the studies of the field actually yield. Hence, the research gap we seek to close is estimating the magnitude of the strength of evidence found in the field.

We understand as \emph{strength of evidence} the probability of a claimed relation being true in reality. We will consider multiple metrics to evaluate that probability under different circumstances. First, we consider the Positive Predictive Value (PPV), that is, the \emph{a posteriori} probability of a relation being true after the study was conducted. The estimate of the PPV and its complement, the False Positive Risk (FPR), are dependent on knowing or assuming the prior probability of before the study. Second, unlike the PPV, the likelihood ratio quantifies the strength of evidence independent of the knowledge of a prior. Finally, we investigate the reverse Bayesian prior, that is, the prior probability one would have needed to achieve a desired fixed false positive risk.
These three perspectives, though all drawn from Bayes' law, offer different lenses to appraise the strength of evidence of relevant user studies.

Clearly, the investigation of positive predicted value and related quantities is not entirely new. Most famously, Ioannidis~\cite{ioannidis2005most} made a convincing case that most published results are false, in general. Others have added to this argument, considering false positive risks or replications in a range of fields~\cite{moonesinghe2007most,wacholder2004assessing} or promoted Bayesian views on statistical testing~\cite{howson2006scientific,colquhoun2014investigation,colquhoun2017reproducibility}. In fact, some of the inspiration for this work is drawn from Ioannidis bold proclamation~\cite{ioannidis2005most} and Colquhoun's thoughts on strength of evidence~\cite{colquhoun2017reproducibility}. In this study, however, we are the first to evaluate the strength of evidence in socio-technical security studies based on an empirical grounding.

In addition, we are interested to what extent the field pays attention to the strength of evidence as a factor to make the decision to cite studies. We thereby ask to what extent the number of citations of studies in the field is correlated to the strength of evidence they provide.

Overall, the strength of evidence evaluation provided in this study offers an empirical scaffolding to make decisions on further studies in the field.

\paragraph{Our Contributions.}
We are the first to offer a systematic evaluation of the strength of evidence in socio-technical security user studies. 
Based on a sizeable empirical sample from a systematic literature review and coded statistical tests.
We provide an assessment of false positive risk based on configurable parameters bias, prior, and effect size.
Further, we evaluate the specific strength of evidence of positive reports in investigated user studies, yielding distributions of likelihood ratios and reverse bayesian prior.
Overall, these contributions yield a comprehensive review of the strength of evidence, creating opportunities to make empirically informed decisions to advance the field.


\section{Background}

\subsection{Null Hypothesis Significance Testing}

\newcommand{\Prob}{\ensuremath{\const{P}}}
\newcommand{\ProbOf}[1]{\ensuremath{\Prob\left[#1\right]}\xspace}
\newcommand{\RevProbOf}[1]{\ensuremath{\Prob^*\left[#1\right]}\xspace}

Null Hypothesis Significance Testing (NHST)~\cite{fisher1925statistical} is a statistical method commonly used to evaluate whether a null hypothesis $H_0$ can be rejected and an alternative hypothesis $H_1$ be considered plausible in its place. Recent reviews of the method include, for instance, the work by Lehmann and Romano~\cite{lehmann2005testing}. Null hypothesis significance testing has often been criticized, in its own right as well as for how scientists have fallen for a range of fallacies~\cite{Nickerson2000null}. Problems with the null hypothesis significance testing have led to a stronger endorsement of estimation theory, that is, relying more on effect sizes and their confidence intervals~\processifversion{DocumentVersionTR}{\cite{colegrave2003confidence,cumming2013understanding,cumming2014new,hoekstra2014robust}}\processifversion{DocumentVersionConference}{\cite{cumming2013understanding}}.

In broad strokes, the method computes a $p$-value, that is, the probability of how likely it us to make observations as extreme or more extreme than the observations made $D$, \emph{assuming the null hypothesis $H_0$ to be true}. Hence, the $p$-value is a conditional probability:
\[ p := \ProbOf{D~|~H_0}. \]

Clearly, the $p$-value does not tell us how likely the alternative hypothesis is after having made the observations of the study. However, misinterpretations of the $p$-value often lead to confusion.

\subsection{Bayes' Law}

Naturally, we are interested in establishing how likely the hypotheses of a study are, \emph{a posteriori} of its observations.
This can be achieved by consulting Bayes' Law. We shall write in a form conducive to our subsequent argument, as promoted by Colquhoun~\cite{colquhoun2017reproducibility}:

\[  \underbrace{\frac{\ProbOf{H_1~|~D}}{\ProbOf{H_0~|~D}}}_{\text{\rm a posteriori odds}}  = 
\underbrace{\frac{\ProbOf{D~|~H_1}}{\ProbOf{D~|~H_0}}}_{\text{\rm likelihood ratio}} \times \underbrace{\frac{H_1}{H_0}}_{\text{\rm prior odds}}. \]
As we can see, the $p$-value $\ProbOf{D~|~H_0}$ is the denominator of the likelihood ratio. The corresponding numerator \ProbOf{D~|~H_1} indicates the probability of observations as extreme or more extreme than the observations made, assuming the alternative hypothesis $H_1$ being true, the statistical power of the test. In general, we subscribe to the Bayesian interpretation of Bayes' law, in which a probability quantifies the belief in a hypothesis.

\subsubsection{Probability Interpretations.}
We distinguish two interpretations for probabilities around the $p$-value, which were discussed by Colquhoun~\cite{colquhoun2017reproducibility}. First, we have the $p$-less-than interpretation, which considers observations as extreme or more extreme as the ones obtained. Under this interpretation the likelihood ratio considered above is computed by the statistical power divided by the $p$-value itself.

Second, in the $p$-equals interpretation, we evaluate the likelihoods exactly at the probability of the observations made. Hence, we consider the probability exactly at the test statistic obtained under the alternative hypothesis divided by the probability at the test statistic under the null hypothesis.

These two interpretations yield different results for the likelihood ratio and related evaluations, where typically, by $p$-value, the $p$-equals interpretation will yield a greater false positive risk than the $p$-less-than interpretation.

\extend{Create graph, e.g., for $t$-test to illustrate the difference between the two interpretations.}

\subsubsection{Positive Predictive Value (PPV) and False Positive Risk (FPR).}

The first quantity we are interested in is the positive predictive value (PPV) $\ProbOf{H_1~|~D}$ and its partner, the false positive risk (FPR) $\ProbOf{H_0~|~D}$.

\[ \Prob_{\const{PPV}} := \ProbOf{H_1~|~D} = \frac{\ProbOf{H_1} \ProbOf{D~|~H_1}}{\ProbOf{H_1} \ProbOf{D~|~H_1} + (1 - \ProbOf{H_1}) \ProbOf{D~|~H_0}} \]

\[ \Prob_{\const{FPR}} := \ProbOf{H_0~|~D} = \frac{(1 - \ProbOf{H_1}) \ProbOf{D~|~H_0}}{\ProbOf{H_1} \ProbOf{D~|~H_1} + (1 - \ProbOf{H_1}) \ProbOf{D~|~H_0}} \]

\subsubsection{Integrating Bias into Estimations.}

Ioannidis~\cite{ioannidis2005most} defined \emph{bias} as ``the combination of various design, data, analysis, and presentation factors that tend to produce research findings when they should not be produced.'' He quantified it as $u$, the proportion of tests that would not have been findings. We use Ioannidis' estimation for PPV under the influences of bias, where $R$ constitutes the prior odds. We express the formula in the probability terminology introduced above.

\[ \Prob_{\const{PPV}, u} := \frac{\ProbOf{D~|~H_1}R + u(1 - \ProbOf{D~|~H_1})R}{R + \ProbOf{D~|~H_0} - \ProbOf{D~|~H_1}R + u - u\ProbOf{D~|~H_0} + u(1 - \ProbOf{D~|~H_1})R} \]

\subsubsection{Likelihood Ratio (LR).}
The likelihood ratio measures the strength of evidence independent from priors and is given by
\[ \vari{LR} := \frac{\ProbOf{D~|~H_1}}{\ProbOf{D~|~H_0}}. \]

\subsubsection{Reverse Bayesian Argument.}
The reverse Bayesian argument aims at computing the prior necessary to achieve a desired fixed false positive risk $\Prob^*_{\const{FPR}}$. This method was originally proposed by Matthews~\cite{matthews2001should} and endorsed by Colquhoun~\cite{colquhoun2017reproducibility}.
We compute the reverse Bayesian prior $\Prob_{\const{RBP}} = \RevProbOf{H_1}$ as follows:

\[ \Prob_{\const{RBP}} := \RevProbOf{H_1} =   \frac{\ProbOf{D~|~H_0} (1 - \Prob^*_{\const{FPR}})}{\ProbOf{D~|~H_0} (1 - \Prob^*_{\const{FPR}}) + \ProbOf{D~|~H_1} \Prob^*_{\const{FPR}}} \]

\extend{Include advantages and disadvantages of the different metrics in Table for TR.}


\section{Related Works}
\label{sec:related}

\subsection{Strength of Evidence in Other Fields}

That most published findings are likely false was prominently discussed by Ioannidis~\cite{ioannidis2005most} in general terms and applicable to any field. That study focused on the estimation of the positive predictive value and offered estimation formula and thresholds for the inclusion of study biases. We adopt Ioannidis estimation methods in this study, as well.

Other studies considered false positive reporting probability and the impact of replications, where we cite two examples~\cite{wacholder2004assessing,moonesinghe2007most} as context.

This study is also related to approaches to estimate likelihood ratios and reverse Bayesian prior. Colquoun~\cite{colquhoun2014investigation,colquhoun2017reproducibility} offered such estimations in the discussion of the $p$-value null hypothesis significance testing and reproducibility. He promoted the use of the reverse Bayesian prior, where the use of the reverse Bayesian argument was originally proposed by Matthews~\cite{matthews2001should}.

\subsection{Appraisals of Cyber Security User Studies}

Cyber security user studies have received a range of appraisals in recent years.
A first step was made by Coopamootoo and Gro{\ss}~\cite{SLR2017}, who conducted a systematic literature review of cyber security user studies from relevant venues in the years 2006--2016.
That study included a coding of nine reporting completeness indicators~\cite{coopamootoo2017CIcodebook}, giving a qualitative overview of scientific reporting. The authors expanded upon said completeness indicators in a design and reporting toolkit~\cite{coopamootoo2017cyber}.

Subsequently, Gro{\ss}~\cite{Gross2019,Gross2019a} built on the same SLR sample to establish the fidelity of statistical reporting.
This study re-computed $p$-values from published test statistics and parameters to find quantitative and decision errors.

Gro{\ss}~\cite{Gross2020,Gross2020a} turned to estimating effect sizes and their confidence intervals of statistics tests in the papers obtained from the 2017 Coopamootoo-Gro{\ss}-SLR. That work further established simulations of statistical power vis-{\`a}-vis specified effect size thresholds, highlighting a power failure. This power failure was deduced by comparison to typical expert recommendations of power thresholds, but did not quantify the actual strength of evidence. In addition, the study showed the presence of statistical biases, such as the publication bias or the winner's curse.

This paper, however, takes a different tack. Though based on the same SLR sample as previous work and considering the same statistical tests as extracted by Gro{\ss}~\cite{Gross2020}, this work focuses on strength of evidence and estimates false positive risk, likelihood ratio and reverse Bayesian prior for the statistical tests reported in papers of the SLR sample.

\section{Aims}
\label{sec:aims}

\subsection{Strength of Evidence}
\begin{researchquestion}[Strength of Evidence]
What is the distribution of the strength of evidence in the field of cyber security user studies?
 \end{researchquestion}
 This study aims at investigating the strength of evidence measured in an empirical evaluation as
 \begin{inparaenum}[(a)] 
    \item Positive Predictive Value (PPV) and its complement False Positive Risk (FPR) (based on assumed priors), 
    \item Likelihood Ratio (LR), and
    \item Reverse-Bayesian Prior (RBP) for a fixed false positive probability.
 \end{inparaenum}

\subsection{Attention to Strength of Evidence}
\begin{researchquestion}[Attention to Strength of Evidence]
To what extent is the attention studies receive (in citations) related to their strength of evidence?
\end{researchquestion}
We investigate this question by evaluating the correlation between the strength of evidence and the number of citations, measured as Average Citations Per Annum (ACPA). We do so by testing the following hypotheses.
\begin{compactdesc}
  \item[$H_{\const{C}, 0}$:] There is no correlation between the strength of evidence (in terms of Reverse Bayesian Prior) and the measured ACPA.
  \item[$H_{\const{C}, 1}$:] The strength of evidence (in terms of Reverse Bayesian Prior) and the measured ACPA are correlated.
\end{compactdesc}

\section{Method}
\label{sec:method}

This study is pre-registered in the Open Science Framework\footnote{\url{https://osf.io/7gv6h/?view_only=222af0e071a94b2482bb8ccb3e1eaa4c}}. 
The statistical estimations are computed with $\textsf{R}$. Statistical tests are computed at a significance level $\alpha = .05$.

\subsection{Sample}
We obtained the sample for this study from prior work. Its original foundation is the 2016/17 Systematic Literature Review (SLR) by Coopamootoo and Gro{\ss}~\cite{SLR2017}. The characteristics of the SLR are also documented by Gro{\ss}~\cite{Gross2019}.

In addition, this work is based on the effect size extraction achieved by Gro{\ss}~\cite{Gross2020}. It contains  $t$-, $\chi^2$-, $r$-, one-way $F$-tests, and $Z$-tests. The effect sizes were extracted based on an automated evaluation by \textsf{statcheck}~\processifversion{DocumentVersionTR}{\cite{epskamp2014statcheck,nuijten2016prevalence,nuijten2017validity}}\processifversion{DocumentVersionConference}{\cite{nuijten2017validity}} as well as manual coding.

\subsection{Procedure}

The study proceeded in the following fashion.
\begin{compactenum}[1.]
  \item We have taken as input a table of coded $p$--values, standardized effect sizes, sample sizes, citations, and year of publication.
  \item We set as parameters 
  \begin{compactenum}[(a)]
    \item effect size thresholds valued at 
      \begin{inparaenum}[(i)]
        \item small: $d = 0.3$, 
        \item medium: $d = 0.5$, 
        \item large: $d = 0.8$;
      \end{inparaenum}
    \item biases valued at 
      \begin{inparaenum}[(i)]
        \item theoretical minimum: $u = .0$,
        \item well-run RCT: $u = .2$,
        \item weak RCT: $u = .3$,
        \item biased study: $u = .8$;
      \end{inparaenum}
    \item priors valued at 
      \begin{inparaenum}[(i)]
        \item ``confirmatory,'' one-to-one ratio of relations being true, $\vari{prior} = .5$,
        \item ``intermediate,'' one-to-four ratio of relations being true, $\vari{prior} = .2$,
        \item ``exploratory,'' one-to-nine ratio of relations being true, $\vari{prior} = .1$;
      \end{inparaenum}
   \end{compactenum}
  \item We established a statistical power simulation based on the parametrized effect size thresholds, incl. a variant with multiple-comparison corrections.
  \item Based on the actual $p$-values in the studies as well as their multiple-comparison adjusted variants, we computed the positive-predictive value (PPV), the false positive risk (FPR), and the reverse Bayesian prior (RBP).
  \item We computed the likelihood ratio in the $p$-less-than interpretation from power and $p$-values.
  \item To assess the relation between attention studies are receiving and their strength of evidence, we computed a hierarchical linear model on strength of evidence by Average Citations per Annum (ACPA), using the study ID as random-effect variable.
  \item Finally, we established the graphs by study and test ID. 
\end{compactenum}


\section{Results}
\label{sec:results}

\subsection{Sample}
The refined sample shown in Table~\ref{tab:sample} includes studies with extractable effect sizes as determined by Gro{\ss}~\cite{Gross2020}.

\begin{table}
\centering
\caption{Sample refinement on SLR papers (as reported in~\cite{Gross2020})}
\label{tab:sample}
\begin{tabular}{lrr}
\toprule
\textbf{Phase}  & Excluded & Retained\\
\midrule
\textit{Source SLR}~\cite{SLR2017}\\
\quad Search results (Google Scholar) &    ---         & 1157\\
\quad Inclusion/Exclusion             & 1011 & 146\\
\midrule
\textit{Refinement in this study}\\
\quad Empirical studies      & 2 & 144\\
\quad With sample sizes      & 21 & 123\\
\quad With extractable tests & 69 & 54\\
\bottomrule
\end{tabular}
\end{table}

The sample diplayed in Table~\ref{tab:sampleTests} includes statistical tests and their effect sizes that could be extracted from the papers in the sample. The final sample includes $431$ statistical tests and their effect sizes.

\newcommand{\sampleTestsRefined}{
\begin{table}[ht]
\centering
\caption{Sample refinement on extracted effect sizes (adapted from~\cite{Gross2020})} 
\label{tab:sampleTests}
\begingroup\footnotesize
\begin{tabular}{lrr}
  \toprule
\textbf{Phase} & Excluded & Retained \\ 
  \midrule
Total effects extracted & 0 & 650 \\ 
   \midrule
 \quad \textsf{statcheck} automated extraction &  &  252 \\
 \quad Test statistic manual coding &   &  89 \\
 \quad Means \& SD manual coding    &   &  309 \\
 \midrule
 \textit{Refinement in Gro{\ss}~\cite{Gross2020}}\\
Violated reporting and assumptions & 219 & 431 \\ 
   \bottomrule
\end{tabular}
\endgroup
\end{table}
}

\sampleTestsRefined

\subsection{False Positive Risk}
\label{sec:fprp}
We examine the distribution of false positive risk (or False Positive Reporting Probability, FPRP), that is, the \emph{a posteriori} probability that the alternative hypotheses of statistical tests are false.

\subsubsection{Fixed Bias and Prior}
We first consider the simulation for a weak random-controlled trial (bias $b=.3$) and an intermediate prior (one in four investigated relations being true, prior = 0.2) presented in Figure~\ref{fig:fprp_WRCT}, the statistics corrected for family-wise multiple comparisons. This plot corresponds to the centre one of subsequent Figure~\ref{fig:fprpDensityMCCBiasByPrior}. Let us unpack this plot step by step as an orientation.

\scatterFPRPintermediateWRCT

The figure depicts the false positive risk as a scatter plot depending of the underlying sample size of the corresponding study.
As we would expect, the false positive risk generally decreases with an increase of the sample size due to the increasing statistical power---up to a point. 

The three colors represent assumed effect size thresholds in the population---small \textcolor{viriyellow}{\CIRCLE}: $d = 0.2$, medium \textcolor{virigreen}{\CIRCLE}: $d = 0.5$, and large \textcolor{viriviolet}{\CIRCLE}: $d = 0.8$. Naturally, the greater the effect size in the population, the less the false positive risk. The lines drawn constitute a \emph{Loess} smoothing of the corresponding scatter plot. We observe that---assuming a setting of a weak random-controlled trials and priors of $0.2$---the false positive risk is at least $60\%$, irrespective of the size of the effect sizes thresholds assumed in the population or the additionally excerted power.

On the right-hand margin of the graph, we included the density of the false positive risks for the SLR sample at hand.
Here, we see that with respect to smaller effect size thresholds, there are clusters of greater false positive risk.

Considering the expectation on number of false positive results, we obtain that of the $142$ MCC-corrected positive reports out of $444$ tests under investigation only an expected $48$ statistically significant results are true in reality, $34\%$. Conversely, $94$ of the positive reports are likely false positives ($66\%$).
\newcommand{\probFalsePositives}{$66\%$}

\subsubsection{Variable Bias and Prior.}
Having evaluated a single parameter set of the simulation, we are now in the position to consider the effects of the parameters bias and prior being varied.


Figure~\ref{fig:fprpDensityMCCBiasByPrior} displays false positive risk graphs for three cases for bias and prior, respectively. 
The graphs are based statistics with family-wise multiple-comparison corrections.
\begin{DocumentVersionTR}
For bias, we have
\begin{inparaenum}[(i)]
  \item Biased Study (bias $u = .8$): Such a study contains a number of biases that affect the study, assumed to be independent of the study results.
  \item Weak Random-Controlled Trial (bias $u = .3$): A random-controlled trial in which biases occured in spite of the intention of a controlled trial.
  \item Well-run Random-Controlled Trial (bias $u = .2$): A random-controlled trial in which biases were minimized, even if not completely eliminated.
\end{inparaenum}
For priors, we consider different levels of confirmatoriness
\begin{inparaenum}[(i)]
  \item Exploratory study ($\vari{prior} = .1$): investigates a larger number of relations, of which only one in nine are true.
  \item Intermediate study ($\vari{prior} = .2$): investigates a moderate number of relations, in which one in four are true.
  \item Confirmatory study ($\vari{prior} = .5$): focuses on confirming relations, with a ratio of one-to-one being true.
\end{inparaenum}
\end{DocumentVersionTR}

\fprpDensityMCCBiasByPrior

First, we observe that the amount of bias present in the study depresses the capacity to reduce the false positive risk with additional power. For biased studies (bias $u = .8$), power in terms of increased sample size, is inconsequential to eliminate false positive risk.
Second, different degrees of confirmatoriness (varying priors) offset the false positive risk, the more confirmatory a study is, the greater a prior a test operates against the less the false positive risk.
Overall, we find that only confirmatory studies ($\vari{prior} = 0.5$) that are either run as well-run RCT ($u = .2$) or weak RCT ($u = .3$) yield a false positive risk less than $50 \%$. \processifversion{DocumentVersionConference}{Appendix~\ref{sec:heatmap} discusses the capacity to gain knowledge with a heatmap.}

\begin{DocumentVersionTR}
\subsubsection{Capacity to Gain Knowledge.}

In Figure~\ref{fig:ppvHeatmap}, we take a different perspective from Figure~\ref{fig:fprpDensityMCCBiasByPrior}. 
Here we consider the simulation of \emph{a posteriori} probability of the alternative hypotheses, as a heatmap faceted by bias and effect size threshold.
This figure is organized by study and conveys the maximal Positive Predictive Value (PPV) the study can achieve with any test conducted. It is ordered by maximal PPV.

\ppvHeatmap

Figure~\ref{fig:ppvHeatmap} shows as yellow and light green a probability that positive reports are more likely than $50\%$. 
The figure can be read as follows: Assuming a certain effect size in the population (say medium, $d = 0.5$), assuming that studies in question were conducted as weakly-run RCT, and assuming a great prior close to $.5$, we would select the centre facet of the plot.
Here we would see that only one fifth ($22\%$) of the studies reached a PPV equal or greater to $50\%$ and only conditioned on a prior close to $.5$.
\end{DocumentVersionTR}

\subsection{Strength Evidence of Positive Reports}
In the following, we focus on positive reports, that is, relations studies reported as statistically significant.

\subsubsection{Likelihood Ratio.}

In Figure~\ref{fig:scatterLR}, we consider the strength of evidence of positive reports quantified as likelihood ratio, that is, the ratio of the report being true in reality by the report being false in reality. This likelihood ratio is independent of the presumed prior.
The likelihood ratio is depicted in a scatter plot by the sample size of the corresponding tests, the left Figure~\ref{fig:scatterLR} containing the original positive reports, the right Figure~\ref{fig:scatterLRMCC} showing the same reports under appropriate family-wise multiple-comparison corrections. Studies are marked by different colors.

\lrScatterCombined

We find in the MCC-corrected Figure~\ref{fig:scatterLRMCC} that most statistically significant results are clustered below a likelihood ratio of $\vari{LR} = 25$. The sample has a median likelihood ratio of $5.21$. That is, for half the positive reports it is approximatly less than five times as likely for the report being true to it being false.
Remarkably, a number of studies yield positive reports with a considerably greater likelinood ratio, that is, a considerable strength of evidence to the positive report made.

\newcommand{\medianLR}{65.38}
\newcommand{\medianLRMCC}{5.21}

\subsubsection{Reverse Bayesian Prior}

In Figure~\ref{fig:rbpScatterCombined}, we evaluate the Reverse Bayesian Prior, that is, prior that one would have needed \emph{a priori} to reach a fixed \emph{a posteriori} false positive risk of $5\%$. In general, one would consider a required prior of greater than $50\%$ as unreasonable, especially in a field that contains copious amounts of exploratory studies. For this evaluation, we fix the effect size threshold to medium ($d = 0.5$) and the bias to $.3$. The figure contains the statistics without and with adjustment for Multiple Comparison Corrections (MCC).

\rbpScatterCombined

Without multiple-comparison corrections, Figure~\ref{fig:scatterRPB} shows ($9\%$) of the $204$ positive reports yield a reverse Bayesian prior greater than $50\%$.
Considerng the upper half of Figure~\ref{fig:scatterRBPMCC}, we observe that approximately half ($48\%$) of the $142$ positive reports left after an adjustment for multiple-comparison corrections show a reverse Bayesian prior greater than or equal to $50\%$.
The closer a positive report is positioned to a prior of zero, the stronger is the strength of evidence speaking for the report, and the more likely the report will advance knowledge.

\subsection{Relation to Citations}
We conducted a hierarchical linear model with the paper ID as random-effect factor to investigate the correlation between the reverse Bayesian prior and a metric of the citations, the average citations per annum. The model is based on $n = 204$ statistically significant reports (after correction for multiple comparison corrections) from 25 papers.
The model did not converge, that is, we could not confirm a correlation between these variables.
Hence, we failed to reject the null hypothesis $H_{\const{C}, 0}$. \processifversion{DocumentVersionConference}{Appendix~\ref{sec:scatterACPA} contains a scatter plot of the variables.}

\begin{DocumentVersionTR}
\scatterACPARBP

Figure~\ref{fig:scatterACPARBP} depicts the association between the reverse Bayesian prior and the citation metric.
It is apparent in th graph that there is no visble correlation between the variables under investigation.
\end{DocumentVersionTR}


\section{Discussion}
\todoB{Difference between probability and likelihood. They are not equivalent.}

\subsection{Why most results of socio-technical user studies are false}

Let us start with the eponymous question of this paper.
While this is generally well understood for any field of science~\cite{ioannidis2005most}, we quantified and contextualized the impact of the adage based on a systematically drawn empirical sample of cyber security user studies.
Based on the sample at hand, we can estimate that for moderate biases ($u = .3$) of weak random controlled trials---arguably already an optimistic assumption---moderate priors ($.2$) and medium effect sizes in the population ($d = .5$) as done in Figure~\ref{fig:fprp_WRCT} of Section~\ref{sec:fprp}, an expected \emph{two thirds} of reported statistically significant results are likely false positives  (\probFalsePositives).

This should give us pause. Well-run random-controlled trials are rare in the field, we often see studies that incur a range of biases in their sampling, weakly randomized or unblinded experiment design, or failure to meet statistical assumptions. Exploratory studies seem to be relatively frequent, often investigating multiple relations, few of which are true in reality. Hence, in many cases we would expect false positives even more numerous.

\subsection{What to take away from the simulations}

The simulation parametrized by bias and prior in Figure~\ref{fig:fprpDensityMCCBiasByPrior} yields a number of important take-aways. Typically, when the question arises what to do to gain more certainty in statistical tests, the answer is having adequate power. Gro{\ss}~\cite{Gross2020} argued the point of power failure of the field in STAST'2020. While the requirement of adequate power---and, thereby, sufficient sample sizes---is a valid point, the simulation teaches us that statistical power only takes us that far. Eventually, the impact of power on the false positive risk is lower-bounded.

Clearly, we cannot change the impact of the prior probability, the \emph{a priori} likelihood of an investigated relation being true. However, it is important to raise the awareness of its effect. Highly exploratory studies, which investigate a large number of relations of which only a few are likely true in reality incur a considerably greater false positive risk than highly confirmatory studies.
While it is a natural sentiment not to take exploratory studies at face value, it is worthwhile to take into account quantitatively the impact of lower priors.

Finally, we find that incurred bias is a crucial factor to consider. As observed in Figure~\ref{fig:fprpDensityMCCBiasByPrior}, increasing bias depresses the capability of studies irrespective of their confirmatoriness or power to yield new knowledge.
Hence, minimizing biases, for instance by conducting systematically sampled, well-run, double-blind random controlled trials, is a premier way to bringing down the false positive risk.
Clearly, these results also stress the importance of well-run replications.

We encourage readers to use the simulation in Figure~\ref{fig:fprpDensityMCCBiasByPrior} as a way of orientating themselves for the appraisal of an individual study or a field. Ask yourself:
\begin{compactenum}
  \item What is the likely ratio of true relations to investigated relations (yielding an estimate of the prior)?
  \item How well are the studies under considerations run? How well do they minimize biases?
  \item What is the likely magnitude of the effects investigated in the population?
  \item What is the effective sample size used for statistical tests.
\end{compactenum}
Based on the former two questions select the facet in Figure~\ref{fig:fprpDensityMCCBiasByPrior} best reflecting the appraisal.
Based on the latter two questions select the appropriate effect size threshold and intersection with the sample size.
This approach offers a rough approximation of the false positive risk to be expected. 
Of course, Ioannidis' simulation equations~\cite{ioannidis2005most} can be used directly to compute the same results.

\begin{DocumentVersionTR}
It is important to realize, as depicted in the heatmap of Figure~\ref{fig:ppvHeatmap}, that depending on the parameters chosen only few studies yield an \emph{a posteriori} probability of the alternative hypothesis greater than $50\%$.
\end{DocumentVersionTR}

\subsection{The impact of strength of evidence}

We observed in Figure~\ref{fig:scatterLRMCC} that positive reports under appropriate multiple-comparison corrections were largely clustered around low likelihood ratios, with a median $\vari{LR} = \medianLRMCC$ against a medium effect size. That means that in the median---irrespective of the prior---a true positive result is five times as likely than a false positive result.

We saw as well that a few studies achieve considerably greater likelihood ratios. We believe that the community would benefit from valuing studies that achieve a great strength of evidence. While independence of the unknown prior makes the likelihood ratio appealing as a metric, it does not frame the results in absolute terms.

Here, the reverse bayesian prior comes into play, that is, what prior probability is required to achieve a fixed false positive risk. In Figure~\ref{fig:scatterRBPMCC}, we presented the reverse bayesian prior of positive reports by the sample sized used.
Nearly half of the tests needed a prior greater than $50\%$ to yield a false positive risk of $5\%$.
Hence we would have been implausibly certain of the relation \emph{a priori} of the study.

\subsection{Attention to strength of evidence}

Our evaluation of the correlation between average citations per annum and strength of evidence were rooted in the aim to estimate whether authors take strength of evidence into account when citing papers.
As we could not show a correlation being present, we have not found evidence of an association.
Hence, we would venture the opinion that the strength of evidence is not a strong factor in deciding to cite another study.

\subsection{Limitations}
\label{sec:limitations}

\subsubsection{Generalizability.}
The study is founded on an existing sample of a systematic literature review (SLR) of the years 2006--2016.
While that sample yields challenges in terms of having been obtained on Google Scholar been restricted to specific venues. Hence, its generalizability to the entire field of socio-technical security user studies is limited. 
However, we believe that the distribution we observe in strength of evidence is not untypical of the field at large.

\subsubsection{Probability Interpretation.} 
The computations in this work as based on the $p$-less-than interpretation of test probabilities. That is, the likelihood ratio, for instance, is computed as the ratio of statistical power ($1-\beta$) by $p$-value itself. Colquhoun~\cite{colquhoun2017reproducibility} made a convincing case that the $p$-equals interpretation is more appropriate for evaluating the strength of evidence of a single test.
The $p$-equals interpretation puts into relation the ordinate of the probability distributions of the null and alternate hypotheses.

At the same time, in the studies of the SLR sample we are often missing the data too compute the $p$-equals interpretation reliably. Hence, even if the preregistration for this study asked for the $p$-equals interpretation as preferred metric, we established the simulations on the $p$-less-than interpretation. For a large number of tests evaluated, this still gives us a conservative estimate:
\begin{inparaenum}[(i)]
  \item Based on Colquhoun's comparative simulations of both interpretation~\cite{colquhoun2017reproducibility}ß, we find that the $p$-equals interpretation yields a greater false positive risk than the $p$-less-than interpretation. Hence, if anything, we are underestimating the false positive probability of statistical tests.
  \item Electing the $p$-less-than interpretation, we obtain a larger sample of tests with likelihood-ratio estimations and, thereby, offer a more reliable sample to estimate the properties of the field.
\end{inparaenum}

\section{Conclusions}

We showed, based on an systematically drawn empirical sample, that most published findings in socio-technical security user studies are false. While this concern has been stated generally for studies of any field~\cite{ioannidis2005most}, we are the first to quantify the strength of evidence and expected number of false positive reports based on an empirical foundation in socio-technical security user studies.

While our simulations depend on external parameters 
\begin{inparaenum}[(i)]
  \item bias, 
  \item prior, and 
  \item effect size threshold in the population,
\end{inparaenum}
which are inherently difficult to estimate accurately, we offer the readers a multi-faceted view of parameter combinations and their consequences. For instance, for the false positive risk, we can estimate make our own assumption on bias and prior present and then consider the consequences of setup.

We also offer investigations of positive reports, that is, results stated as statistically significant, showing the distribution of likelihood ratios and reverse Bayesian prior. These simulations establish an appraisal of the strength of evidence while being independent from an unknown prior.

Our results raise caution about believing positive reports out of hand and sensitize towards appraising the strength of evidence found in studies under consideration. Our work makes the case to drive biases down as a factor of experiment design and execution too often neglected in this field. Finally, we believe that we can see that strength of evidence is receiving little attention as a factor in the decision to cite publications, raising the awareness of including the strength-of-evidence consideration into the reporting recommendations for authors and reviewing recommendations for gatekeepers.

\section*{Acknowledgment}

Early aspects of this study were in parts funded by the UK Research Institute in the Science of Cyber Security (RISCS) under a National Cyber Security Centre (NCSC) grant on ``Pathways to Enhancing Evidence-Based Research Methods for Cyber Security.'' Thomas Gro{\ss} was funded by the \CASCAde.

\balance

\bibliographystyle{splncs04}
\bibliography{methods_resources,slr_user_studies,stat_check,slr,r_tools,slr_meta,guides}

\begin{appendix} 
\begin{DocumentVersionConference}
\section{Capacity to Gain Knowledge}
\label{sec:heatmap}

In Figure~\ref{fig:ppvHeatmap}, we take a different perspective from Figure~\ref{fig:fprpDensityMCCBiasByPrior}. 
Here we consider the simulation of \emph{a posteriori} probability of the alternative hypotheses, as a heatmap faceted by bias and effect size threshold.
This figure is organized by study and conveys the maximal Positive Predictive Value (PPV) the study can achieve with any test conducted. It is ordered by maximal PPV.

\ppvHeatmap

Figure~\ref{fig:ppvHeatmap} shows as yellow and light green a probability that positive reports are more likely than $50\%$. 
The figure can be read as follows: Assuming a certain effect size in the population (say medium, $d = 0.5$), assuming that studies in question were conducted as weakly-run RCT, and assuming a great prior close to $.5$, we would select the centre facet of the plot.
Here we would see that only one fifth ($22\%$) of the studies reached a PPV equal or greater to $50\%$ and only conditioned on a prior close to $.5$. 
\end{DocumentVersionConference}

\begin{DocumentVersionConference}
\section{Association Between RBP and ACPA}
\label{sec:scatterACPA}
\scatterACPARBP

Figure~\ref{fig:scatterACPARBP} depicts the association between the reverse Bayesian prior and the citation metric.
It is apparent in th graph that there is no visble correlation between the variables under investigation.
\end{DocumentVersionConference}
\end{appendix}

\end{document}